\pdfoutput=1
\documentclass{article}
\usepackage[sort,authoryear]{natbib}
\usepackage{mathtools}
\usepackage{amssymb}
\usepackage{amsthm}
\usepackage{enumitem}
\usepackage[margin=1in]{geometry}
\usepackage{graphicx}
\usepackage{subcaption}
\usepackage[utf8]{inputenc}
\usepackage[english]{babel}
\usepackage{xcolor}
\usepackage[colorlinks = true, citecolor = blue, urlcolor = blue]{hyperref}
\usepackage{indentfirst}
\usepackage{apptools}
\usepackage{cleveref}
\usepackage{algorithm}
\usepackage[noend]{algpseudocode}

\usepackage{fancyhdr}
\graphicspath{{fig/}}

\AtAppendix{\counterwithin{lemma}{section}}
\AtAppendix{\counterwithin{proposition}{section}}

\DeclarePairedDelimiter\floor{\lfloor}{\rfloor}

\DeclareMathOperator*{\argmin}{argmin}

\newtheorem{lemma}{Lemma} 
\newtheorem{proposition}{Proposition}
\newtheorem{theorem}{Theorem}
\newtheorem{corollary}{Corollary}
\newtheorem{assumption}{Assumption}

\newcommand{\Cov}{\textup{Cov}}
\newcommand{\Var}{\textup{Var}}
\newcommand{\Proj}{\textup{Proj}}
\newcommand{\Span}{\textup{Span}}
\newcommand{\Diag}{\textup{Diag}}
\newcommand{\tr}{\textup{Tr}}
\newcommand{\rank}{\textup{rank}}

\newcommand{\0}{\boldsymbol{0}}
\newcommand{\1}{\boldsymbol{1}}
\newcommand{\E}{\boldsymbol{E}}
\newcommand{\A}{\boldsymbol{A}}
\newcommand{\B}{\boldsymbol{B}}
\newcommand{\C}{\boldsymbol{C}}
\newcommand{\D}{\boldsymbol{D}}
\newcommand{\N}{\boldsymbol{N}}
\newcommand{\M}{\boldsymbol{M}}
\newcommand{\I}{\boldsymbol{I}}
\newcommand{\X}{\boldsymbol{X}}
\newcommand{\W}{\boldsymbol{W}}
\newcommand{\V}{\boldsymbol{V}}
\newcommand{\U}{\boldsymbol{U}}
\newcommand{\bSigma}{\boldsymbol{\Sigma}}
\newcommand{\bOmega}{\boldsymbol{\Omega}}
\newcommand{\R}{\mathbb{R}}
\renewcommand{\vec}{\textup{vec}}
\renewcommand{\H}{\boldsymbol{H}}
\renewcommand{\S}{\mathbb{S}}
\renewcommand{\L}{\boldsymbol{L}}

\providecommand{\keywords}[1]
{
  \textbf{\textit{Keywords---}} #1
}

\newcommand{\footremember}[2]{%
    \footnote{#2}
    \newcounter{#1}
    \setcounter{#1}{\value{footnote}}%
}
\newcommand{\footrecall}[1]{%
    \footnotemark[\value{#1}]%
} 

\definecolor{revision}{rgb}{0.0, 0.0, 0.0}
\definecolor{revision2}{rgb}{0.0, 0.0, 0.0}

\title{A Statistical View of Column Subset Selection}
\author{Anav Sood \footremember{stats}{Department of Statistics, Stanford University, USA} \footremember{address}{Address for correspondence: Sequoia Hall, 390 Jane Stanford Way, Stanford, CA 94305, USA.
 Email: \href{mailto:anavsood@stanford.edu}{anavsood@stanford.edu}.} \and Trevor Hastie \footrecall{stats}{}}
\date{October 7th, 2024}
\begin{document}

\maketitle

\begin{abstract}
We consider the problem of selecting a small subset of representative variables from a large dataset. In the computer science literature, this dimensionality reduction problem is typically formalized as Column Subset Selection (CSS). Meanwhile, the typical statistical formalization is to find an information-maximizing set of Principal Variables. This paper shows that these two approaches are equivalent, and moreover, both can be viewed as maximum likelihood estimation within a certain semi-parametric model. \textcolor{revision2}{Within this model, we establish suitable conditions under which the CSS estimate is consistent in high dimensions, specifically in the proportional asymptotic regime where the number of variables over the sample size converges to a constant.} Using these connections, we show how to efficiently (1) perform CSS using only summary statistics from the original dataset; (2) perform CSS in the presence of missing and/or censored data; and (3) select the subset size for CSS in a hypothesis testing framework. 
\end{abstract}

\keywords{column subset selection, interpretable dimensionality reduction, principal components analysis, principal variables, probabilistic modeling, high-dimensional statistics}

\section{Introduction}
\label{sec:intro}
In modern data applications, it is common to reduce a large dataset into a smaller one, either for storage efficiency or ease of downstream analysis. As methods become more complex, it is increasingly \textcolor{revision2}{desirable} for this reduced dataset to be interpretable. 

In this paper, we consider the dimensionality reduction task of selecting a subset of variables that is most representative of the entire dataset. We establish an equivalence between two popular methods for this task, \emph{Column Subset Selection (CSS)} and \emph{Principal Variables}, and provide a generative view of both methods. 

\subsection{Motivation}

The textbook approach to dimensionality reduction is Principal Components Analysis (PCA) \citep{Jolliffe2002}. PCA projects the observed data onto a small set of variance-maximizing axes. Ultimately, the reduced dataset consists of linear combinations of the original variables, and these resulting linear combinations are referred to as principal components. 

The diverse range of perspectives on PCA has fostered many useful theoretical and methodological insights. For example, although PCA’s objective is typically written in terms of unit-level data, it can also be written solely in terms of the data’s covariance \citep{Pearson}. This covariance characterization turns out to provide a more fundamental view of the problem, and \cite{Okamoto} strengthen PCA's theoretical foundation by using it to establish the method's simultaneous optimality for a large class of objectives. Another example is probabilistic PCA (PPCA), a generative variant of PCA developed in \cite{Tipping}. PPCA broadens PCA’s scope by enabling the application of standard statistical tools (e.g. likelihood ratio testing, expectation-maximization, Bayesian inference).

Unfortunately, the linear combinations that PCA outputs are typically dense and difficult to interpret. The field of interpretable dimensionality reduction has emerged largely in response to this phenomenon, and it is spearheaded by methods that output sparse linear combinations of the original variables  \citep{d'Aspremont, Jolliffe2003, Witten, Zou}. These sparse variants of PCA provide more interpretable results by simplifying the relationship between the reduced and original dataset. Methods like CSS and Principal Variables that select a subset of the original variables simplify this relationship as much as possible and, consequently, can be viewed as PCA’s sparsest and most interpretable alternatives. 

This paper aims to provide for CSS the same diverse set of viewpoints that are available for PCA. In particular, we show that Principal Variables solves the exact same problem as CSS, but is characterized in terms of covariances rather than unit-level data, and we provide a generative variant of CSS that facilitates the application of statistical tools. These viewpoints of CSS lead to novel methodology for a number of open problems in interpretable dimensionality reduction.  

\subsection{Background}

In what follows, we give a detailed description of CSS, Principal Variables, and the relationship (or lack thereof) between their corresponding literatures. 

In the CSS problem we aim to choose $k$ of $p$ columns from a data matrix $\X \in \R^{n \times p}$ that best linearly reconstruct the rest:
\begin{equation}\label{eq:css}
\argmin_{S \subseteq [p]: |S| = k} \min_{\B \in  \R^{k \times p}} \|\X -  \X_{\bullet S} \B\|_F^2
\end{equation}
Here, $\X_{\bullet S}$ denotes the sub-matrix of $\X$ containing columns with indices in $S$. While computing the exact optimum in \eqref{eq:css} is NP-complete \citep{Shitov}, a vast approximation literature \citep{Boutsidis2009, Civril,  Deshpande, Drineas, Farahat, Ordozgoiti2018, Tropp} has produced tractable algorithms with successful applications in epidemiology \citep{Fink, Nowakov}, networking \citep{Tripathi}, business \citep{Boutsidis2008}, imaging \citep{Kromer, Strauch, WangC}, and low-rank matrix approximation \citep{Boutsidis2011,  Dan}. 

\textcolor{revision2}{Prior to gaining popularity in computer science, the CSS problem first garnered attention in the numerical linear algebra community. Many foundational works in numerical linear algebra, past and present, study criteria similar to CSS \citep{Bischof, Baker, Chan1987, Chan1994, Chandrasekaran, Chen, Foster, Golub, Gu1996, Gu2004, Hong, Pan1999, Pan2000, Stewart1998, Stewart1999QLP, Stewart1999sparse}. The articles referenced in the previous paragraph extend this line of research by focusing specifically on the CSS problem and providing either improved theoretical guarantees or superior empirical performance for the objective \eqref{eq:css}.}

Principal Variables \citep{McCabe} puts a non-degenerate $N(0, \bSigma)$ model on the observed data and suggests selecting a subset of ``principal variables'' according to various information criteria. The most popular of the four criteria suggests selecting the subset $S$ that solves 
\begin{equation}\label{eq:mccabe_pv}
\argmin_{S \subseteq [p]: |S| = k} \tr(\bSigma_{-S} - \bSigma_{-S, S}\bSigma_{S}^{-1}\bSigma_{S, -S})
\end{equation}
where $-S$ is the complement of $S$ \textcolor{revision2}{and $\bSigma_{S}$ and $\bSigma_{-S}$ are the principal submatrices of $\bSigma$ whose rows and columns are subseted by $S$ and $-S$ respectively} . In practice, an estimate $\hat{\bSigma}$ of the covariance matrix is used in \eqref{eq:mccabe_pv}. Like the CSS literature, the Principal Variables literature has produced approximation algorithms \citep{Brusco, Camida,  Guo, Masaeli, Wei} that have enjoyed application in a number of fields, including psychology \citep{Fehrman, Ladd}, imaging \citep{Chang}, and the natural sciences  \citep{Eitrich, Isaac}. 

Expectedly, each thread of literature has something fruitful to offer: principal variables have strong statistical motivation and practitioners can find them in settings where only covariance or correlation estimates are available. We provide examples of such settings in \Cref{sec:equivalence}. The CSS literature places larger emphasis on providing algorithms with theoretical guarantees and minimal computational and storage complexity. 

Despite similarities between the methods, the CSS and Principal Variables literatures have existed entirely independently. For example, \cite{Masaeli} pose a regularized matrix reconstruction problem in the style of \eqref{eq:css}, but cites Principal Variables as motivation and makes no reference to any CSS literature.

\subsection{Our Contributions}

\textcolor{revision2}{In this paper}, we establish an equivalence between CSS and Principal Variables, and via this equivalence, provide a novel generative view of both methods. In particular, we show that performing CSS on $\X$ is equivalent to finding the principal variables according to $\bSigma = \X^{\top} \X/n$, and both are equivalent to maximum likelihood estimation in the following semi-parametric model:
\begin{align}
\begin{split}
&X_S \sim F\\
&X_{-S} \mid X_S \sim N(\mu_{-S} + \W(X_S - E_{F}[X_{S}]), \sigma^2 \I_{p-k}). 
\end{split} \label{eq:pcss_model}
\end{align}
\textcolor{revision}{
In the model \eqref{eq:pcss_model} we sample $k$ principal variables $X_S$ from some unrestricted distribution $F$, and then generate the remaining variables $X_{-S}$ as linear combinations of these principal variables plus spherical Gaussian noise. The matrix $\W$ is the coefficient matrix from the population regression of $X_{-S}$ on $X_{S}$, and $\sigma^2$ is the residual variance.} \textcolor{revision2}{We also show that, under mild conditions, the CSS solution \eqref{eq:css} is consistent for the principal variable set $S$ in high dimensions, specifically in the proportional asymptotic regime where the number of variables $p$ and sample size $n$ both grow to infinity with $p/n$ approaching a constant.}

This equivalence and generative view suggest new methodology for several problems in interpretable dimensionality reduction: \newline 

\noindent \textbf{Scalable CSS without unit-level data:} We introduce fast algorithms for performing CSS when only summary statistics are available. To demonstrate our algorithms' efficacy, we provide a real world example where BlackRock, our industry affiliate, wants to perform CSS but only has access to a covariance estimate. Our insights enable us to perform CSS in this setting, and our novel algorithms enable us to run the appropriate experiments, which would take days using current methods, in less than three minutes. \newline 

\noindent \textbf{CSS with missing and/or censored data:} The equivalence between CSS and Principal Variables suggests a natural workflow for performing CSS when data is missing and/or censored: model the missingness and/or censorship, produce a covariance estimate that is reasonable under this model, and perform CSS using this covariance estimate. As an illustrative example, we use this workflow to propose a simple and novel procedure for performing CSS with missing-at-random data. In a simulation study where we sample data from the model \eqref{eq:pcss_model} and omit entries at random, our method always selects the correct subset while the best competing method selects it less than half the time. We also confirm that our method outperforms existing methods on a real dataset. \newline 

\noindent \textbf{Selecting the Subset Size for CSS:} Supposing we observe data drawn from a generalization of the model \eqref{eq:pcss_model}, we propose a novel and theoretically motivated procedure for selecting an appropriate subset size. We first verify that our procedure performs well in some difficult simulated settings, and we then demonstrate how it can be used to shorten surveys in an automated way. As an example, we apply our procedure to the Big Five Inventory (BFI), a 44-item personality questionnaire \citep{John}. In less than a tenth of a second, it suggests cutting the survey in more than half. Importantly, the questions our procedure selects appear to sufficiently capture the relevant information in the survey.\newline 

We provide a python package, \texttt{pycss}, that allows practitioners to use our methods and notebooks that allow for easy replication of our results at \texttt{\href{https://github.com/AnavSood/CSS}{github.com/AnavSood/CSS}}. 

\subsection{Notation and Preliminaries}

We use the following linear algebra notation. All matrices are bolded capital letters. \textcolor{revision}{We use $||\A||_F$ to denote the Frobenius norm of the matrix $\A \in \R^{p \times q}$. We denote the Moore-Penrose inverse, or pseudo-inverse, of $\A$ by $\A^+$, and the $i$th row and $j$th column of $\A$ by $\A_{i \bullet}$ and $\A_{\bullet j}$ respectively}. For subsets $S \subseteq [p] = \{1, \dots, p \}$ and $T \subseteq [q]$, $\A_{S, T}$ is the $|S| \times |T|$ sub-matrix of $\A$ with rows and columns indexed by $S$ and $T$ respectively. \textcolor{revision}{If we want to index just the rows (resp. just the columns), we use $\A_{S \bullet} \in \R^{|S| \times q}$ (resp. $\A_{\bullet T} \in \R^{p \times |T|}$).} Indexing always precedes matrix operations, e.g., $\A^{+}_{S,T} = (\A_{S,T})^{+}  \neq (\A^{+})_{S,T}$. \textcolor{revision}{When $\A \in \R^{p \times p}$ is square, we let $\A_S \in \R^{|S| \times |S|}$ denote the sub-matrix with rows and columns both indexed by $S$. We denote the determinant of a square matrix $\A$ by $|\A|$.} For symmetric square matrices $\A$ and $\B$, we write $\A \succeq \B$ (resp. $\A \succ \B$) if $\A - \B$ is positive semi-definite (resp. positive definite), and we denote the set of positive-semi definite matrices as $\S^{p \times p}_{+}$. The $p \times p$ identity matrix is $\I_p$. \textcolor{revision}{For a vector $x \in \R^p$, we use $\|x\|_{2}$ to denote the $L^2$ norm of $x$. The vector $x_{S} \in \R^{|S|}$ is the sub-vector of $x$ with entries indexed by $S$}. When appropriate, $0$ and $1$  (resp. $\0$ and $\1$) refer to a vector (resp. matrix) of zeros and ones. 

We use the following statistics notation. We typically denote random vectors as unbolded capital letters, e.g., $X \in \R^{p}, Y \in \R^{q}$, and $X \sim (\mu, \bSigma)$ means that $X$ has mean $\mu$ and covariance $\bSigma$. Generally $\bSigma_{Y}$ is $\Cov(Y)$, but we sometimes write $\bSigma$ for $\Cov(X)$ specifically (when unambiguous). Samples drawn from a distribution are assumed independent and identically distributed (i.i.d). A model refers to a family of distributions. 

We will assume that all random vectors $X \in \R^p$ we consider satisfy $E[\|X\|_2^2] < \infty$. 

All proofs are in the appendix. 

\section{The Equivalence of CSS and Principal Variables}
\label{sec:equivalence}
In this section, we establish a mathematical equivalence and conceptual connection between CSS and Principal Variables. \textcolor{revision2}{This equivalence is not mathematically difficult to establish, but it unlocks new perspectives that enable us to design novel and useful methodology for several problems in interpretable dimensionality reduction.} Henceforth, selecting principal variables using $\bSigma$ refers to the following subset search:
\begin{equation}
\label{eq:pv}
\argmin_{S \subseteq [p]: |S| = k} \tr(\bSigma - \bSigma_{\bullet S}\bSigma_{S}^{+}\bSigma_{S \bullet} ).
\end{equation}
The subset search \eqref{eq:pv} generalizes the original Principal Variables criterion \eqref{eq:mccabe_pv} by allowing $\bSigma$ to be singular. When $\bSigma$ is non-singular, \textcolor{revision2}{Cauchy's interlacing theorem \citep{Bhatia} implies that $\bSigma_S$ is always also non-singular}, so $\bSigma^{+}_{S} = \bSigma^{-1}_{S}$ and our new search problem 
\eqref{eq:pv} reduces to the original search problem \eqref{eq:mccabe_pv}.

\Cref{prop:css_equiv_pv} establishes an exact mathematical equivalence between CSS \eqref{eq:css} and Principal Variables \eqref{eq:pv}. It tells us explicitly how we can perform CSS by instead searching for principal variables, and vice-versa. 
\begin{proposition}[CSS and Principal Variables are equivalent]
\label{prop:css_equiv_pv}
Consider a matrix $\X \in \R^{n \times p}$ and define $\hat{\bSigma} =  \X^\top\X/n$. For any size-$k$ subset $S \subseteq [p]$, the CSS objective \eqref{eq:css} with $\X$ and the Principal Variables objective \eqref{eq:pv} with $\hat{\bSigma}$ are equal:
\begin{equation*}
\min_{\B \in  \R^{k \times p}} \frac{1}{n}\|\X - \X_{\bullet S}\B\|_{\textcolor{revision}{F}}^2 = \tr(\hat{\bSigma} - \hat{\bSigma}_{\bullet S} \hat{\bSigma}_{S}^{+}\hat{\bSigma}_{S \bullet}).
\end{equation*}
\end{proposition}
 
One important consequence of \Cref{prop:css_equiv_pv} is that performing CSS on a centered data matrix $\X$ is identical to selecting principal variables using the sample covariance. Conceptually, this suggests viewing CSS as an estimation problem where the estimand is the population principal variable set (i.e., the principal variable set one would select using $\bSigma$). This draws a clear parallel between CSS and PCA: both methods seek to estimate some function of the population covariance, and the typical unit-level data characterization of each method does so by plugging in the sample covariance as an estimate.

In what follows, we fully develop this point-estimation view of CSS. Consider the typical CSS use case where $\X \in \R^{n \times p}$ is a data matrix whose rows are centered i.i.d samples of a $p$-dimensional random vector $X \sim (\mu, \bSigma)$. In this setting, it is immediate that the CSS solution \eqref{eq:css} is the plug-in estimate for the estimand 
\begin{equation}\label{eq:pop_css_centered}
\argmin_{S \subseteq [p]: |S| = k} \min_{\B \in  \R^{p \times k}} E[\|(X - \mu) -  \B (X_S -\mu_{S})\|_2^2].
\end{equation}
\Cref{prop:pop_css_equiv_pv} confirms that the population CSS subset search \eqref{eq:pop_css_centered} is equivalent to selecting principal variables using $\bSigma$. Furthermore, it demonstrates that, even when $X$ does not follow a non-degenerate Gaussian distribution, the subset selected by Principal Variables is characterized by an interpretable and desirable optimality criterion: optimal linear reconstruction of the remaining variables (after centering). 

\begin{proposition}[The CSS estimand]
\label{prop:pop_css_equiv_pv}
Consider a $p$-dimensional random vector $X \sim (\mu, \bSigma)$. For any size-$k$ subset $S \subseteq [p]$, the population CSS problem \eqref{eq:pop_css_centered} has the same objective as the Principal Variables objective with $\bSigma$:
\begin{equation*}
\min_{\B \in  \R^{p \times k}} E[\|(X - \mu) - \B (X_{S} -\mu_{S}) \|_{2}^2] = \tr(\bSigma - \bSigma_{\bullet S}\bSigma_{S}^{+}\bSigma_{S \bullet}).
\end{equation*}
\end{proposition}

Characterizing CSS in terms of covariances and reframing it as an estimation problem yields a number of benefits and insights:

\begin{enumerate}
    \item A more general class of estimators for the population CSS solution \eqref{eq:pop_css_centered} results from solving \eqref{eq:pv} with any estimate $\hat{\bSigma}$ of $\bSigma$, not just sample covariance. Henceforth we will refer to solving \eqref{eq:pv} with $\hat{\bSigma}$ as performing CSS with the covariance $\hat{\bSigma}$.
    \item It is immediately clear that practitioners can perform CSS in settings where they cannot access unit-level data but can access covariance or correlation estimates. For example, only the correlation matrix was available for four of the nine real data examples in \cite{Camida}, and in genome wide association studies practitioners often work with correlation estimates in place of inaccessible unit-level data \citep{Mak, Zihua}.
    \item Characterizing CSS as an estimation problem stresses the importance of selecting a subset that generalizes well. This is in line with recent work from \cite{Ordozgoiti2019} that emphasizes CSS's out-of-sample performance and examines the benefits of performing CSS with $\ell_2$ regularization. Our equivalence suggests performing CSS with regularized covariance estimates as an alternative regularization scheme. 
    
    \item \textcolor{revision}{Similarly, a practitioner may be interested in the quality of their regression coefficient estimates from the regression of the remaining variables on the selected subset (i.e., the minimizing $\B$ in \eqref{eq:css}). Using our covariance view, we show in \Cref{sec:add_theory_error} that the error of these coefficient estimates is well-behaved, even when the practitioner selects a subset via a complex CSS algorithm.}
    
    \item The NP-completeness of CSS implies the NP-completeness of solving \eqref{eq:pv}. Equivalently, if $X \sim N(0, \bSigma)$, $\bSigma \succeq 0$, finding a size $k$ subset $S \subset [p]$ that minimizes $\tr(\Cov(X|X_{S}))$ is NP-complete. 
    \item If $\X \in \R^{n \times p}$ has rank $r \ll \min(n, p)$, we can find $\X_{r} \in \R^{r \times p}$ such that $\X_{r}^\top\X_{r} = \X^\top\X$ via the singular value decomposition (SVD) of $\X$. \textcolor{revision}{Because $\X^T\X = \X_{r}^\top\X_{r}$, \Cref{prop:css_equiv_pv} tells us that the CSS objective value for any subset $S$ is the same for $\X$ and $\X_r$, and thus performing CSS on $\X_r$ is equivalent to performing CSS on $\X$. The time complexity of finding $\X_r$ is $O(npr)$, and computing and then performing CSS on $\X_r$ may be much faster than performing CSS directly on $\X$.}
\end{enumerate}

\section{A Generative View of CSS}
\label{sec:gen_view}

In this section we show that CSS arises naturally from a certain probabilistic model and then discuss a useful generalization of this model. 

\subsection{CSS as Maximum Likelihood Estimation}

In what follows, we show that CSS and Principal Variables can both be viewed as maximum likelihood estimation within a certain semi-parametric model.  We call this viewpoint \emph{Probabilistic Column Subset Selection (PCSS)}. As the name suggests, PCSS provides a generative view of CSS the same way that PPCA does for PCA.  

We introduce the $k$-dimensional \emph{PCSS model}. It posits that each data vector $X \in \R^p$ arises from the following two-step sampling procedure:
\begin{align}
\tag{\ref{eq:pcss_model}, revisited}
\begin{split}
&X_S \sim F\\
&X_{-S} \mid X_S \sim N(\mu_{-S} + \W(X_S - E_{F}[X_{S}]), \sigma^2 \I_{p-k}). 
\end{split} 
\end{align}
In words, we first sample the $k$ principal variables $X_S$ from some distribution $F$, and then generate the remaining variables $X_{-S}$ as linear combinations of these principal variables plus spherical Gaussian noise. The parameters of this model are the set of principal variables $S$, their distribution $F$, the remaining variables' mean $\mu_{-S}$, the regression coefficients $\W$, and the residual variance $\sigma^2 > 0$. The size of the principal variable set, $k$, is fixed and not a parameter in the model. 

The relationship between PCSS and CSS is explained by \Cref{thm:css_is_mle}. It says that, in our PCSS model, any CSS solution is a maximum likelihood estimate (MLE) for $S$ over a non-parametric family of principal variable distributions $F$.  

\begin{theorem}[CSS solution is an MLE]
\label{thm:css_is_mle}
    Let $x^{(1)}, \dots, x^{(n)} \in \R^p$ be samples with sample covariance $\hat{\bSigma}$. For any $M <\infty $, let $\mathcal{P}_M$ be the set of distributions in the $k$-dimensional PCSS model \eqref{eq:pcss_model} where $F$ admits a probability density bounded by $M$. Then, so long as no $k$ variables achieve perfect in-sample linear reconstruction of the remaining variables, any CSS solution
    \begin{equation*}
    \hat{S} \in \argmin_{U \subset [p], |U| = k} \tr(\hat{\bSigma} - \hat{\bSigma}_{\bullet U}\hat{\bSigma}^{+}_{U}\hat{\bSigma}_{U\bullet})
    \end{equation*}
    is a maximum likelihood estimator for $S$ in the model $\mathcal{P}_M$. Moreover, each $\hat{S}$ is a maximum hybrid likelihood estimator (see \citep[Equation 9.1]{Owen}) for $S$ in the original model \eqref{eq:pcss_model} where $F$ is completely unrestricted.
\end{theorem}

The density bound $M < \infty$ is only in place to ensure that the likelihood is finite. Otherwise every subset $S$ results in infinite likelihood. Likewise, if some size-$k$ subset of variables achieves perfect in-sample linear reconstruction of the remaining (and thus attains the minimum CSS objective value of zero), then the likelihood for this subset tends to $\infty$ as we send $\sigma^2 \rightarrow 0$ and technically no MLE exists. 

\textcolor{revision2}{\Cref{thm:css_is_mle} tells us that any CSS solution is a MLE under the model \eqref{eq:pcss_model}, but since the model \eqref{eq:pcss_model} is not fully parametric, it is not immediately clear that the MLE is a favorable estimate. The usual desirable properties of the MLE, e.g., consistency, are not automatically guaranteed.}

\textcolor{revision2}{\Cref{thm:high_dim_consistency} alleviates these concerns by establishing that, in the high-dimensional regime where tools like CSS are typically applied, the MLE subset $\hat{S}$ from \Cref{thm:css_is_mle} is consistent for the population subset $S$. Further, it tells us that this population subset minimizes the population CSS objective \eqref{eq:pv}.}
\textcolor{revision2}{
\begin{theorem}[High-dimensional consistency of CSS]
    \label{thm:high_dim_consistency}
    Suppose we observe $n$ samples $x^{(1)}, \dots, x^{(n)} \in \mathbb{R}^{p}$ from a distribution $P$ in the $k$-dimensional PCSS model, and consider the proportional asymptotic regime where $p/n$ converges to a constant as $n, p \rightarrow \infty$ while $k$ remains fixed. If, as $n, p \rightarrow \infty$, the data generating distribution satisfies 
    \begin{enumerate}
        \item (Light tails) The sub-Gaussian norm (see \citep[Definition 2.5.6]{Vershynin}) of each variable is bounded above by a constant independent of $p$ and $n$,
        \item (Imperfect reconstruction) The minimum eigenvalue of the principal variables' covariance $\Cov_P(X_S)$ and the residual variance $\sigma^2$ are both bounded below by a positive constant independent of $p$ and $n$,
        \item (Important principal variables) For some $\delta > 0$, at least $\Omega(p^{1/2 + \delta})$ entries in each column of the regression coefficients $\W$ are bounded away from zero by some constant independent of $p$ and $n$, 
    \end{enumerate}
    then the MLE subset $\hat{S}$ from \Cref{thm:css_is_mle} is consistent for the population subset $S$ (i.e., $\lim_{n\rightarrow \infty} P(\hat{S} = S) = 1$), which eventually minimizes the population CSS objective \eqref{eq:pv}. 
\end{theorem}}

\textcolor{revision2}{A closer examination of \Cref{thm:high_dim_consistency}'s proof actually tells us that, under the three conditions, the probability that $\hat{S}$ does not equal $S$ shrinks to zero at a super-polynomial rate. The first two conditions of the theorem are standard regularity conditions. They ensure that the in-sample CSS objective is well behaved for all size-$k$ subsets. The third condition is a strong identifiability condition for our problem. It ensures that, compared to other size-$k$ subsets $U$, the population subset $S$ results in lower reconstruction error for sufficiently many (but still a vanishing proportion) of the variables in  $-S \setminus U$. Without some such condition, the population subset $S$ may not even be unique. For the interested reader, \Cref{sec:high_dim_consistency_appdx} identifies a more general high-dimensional setting where the CSS estimate \eqref{eq:css} is consistent for population CSS solution, but at the expense of giving a less interpretable statement.}

\subsection{The Subset Factor Model}

By generalizing the probabilistic model \eqref{eq:pcss_model} in which the CSS solution is the MLE, we can motivate subset versions of other existing probabilistic methods.

In this vein, we introduce the $k$-dimensional \emph{subset factor model}, a generalization of our $k$-dimensional PCSS model that drops all distributional assumptions and allows the covariance of $X_{-S} \mid X_S$ to be a diagonal matrix $\D \succ \0$: 
\begin{align}
\begin{split}
&X_S \sim F \qquad \epsilon \sim (0, \D) \qquad \epsilon_1 \perp , \dots, \perp \epsilon_{p-k} \qquad \Cov(X_S, \epsilon) = \0 \\
&X_{-S} = \W(X_{S} - E_{F}[X_S]) + \mu_{-S} + \epsilon.
\end{split} \label{eq:subset_factor_model}
\end{align}
Our $k$-dimensional  subset factor model has a close relationship with the $k$-dimensional factor model \citep[Section 9.2.1]{Mardia}
\begin{align}
\begin{split}
&Z \sim (0, \I_k) \qquad \epsilon \sim (0, \D) \qquad \epsilon_1 \perp , \dots, \perp \epsilon_{p} \qquad \Cov(Z, \epsilon) = \0 \\
&X = \W Z + \mu + \epsilon,
\end{split} \label{eq:factor_model}
\end{align}
where again $\D \succ \0$. In the $k$-dimensional factor model, each variable $X_i$ is a linear combination of latent common factors $Z \sim (0, \I_k)$ plus a mean-zero  unique factor $\epsilon$ and a non-random shift $\mu_i$. In our $k$-dimensional subset factor model, an unknown subset of the observed variables act as common factors (once centered), and the remaining variables are similarly linear combinations of these common factors plus a unique factor and non-random shift. Moreover, by dropping distributional assumptions and allowing a previously spherical conditional covariance to be diagonal, our subset factor model generalizes our PCSS model the same way the factor model generalizes the model used for PPCA.

\textcolor{revision}{The subset factor model \eqref{eq:subset_factor_model} arises naturally from  a hybridization of PCA and Factor Analysis that aims to maintain strengths from both methods. To see how, we must first discuss the advantages and disadvantages of PCA and Factor Analysis. Factor Analysis enjoys two major benefits over PCA. First, the factor model \eqref{eq:factor_model} provides an interpretable generative model for the data. In contrast, PCA has no associated generative model. Second, in the factor model, the regression of $X\in \R^p$ on its lower dimensional representation $Z \in \R^k$ has uncorrelated residuals, meaning that $Z$ accounts for all the common variation in $X$. If these residuals were correlated, i.e., they jointly varied about their means in some structured way, that would mean there may be additional meaningful trends in $X$ that $Z$ does not account for. This is the case in PCA, where the regression of $X$ on its top $k$ principal components typically yields correlated residuals. Due to the fact that the lower dimensional representation $Z$ is latent and unobserved, however, Factor Analysis suffers from a slew of unidentifiability and indeterminancy issues \citep{Anderson1956, Guttman, Steiger1994}. Proponents of components-based methods are particularly critical of factor indeterminacy, the fact that the common factors are not uniquely determined, even up to rotation \citep{Schonemann1976, Schonemann1978, Steiger1979,  Velicer}. In contrast, components (linear combinations of the observed variables) are always observed and determined, and they therefore do not suffer from any of the same issues.}

\textcolor{revision}{A natural attempt to hybridize PCA and Factor Analysis in a way that inherits strengths from both methods is to assume a factor model where the factors are themselves components.} Then, via an interpretable generative model, we can recover determined factors that account for all the common variation in $X$. \textcolor{revision}{This idea has been of great historical interest, but it has repeatedly been emphasized that it is impossible to do so when all the unique factors $\epsilon_i$ are required to have positive variance \citep{Wilson, Guttman, Schonemann1976}.  However, if we allow some unique factors to have zero variance, \Cref{thm:compromise} tells us that, surprisingly, this is equivalent to assuming a subset factor model.}

\begin{theorem}[The subset factor model compromise]
    \label{thm:compromise}
    Suppose $X \sim (\mu, \bSigma)$, $\bSigma \succ \0$ follows the $k$-dimensional factor model \eqref{eq:factor_model} where the unique factors $\epsilon_i$ are allowed to have zero variance. Then the common factors $Z$ can be written as components (i.e., $Z = \B(X - \mu)$ for some $\B \in \R^{k \times p}$) if and only if $X$ follows the $k$-dimensional subset factor model \eqref{eq:subset_factor_model}.
\end{theorem}

\section{Applications}
\label{sec:applications}
In this section we illustrate some useful applications of the viewpoints we provided earlier.  
\subsection{Scalable CSS without Unit-level Data}
\label{sec:scalable_css}
Our first application is scalable CSS without unit-level data. We provide novel algorithms that allow us to efficiently perform CSS when we can only access covariance estimates and a real data example that illustrates these algorithms' usefulness. These algorithms are made possible by the equivalence drawn in \Cref{sec:equivalence}. We also discuss some other reasonable approaches that perform surprisingly poorly on our real data example.  

\subsubsection{Algorithms}
\label{sec:algorithms}

As it stands, it is unclear how to efficiently perform CSS with just a covariance $\bSigma$. The Principal Variables literature suggests some sophisticated approximation algorithms, but it lacks efficient implementations of them. The CSS literature provides efficient approximation algorithms, but it is unclear how to apply them when we only have covariance estimates. \Cref{prop:css_equiv_pv} implies that we could find $\X$ such that $\bSigma = \X^\top\X/n$ and apply existing CSS algorithms to $\X$. Unfortunately, factoring $\bSigma$ in this way has larger time complexity than some CSS algorithms themselves.

Hence, we derive versions of the greedy algorithm from \cite{Farahat} and the swapping algorithm from \cite{Ordozgoiti2018} that take a covariance $\bSigma$ as input. These two algorithms \textcolor{revision2}{are the empirically} best-performing and fastest algorithms in the CSS literature, and they can be easily modified to perform subset search according to other objectives as well, \textcolor{revision}{which will be} \textcolor{revision2}{crucial} \textcolor{revision}{for our application in \Cref{sec:model_selection}}. Working directly with $\bSigma$ \textcolor{revision2}{allows for significantly} simpler presentation, \textcolor{revision2}{derivation, and modification} of these algorithms, and our new algorithms maintain the same time complexity as the originals when $n \geq p$. For the interested reader, we discuss how to modify our algorithms to perform subset search according to the other three Principal Variables criteria in \Cref{sec:other_algs_appdx}. \textcolor{revision}{We also provide some stylized settings where we can guarantee that our algorithms find the optimal subset in \Cref{sec:add_theory_subset}}. 

The key insight in deriving our algorithms is writing the CSS objective in terms of residual covariance matrices. For two mean-zero random vectors $X \in \R^{p}$ and $Y \in \R^{q}$ we define the residual from the regression of $Y$ on $X$ as 
\begin{equation*}
    R(Y, X) = Y - \B^* X \qquad \B^* \in \argmin_{\B \in \R^{q \times p}} E[\|Y - \B X \|_2^2].
\end{equation*}
The discussion in \Cref{sec:hilbert_space_appdx} establishes that $R(Y, X)$ is well defined. 

\Cref{lem:residual_cov_update} summarizes what we need to know about residual covariances. In its statement, we adopt the convention that $0 \cdot \infty = 0$.
\begin{lemma}[Efficient residual covariance update]
\label{lem:residual_cov_update}
    Consider a random vector $X \sim (0, \bSigma)$ and a subset $U \subset [p]$ that does not contain the $i$th variable. Then the residual covariance for the subset $U$ is given by $\bSigma_{R(X, X_{U})} = \bSigma - \bSigma_{\bullet U}\bSigma_{U}^{+} \bSigma_{U\bullet}$, and the residual covariance for the subset $U \cup \{i\}$ is given by 
    \[\bSigma_{R(X, X_{U \cup \{i\}})} = \bSigma_{R(X, X_U)}  - \frac{\beta \beta^\top}{\beta_i} \cdot I_{\beta_i > 0}\]
    where $\beta = (\bSigma_{R(X, X_U)})_{\bullet i} =  \bSigma_{\bullet i} - \bSigma_{\bullet U}\bSigma_{U}^{+}\bSigma_{U, i}$. That is, we can update $\bSigma_{R(X, X_{U})}$ when adding variables to $U$ in $O(p^2)$ time, and removing variables from $U$ in $O(p^2)$ time if $\bSigma_{U}^{+}$ is known. 
\end{lemma}
\Cref{lem:residual_cov_update} illustrates that CSS amounts to finding a subset $S$ that minimizes $\tr(\bSigma_{R(X, X_{S})})$, and it also allows us to quickly update $\tr(\bSigma_{R(X, X_{S})})$ when we add or remove variables from $S$. Define the function  
\begin{equation}
\label{eq:alg_function}
    f(i, \A) = -\|\A_{\bullet i}\|_2^2/\A_{ii} \cdot I_{\A_{ii} > 0}.
\end{equation}
Considering some $U \subset [p]$, computations in \Cref{sec:alg_correctness_appdx} show that any $i \not \in U$ which minimizes $f(i, \bSigma_{R(X, X_{U})})$ also minimizes $\tr(\bSigma_{R(X, X_{U  \cup \{i\}})})$. This is sufficient for establishing the correctness of the greedy and swapping algorithms we present below. \newline 

\noindent \textbf{Greedy Subset Selection:} Our greedy algorithm (\Cref{alg:greedy}) starts with the empty subset $S^{(0)} = \{ \}$ and, on iteration $t$, finds $S^{(t)}$ by adding the variable to $S^{(t-1)}$ that minimizes $\tr(\bSigma_{R(X, X_{S^{(t)}})})$. It terminates after $k$ iterations and returns $S^{(k)}$. Alternatively, we can allow it to iterate until $\tr(\bSigma_{R(X, X_{S^{(t)}})})$ is sufficiently small. \Cref{alg:greedy}'s time complexity is $O(p^2k)$. A nice property of greedy subset search is that greedily selected subsets of different sizes are nested: upon finding the greedily selected size-$k$ subset, we have necessarily found the greedily selected subsets of size $k-1, \dots, 1$ in the process. \newline 

\begin{algorithm}[H]
\begin{algorithmic}
\Require covariance matrix $\bSigma$, number of columns to select $k$

\vspace{.15cm}

\Procedure{greedy\_subset\_selection($\bSigma, k$)}{} 

\State $S^{(0)} \gets \{ \}$ 
\State $\bSigma_{R(X, X_{S^{(0)}})} \gets \bSigma$ 
\For{$t= 1, \dots, k$}
    \State $i^* \gets \argmin_{i \not \in S^{(t-1)}} f(i, \bSigma_{R(X, X_{S^{(t-1)}})}) $   \Comment{$f$ is from (\ref{eq:alg_function}), arbitrarily break ties}
    \State $S^{(t)} \gets S^{(t-1)} \cup \{i^*\} $ 
    \State Compute $\bSigma_{R(X, X_{S^{(t)}})}$ from $\bSigma_{R(X, X_{S^{(t-1)}})}$ using \Cref{lem:residual_cov_update}
\EndFor
\State \textbf{return:} $S^{(k)}$
\EndProcedure
\end{algorithmic}
\caption{Greedy Subset Selection}
\label{alg:greedy}
\end{algorithm}

\noindent \textbf{Subset Selection via Swapping}: On each iteration of our swapping algorithm (\Cref{alg:swap}) we loop through our currently selected subset $S$ and swap each selected variable out for the variable that most reduces $\tr(\bSigma_{R(X, X_{S})})$. If no swap reduces $\tr(\bSigma_{R(X, X_{S})})$ for a particular selected variable, then we keep that variable in $S$. This is reminiscent of coordinate descent, a procedure that searches for a minimum of $f: \R^q \rightarrow \R$ by iteratively optimizing each coordinate while holding the others fixed. \Cref{alg:swap} terminates after an iteration where no swaps are made, at which point the algorithm has converged to a kind of local optimum. Since each swap strictly reduces the objective value, \Cref{alg:swap} must converge in finite time. Its time complexity is $O(p^2k)$ per iteration. In practice, we run the algorithm multiple times with different random initializations and take the best resulting subset.  

\begin{algorithm}[H]
\begin{algorithmic}
\Require covariance matrix $\bSigma$, number of columns to select $k$, initial selected subset $S \in [p]^k$ 

\vspace{.15cm}

\Procedure{swapping\_subset\_selection($\bSigma, k, S$)}{} 
\State $S_{copy} \gets 0 \in \R^{k}$
\State $\bSigma_{S}^{+} \gets$ compute pseudo-inverse of $\bSigma_{S}$
\State $\bSigma_{R(X, X_{S})} \gets \bSigma - \bSigma_{\bullet S} \bSigma_{S}^{+} \bSigma_{ S \bullet}$  
\While{$S \neq S_{copy}$}
    \State $S_{copy} \gets S$
    \For{$j=1, \dots, k$}
        \State $U \gets S_{-j}$  \Comment{ $S_{-j} \in \R^{k-1}$ is $S$ with $S_j$ removed}
        \State $\bSigma_{U}^{+} \gets$ compute $\bSigma_{U}^{+}$ from $\bSigma_{S}^{+}$ using \citep[3.2.7]{Petersen} \Comment{Takes $O(k^2)$ time}
        \State $\bSigma_{R(X, X_{U})}\gets  $ compute $\bSigma_{R(X, X_{U})}$ from $\bSigma_{R(X, X_{S})}$ using \Cref{lem:residual_cov_update}
        \If{ $S_j \not \in \argmin_{i \not \in U} f(i, \bSigma_{R(X, X_{U})} )  $ }  \Comment{$f$ is from (\ref{eq:alg_function})}
            \State $i^{*} \gets \argmin_{i \not \in U} f(i, \bSigma_{R(X, X_{U})} )$  \Comment{Arbitrarily break ties}
            \State $S_j \gets i^*$  \Comment{This alters $S$ in place}
            \State $\bSigma_{S}^{+} \gets$ compute $\bSigma_{S}^{+}$ from $\bSigma_{U}^{+}$ using \citep[3.2.7]{Petersen} \Comment{Takes $O(k^2)$ time}
            \State $\bSigma_{R(X, X_{S})}\gets  $ compute $\bSigma_{R(X, X_{S})}$ from $\bSigma_{R(X, X_{U})}$ using \Cref{lem:residual_cov_update}  
        \EndIf
    \EndFor
\EndWhile
\State \textbf{return:} $S$
\EndProcedure
\end{algorithmic}
\caption{Subset Selection by Swapping}
\label{alg:swap}
\end{algorithm}

\subsubsection{An Example from BlackRock}
\label{sec:blackrock}

We provide a real world application where BlackRock, an industry affiliate of ours, seeks to perform CSS and only has access to a covariance estimate. BlackRock is currently the largest asset manager in the world. For proprietary reasons, the data has been anonymized.

BlackRock has a collection of $p = 774$ thematic investment portfolios and would like to find a small subset of them that are representative of the remaining. Specifically, they are interested in performing CSS. Thematic portfolios invest in assets that are tied together by an interpretable common thread, e.g., renewable energy companies or start-ups in India. It is important to select a subset of portfolios (rather than find linear combinations of them) so that this thematic structure is maintained. 

\begin{figure}
\centering
\includegraphics[scale=0.65]{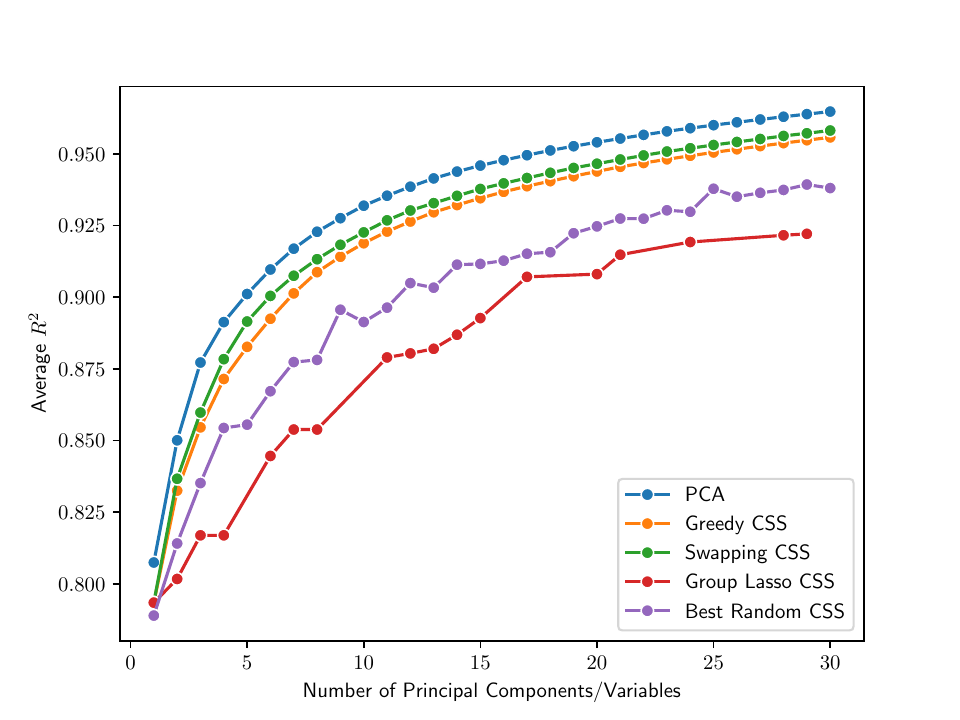}
\caption{For the BlackRock example discussed in \Cref{sec:blackrock}, the average $R^2$ from the regression of each variable on the selected subset (resp. principal components) versus increasing subset size (resp. number of principal components).}
\label{fig:BlackRock}
\end{figure}

BlackRock does not have unit-level returns for each portfolio, but they do have a covariance estimate. After standardizing the returns of each portfolio to have unit-variance, we perform CSS using our greedy and swapping algorithms for $k=1, \dots, 30$. For our swapping algorithm, we try 25 random initializations. The results are presented in \Cref{fig:BlackRock}, where we plot the (estimated) average $R^2$ from the regression of each individual portfolio on the selected subset. For the size six subset found by our swapping algorithm, this average $R^2$ is above 90\%. The swapping algorithm consistently outperforms the greedy one, and for smaller subset sizes it is able to get the same performance with one less variable. For sake of comparison, we also plot the average $R^2$ from the regression of each portfolio on the top $k$ principal components. The principal components maximize this $R^2$ among all possible linear combinations of the portfolios. We find that the performance gap between PCA and CSS is well worth the trade-off in interpretability. For smaller subset sizes, CSS achieves the same performance as PCA while using a subset size that is is only one larger than the number of principal components. 

It is important to use efficient algorithms for problems of this scale. The Principal Variables literature, which mainly considers $p < 100$, suggests recomputing the objective from scratch for each new considered subset \citep{Camida}. For the BlackRock data, using Algorithm \ref{alg:greedy} to greedily select a size-$30$ subset takes less than a tenth of a second, while this naive approach takes more than six minutes. For subset selection via swapping, Algorithm \ref{alg:swap} typically runs in less than half a second for one initialization, while this naive approach often takes over an hour (depending on the number of iterations until convergence). We remark that although the swapping algorithm finds better subsets, the greedy algorithm has comparable performance and is notably faster, especially considering it only needs to be run once for the largest $k$ of interest.  

\subsubsection{Other Approaches}

Another approach to finding approximate CSS solutions is formulating a convex relaxation of the original objective \citep{Balzano, WangY, Masaeli}. The group lasso \citep{Yuan} provides one such natural convex relaxation:
\begin{equation}
\label{eq:group_lasso}
\argmin_{\B \in  \R^{p \times p}} \|\X -  \X \B\|_F^2 + \lambda \sum_{i=1}^p \|\B^{\textcolor{revision}{T}}_{i\bullet} \|_2
\end{equation}
For large values of $\lambda$, the solution to \eqref{eq:group_lasso} has many zeroed-out rows, i.e., many variables are not used in the reconstruction.  This suggests the following procedure for CSS: tune $\lambda$ in \eqref{eq:group_lasso} to yield exactly $k$ non-zero rows and select the variables corresponding to these rows. Another approach from \cite{Masaeli},  Convex Principal Feature Selection (CPFS), solves \eqref{eq:group_lasso} but with an $\ell_{\infty}$ penalty on the rows of $\B$ instead of a $\ell_2$ one.

We can characterize \eqref{eq:group_lasso} solely in terms of covariances and therefore also apply this group lasso approach to BlackRock's data. Defining $\hat{\bSigma} = \X^T\X/n$, \eqref{eq:group_lasso} is equivalent to 
\begin{equation*}
\argmin_{\B \in  \R^{p \times p}} \tr(\hat{\bSigma} - 2\B^T\hat{\bSigma} + \B^{T}\hat{\bSigma}\B) + \frac{\lambda}{n} \sum_{i=1}^p \|\B^{\textcolor{revision}{T}}_{i\bullet} \|_2
\end{equation*}

Surprisingly, on BlackRock's data, this group lasso approach performs worse than selecting subsets at random! \Cref{fig:BlackRock} displays the group lasso's performance as well as the best performance from 25 randomly selected subsets. The subset chosen by the group lasso performs notably worse than the best random subset. We were unable to tractably apply CPFS to the whole dataset, so we tried it on a random sub-sample of one hundred portfolios. On this sub-sample, CPFS did even worse than the group lasso (results not shown). Empirically, we found that the group lasso and CPFS systematically pulled in variables that were highly correlated with those that had already been selected. Our results are another example of convex variable selection methods performing poorly in settings with high correlation \citep{Freijeiro}.

\subsection{CSS with Missing and/or Censored Data}
\label{sec:missing_data}

The equivalence drawn in \Cref{sec:equivalence} suggests a natural workflow for performing CSS when some data is missing and/or censored: model the missingness and/or censorship, produce a covariance estimate that is reasonable under this model, and perform CSS using this covariance estimate. 

As an illustrative example of this workflow, we use it to suggest a procedure for performing CSS with missing-at-random data. Suppose we observe i.i.d samples $x^{(1)}, \dots, x^{(n)} \in \R^p$ and some values are missing-at-random. Let the set $\mathcal{I}_{r}$ be the indices of samples for which the $r$th variable is not missing, and $\mathcal{I}_{rs}$ be the indices of samples for which the $r$ and $s$th variables are not missing. To get a reasonable covariance estimate $\hat{\bSigma}$, we estimate the pairwise covariances using the not-missing values and then project onto the positive semi-definite cone:
\begin{equation*}
    \hat{\bSigma} = \argmin_{\M \in \S_{+}^{p \times p}} \|\M - \hat{\boldsymbol{\Psi}} \|_F^2 \qquad  \hat{\boldsymbol{\Psi}}_{rs} = \frac{1}{|\mathcal{I}_{rs}|} \sum_{i \in \mathcal{I}_{rs}} \left(x_r^{(i)} - \frac{1}{|\mathcal{I}_{r}|}\sum_{j \in \mathcal{I}_{r}} x_r^{(j)}\right) \left(x_s^{(i)} - \frac{1}{|\mathcal{I}_{s}|}\sum_{j \in \mathcal{I}_{s}} x_s^{(j)}\right)
\end{equation*}
Projecting a symmetric matrix onto the positive semi-definite cone amounts to setting its negative eigenvalues to zero. Finally, we apply the algorithms from \Cref{sec:algorithms} to $\hat{\bSigma}$.

Using both simulated and real data, we demonstrate that our procedure outperforms the two existing methods for CSS with missing data. Initial work on this problem by \cite{Balzano} uses a greedy Block Orthogonal Matching Pursuit (BOMP) algorithm to approximately solve a CSS-like objective that ignores the missing values. In follow-up work, \cite{WangY} mask the missing values and solve a group lasso penalized problem similar to \eqref{eq:group_lasso}. Recalling our PCSS model \eqref{eq:pcss_model}, we remark that one could put a parametric assumption on the principal variable distribution $F$ and then apply the expectation-maximization (EM) algorithm to handle missingness. We found this to be too sensitive to the distributional assumptions on $F$ and the EM algorithm's initialization to work well in practice. 

\subsubsection{A Simple Simulation Study}
\label{sec:missing_sim}
We run a simulation study to evaluate the quality of each method. In each trial, we draw $n=200$ samples from a particular distribution in our PCSS model \eqref{eq:pcss_model} and randomly omit each observed value with probability $0.05$. The distribution we sample from has $p = 20$ unit-variance variables and a ground truth subset size of $k=4$. It is described in full detail in \Cref{sec:missing_data_dist_appdx}.

For each trial, we select a size-four subset using each method. We also select a random size-four subset as a benchmark. For our method, we use our swapping algorithm and try ten random initializations. For the group lasso, we find the regularization strength that selects a size-four subset via a binary search. We run one thousand trials and use three metrics to measure the quality of each method: (1) whether or not the selected subset $\hat{S}$ is exactly equal to the ground truth subset $S$; (2) the size of the intersection of $\hat{S}$ and $S$; and (3) the selected subset's population CSS objective $\tr(\bSigma - \bSigma_{\bullet\hat{S}}\bSigma_{\hat{S}}^{-1} \bSigma_{\hat{S}\bullet})$, where $\bSigma$ is the population covariance. Recovery of the ground truth subset $S$ is indeed a reasonable performance metric as, in our example, $S$ achieves the lowest population CSS objective among all size-$4$ subsets.  

Table \ref{table:missing_sim} documents the results from our simulation. Our method selects the correct subset every time while BOMP selects it less than half the time and group lasso fails to select it even once. As expected, our method also achieves the lowest average CSS objective value by a significant amount.

\begin{table}
\centering
\begin{tabular}{ |p{2cm}||p{4cm}|p{4.2cm}|p{2.5cm}|}
 \hline
Method & \% Corr. Subset Selected & \# Corr. Variables Selected & CSS Obj.   \\
 \hline
 Our Method  &  $\boldsymbol{1.000}$  &  $\boldsymbol{4.000 \pm 0.000}$   & $\boldsymbol{2.400 \pm 0.000}$ \\
 BOMP &   $0.456$  &  $2.191 \pm 0.056$   & $3.223 \pm 0.025$\\
 Group Lasso &  $0.000$  &  $1.713 \pm 0.016$  & $5.806 \pm 0.037$\\
 Random & $0.001$ & $0.804 \pm 0.023$ & $6.065 \pm 0.045$ \\
 \hline
\end{tabular}
\caption{The proportion of times the correct subset was selected (\% Corr. Subset Selected), the average size of the intersection of the selected subset and correct subset (\# Corr. Variables Selected), and the population CSS objective (CSS Obj.) for the thousand simulated trials described in \Cref{sec:missing_sim}. The $\pm$ reports one standard error. Bold indicates best performance.}
\label{table:missing_sim}
\end{table}

\subsubsection{An Example with Ozone Level Detection Data}
\label{sec:missing_real}

We consider the Ozone Level Detection Dataset from the UCI machine learning repository \citep{Dua}. The dataset has $n=2535$ samples of $p=73$ variables. So that we can compute a ground truth CSS objective value for any selected subset, we drop samples with already missing values and are left with a sample size of $n=1847$. 

One hundred times, we randomly omit observed values with probability $q=0.05, 0.1, 0.2$ and select a size $k=5, 10, 20$ subset using our method and BOMP. For our method, we use our swapping algorithm and only try one random initialization. We neglect the group lasso approach because it is too slow for a problem of this scale. Given its noted bad performances, we expect it would have done poorly. To measure a selected subset's quality, we compute the CSS objective value it would have attained on the fully observed dataset. As a benchmark for how well we can expect to do with no missing data, we run our swapping algorithm on the fully observed dataset with one hundred different random initializations and record the best result. 

\begin{figure}
\centering
\includegraphics[scale=0.4]{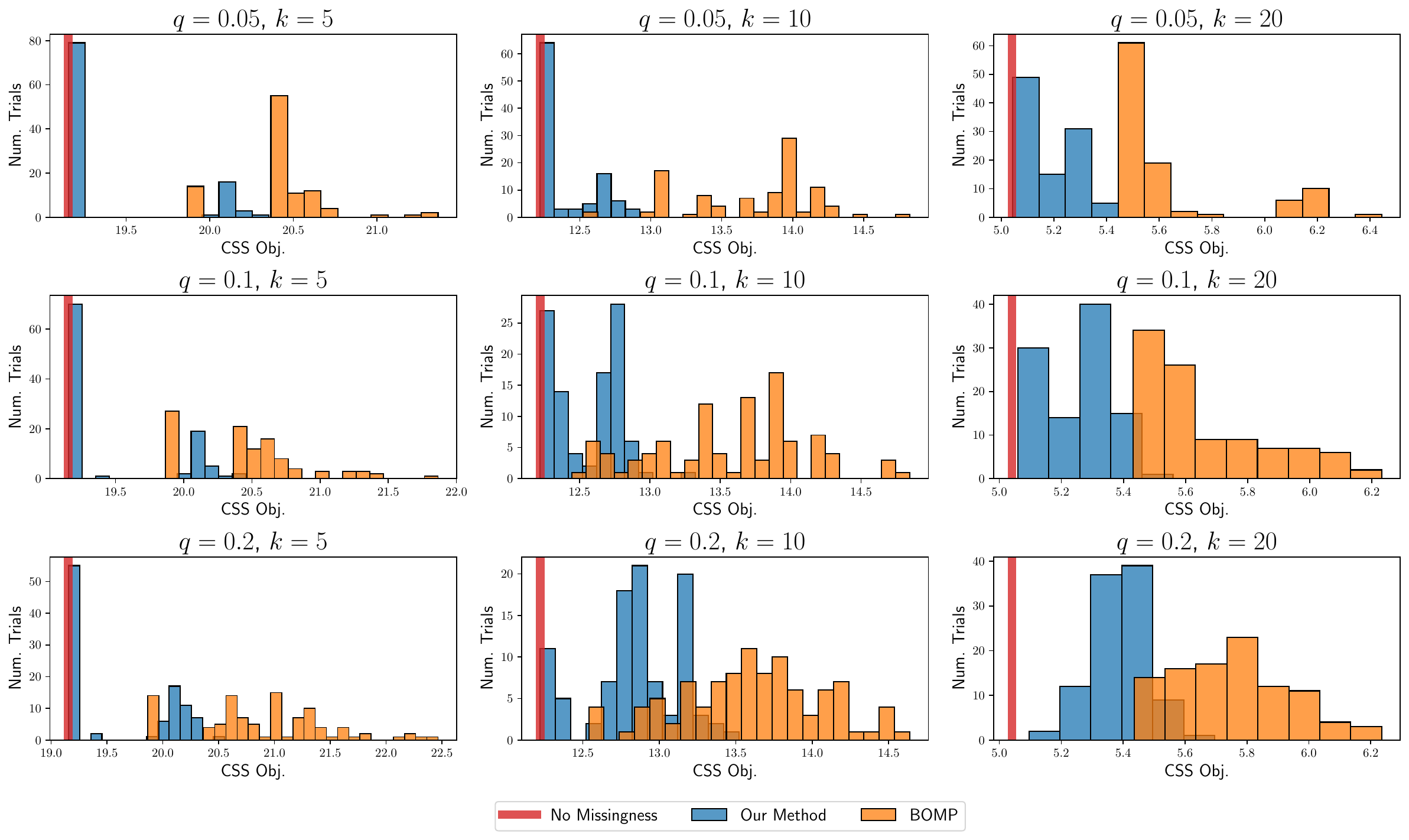}
\caption{For different amounts of missingness $q$ and subset sizes $k$, the selected subset's CSS objective value on the fully observed Ozone Level Detection data over the hundred trials described in \Cref{sec:missing_real}. The red line marks the CSS objective value attained by the subset that was selected using the fully observed data.}
\label{fig:ozone_results}
\end{figure}

We provide the results from our experiment in \Cref{fig:ozone_results}. Our method significantly outperforms BOMP in every setting and performs similarly to our no-missing-data benchmark when $k$ and $q$ are not too large. Also, our method's performance appears to be less variable than BOMP's. 

\subsection{Selecting the Subset Size for CSS} \label{sec:model_selection}

Our final application is, given some observed data, determining the smallest subset size needed to capture the data's underlying structure. Our formalization of this problem relies entirely on the generative view discussed in \Cref{sec:gen_view}, and the procedure we propose showcases how this generative view facilitates the application of statistical tools in subset selection problems. 

Formally, we aim to find $k^*$, the smallest non-negative integer $k$ for which the $k$-dimensional subset factor model \eqref{eq:subset_factor_model} holds. Recall that the $k$-dimensional subset factor model consists of distributions satisfying \eqref{eq:subset_factor_model} for some size-$k$ factor set $S$:
\begin{align}
\tag{\ref{eq:subset_factor_model}, revisited}
\begin{split}
&X_S \sim F \qquad \epsilon \sim (0, \D) \qquad \epsilon_1 \perp , \dots, \perp \epsilon_{p-k} \qquad \Cov(X_S, \epsilon) = \0 \\
&X_{-S} = \W(X_{S} - E_{F}[X_S]) + \mu_{-S} + \epsilon.
\end{split} 
\end{align}
We focus on the subset factor model over the PCSS model \eqref{eq:pcss_model} because it is more general and hence more representative of the structure we expect to see in real data. An interested reader can find the analogous discussion for the PCSS model in \Cref{sec:model_selection_pcss_appdx}.  Prior to proceeding, we point out that subset factor models of increasing subset size are nested. This is because any distribution satisfying \eqref{eq:subset_factor_model} for a set $S$ also satisfies \eqref{eq:subset_factor_model} for any superset of $S$. 

Our approach is to output $\hat{k}$, the smallest $k$ for which we cannot falsify that the data generating distribution belongs to the $k$-dimensional subset factor model. Formally, $\hat{k}$ is the smallest $k$ for which we fail to reject the null $H_{0, k}: |S| \leq k$ that the data is drawn from a distribution 
satisfying \eqref{eq:subset_factor_model} for some size-$k$ set $S$.  

To start, we present an idealized procedure that guarantees a notion of error control. This ideal procedure controls the probability that the suggested subset size is too large, i.e., that $\hat{k} > k^*$.  Unfortunately, the procedure is computationally intractable, and we ultimately use it to motivate a tractable procedure that performs well in practice. A more technical and detailed discussion of this ideal procedure can be found in \Cref{sec:model_selection_appdx}.

For a fixed $k$, our idealized procedure rejects the null $H_{0, k}: |S| \leq k$ when the statistic $T_k$ is large:
\begin{equation}
\label{eq:test_stat}
T_k = \min_{\substack{U \subset [p]: |U| = k}} T(U),  \qquad  T(U) = n \log \left(\frac{|\Diag(\hat{\bSigma}_{-U} - \hat{\bSigma}_{-U, U} \hat{\bSigma}_{U}^{+}\hat{\bSigma}_{U, -U})|}{|\hat{\bSigma}_{-U} - \hat{\bSigma}_{-U, U} \hat{\bSigma}_{U}^{+}\hat{\bSigma}_{U, -U}| } \right) 
\end{equation}
The statistic $T_k$ is the generalized likelihood ratio test statistic for testing the null that the data is drawn from a Gaussian distribution in the $k$-dimensional subset factor model (see  \Cref{sec:model_selection_test_stat_appdx} for details). Although the test statistic $T_k$ is designed for Gaussians, \textcolor{revision2}{we will be able to make use of it even in a totally non-parametric setting}. Intuitively, $T_k$ is large when the residual covariance from the regression of $X_{-U}$ on $X_{U}$ is highly non-diagonal for every size-$k$ subset $U$. If this is the case, the $k$-dimensional subset factor model cannot possibly hold. 

In what follows, we determine an appropriate critical value for the aforementioned test. By definition, the null $H_{0, k} : |S| \leq k$ is true exactly when $k \geq k^*$. Bearing this in mind, fix some $k \geq k^*$. \textcolor{revision2}{First, we imagine restricting} the subset factor model so that the unique factors $\epsilon_j$ are Gaussian and independent of the principal variables $X_S$, \textcolor{revision}{which we assume have a density}. In this simpler setting, whenever the data generating distribution satisfies \eqref{eq:subset_factor_model} for some size-$k$ set $S$, the statistic $T(S)$ has distribution exactly 
\begin{equation}
\label{eq:null_dist}
    n \sum_{j=2}^{p-k} \log \left(1 + \frac{\tilde{\chi}^{2}_{j - 1}}{\chi^2_{n- k - j}} \right),
\end{equation}
where $\{ \chi^2_{\ell} \}, \{ \tilde{\chi}^{2}_{\ell}\}$ are mutually independent chi-squared random variables with degrees of freedom specified by their subscript. Thus, rejecting \textcolor{revision2}{ $H_{0, k} : |S| \leq k$ } when $T_k =\min_{U: |U| = k} T(U)$ is larger than the critical value 
\begin{equation}
\label{eq:crit_value}
Q_{n,p,k}(1 - \alpha) = \textup{Quantile} \left\{ 1 - \alpha, n \sum_{j=2}^{p-k} \log \left(1 + \frac{\tilde{\chi}^{2}_{j - 1}} 
{\chi^2_{n- k - j}}\right) \right\}
\end{equation}
\textcolor{revision2}{results in exact finite sample type I error control at level $\alpha$. Surprisingly, even when we make no parametric assumptions, this same critical value guarantees asymptotic type I error control.}

\Cref{thm:error_control} establishes that, by using the above critical value, our ideal procedure does not suggest a subset size that is too large with high probability. 

\begin{theorem}[Error control]
\label{thm:error_control}
    Consider $n > p$ samples $x^{(1)}, \dots, x^{(n)} \in \R^p$ from a distribution $P$ and let $k^*$ be the smallest non-negative integer $k$ for which $P$ belongs to the $k$-dimensional subset factor model \eqref{eq:subset_factor_model}. If $\hat{k}$ is the smallest $k$ for which the test statistic $T_k$ from \eqref{eq:test_stat} is not strictly larger than the critical value $Q_{n, p, k}(1 - \alpha)$ defined in \eqref{eq:crit_value}, then 
    \begin{equation*}
        \limsup_{n \rightarrow \infty} P(\hat{k} > k^*) \leq \alpha.
    \end{equation*}
    That is, our ideal procedure suggests a subset size that is too large with probability at most $\alpha$ in large samples. Furthermore, if we restrict the subset factor model \eqref{eq:subset_factor_model} to have Gaussian unique factors $\epsilon_j$ that are independent of the principal variables $X_S$, then the above conclusion holds in finite samples, i.e., $P(\hat{k} > k^*) \leq \alpha$. 
\end{theorem}

With the result of \Cref{thm:error_control} in mind, we make two remarks about the output $\hat{k}$ of our ideal procedure. First, if the test statistic $T_k$ typically exceeds the critical value $Q_{n, p, k}(1-\alpha)$ whenever $k < k^*$ (i.e., the test is powerful for the nulls $H_{0, k}: |S| \leq k$ ,  $k < k^*$), our ideal procedure should return $k^*$ with high probability. Second, whenever our ideal procedure returns $\hat{k} > k^*$, the outputted $\hat{k}$ should seldom be much larger than $k^*$. Since $T_k$ minimizes $T(\cdot)$ over all size-$k$ subsets, the proposed critical value becomes increasingly conservative when the data generating distribution satisfies \eqref{eq:subset_factor_model} for many size-$k$ subsets. As $k$ exceeds $k^*$, the proportion of size-$k$ subsets for which \eqref{eq:subset_factor_model} is satisfied grows quickly to one. Also, for $k$ up to $\floor{(k^* + p)/2}$, there exist subsets $S$ for which \eqref{eq:subset_factor_model} is satisfied with progressively larger symmetric differences, and their corresponding statistics $T(S)$ are accordingly less positively correlated. These facts combined suggest that the probability of rejecting $H_{0, k}: |S| \leq k$ shrinks quickly as $k$ increases past $k^*$. 

Since computing the test statistic $T^*_k$ requires an exhaustive subset search, our idealized procedure is admittedly computationally intractable. Still, we can use it as motivation to design a tractable procedure that works well in practice. 

Our practical approach is to find a size-$k$ subset $\hat{S}$ that approximately minimizes $T(\cdot)$, and then to reject the null $H_{0, k}: |S| \leq k$ when $T(\hat{S})$ exceeds the same critical value $Q_{n, p, k}(1-\alpha)$ as before. We again output the smallest $k$ for which we fail to reject. Our approach is analogous to how, when the likelihood is not convex, a statistician may use an approximately maximized likelihood to run a likelihood ratio test (e.g., when selecting the number of factors to use in a factor model \citep{Beaujean}). \Cref{sec:other_algs_appdx} describes how to modify  \Cref{sec:algorithms}'s algorithms to search for this subset $\hat{S}$.

While our practical procedure does not guarantee error control, we still expect it to output $\hat{k}$ that will rarely be much larger than $k^*$. Again fix $k \geq k^*$. So long as our algorithms happen upon just one of the size-$k$ subsets for which \eqref{eq:subset_factor_model} is satisfied, we expect that $T(\hat{S})$ will be larger than $Q_{n, p, k}(1-\alpha)$ with probability at most $\alpha$ (and much less when $k$ is larger than $k^*$). As mentioned before, the proportion of size-$k$ subsets for which \eqref{eq:subset_factor_model} is satisfied quickly approaches one as $k$ increases, and it thus becomes correspondingly more likely that our approximation algorithms will happen upon such a subset.

Ultimately, if our procedure suggests selecting a size-$\hat{k}$ subset, we suggest selecting the size-$\hat{k}$ subset $\hat{S}$ that approximately minimizes $T(\cdot)$. This subset results in the most diagonal residual covariance we can find, and is thus the subset that best fits the data as per the subset factor model. Also, upon restricting the subset factor model to multivariate Gaussian distributions, $\hat{S}$ acts as an approximation for the MLE of $S$ (see \Cref{sec:model_selection_selected_subset_appdx} for details). 

In what follows, we verify that our practical procedure suggests reasonable subset sizes in some difficult simulated settings \textcolor{revision2}{where the number of samples $n$ is not much larger than the number of variables $p$}. Then, we demonstrate how it can be applied to dramatically reduce the length of a personality survey. 

\subsubsection{A Simulated Example}
\label{sec:model_selection_sim}

We consider a difficult simulated example that reflects the real data example to be presented in \Cref{sec:model_selection_survey}. For each trial, we draw $n=200$ samples from a mean-zero distribution in the subset factor model  \eqref{eq:subset_factor_model} with $p = 50$ variables and a size $k = 20$ population factor set $S$. For context, there are over 47 trillion size-20 subsets of 50 variables. We consider two cases: one where the unique factors are Gaussian, and one where they are a collection of independent Rademacher, $t$-distributed, and centered exponential random variables. Within both cases we consider distributions with varying amounts of ``signal'', which we define to be the average population $R^2$ from the regression of the variables not in $S$ on those in $S$. \Cref{sec:model_selec_dist_appdx} provides exact descriptions of the distributions we use, and \Cref{sec:model_selec_results_appdx} documents additional results for other sample sizes $n$.

For each trial we apply our procedure at level $\alpha=0.05$ and record (1) the size of the selected subset $\hat{S}$; (2) the size of the intersection of $\hat{S}$ and $S$; and (3) the population sum of square canonical correlations between the variables in $\hat{S}$ and $S$. The sum of square canonical correlations measures the degree to which the variables in $S$ and $\hat{S}$ span similar subspaces (see Section 3.1 of \cite{Schneeweiss} for more detail), and it attains its maximum value of $k=20$ when $S \subseteq \hat{S}$. To search for the subset that minimizes $T(\cdot)$ we use our swapping algorithm and try ten random initializations. \textcolor{revision}{Separately, we tried searching for the minimizing subset via a forward-backward algorithm. Although this approach is faster, the results were appreciably worse. These additional results as well as a description of our forward-backward algorithm can be found in \Cref{sec:forward_backward_appdx}.} 

\begin{figure}
  \centering
  \begin{subfigure}{\textwidth}
    \centering
    \includegraphics[scale=0.315]{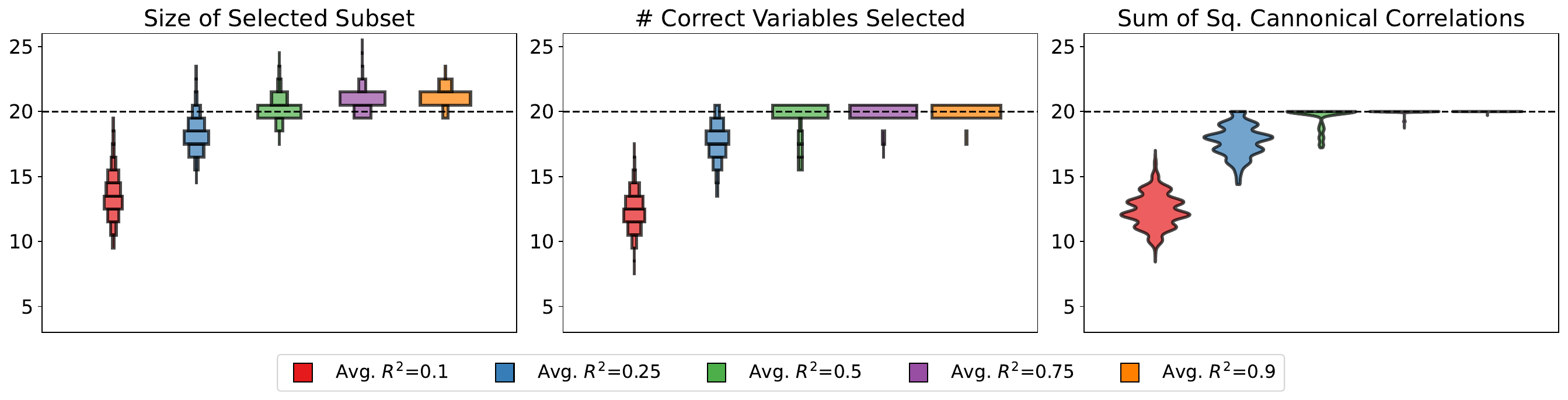}
    \caption{Gaussian unique factors}
  \end{subfigure}
  \bigskip 
  \begin{subfigure}{\textwidth}
    \centering
    \includegraphics[scale=0.315]{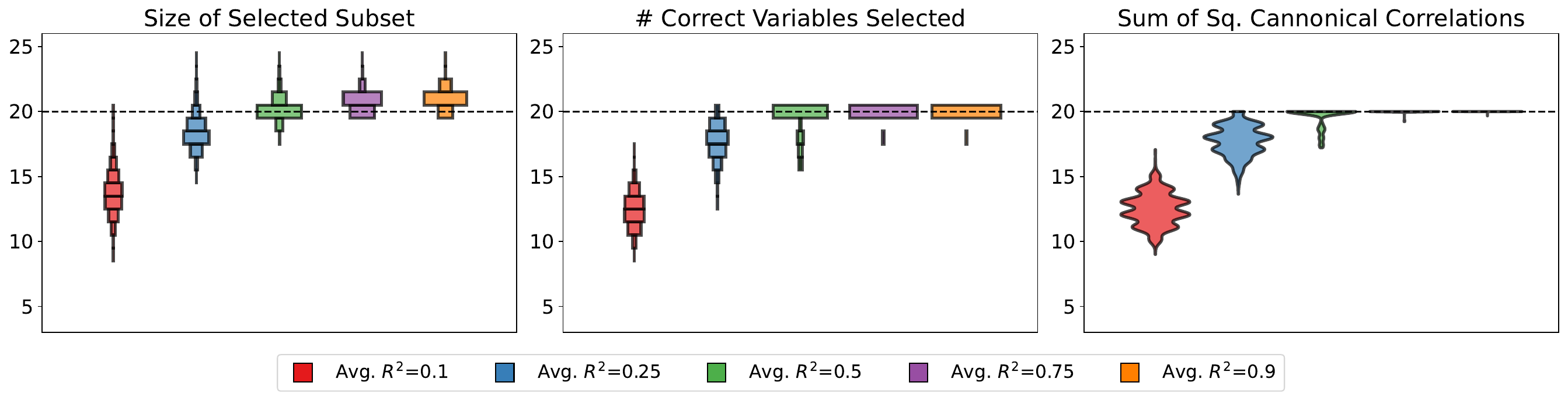}
    \caption{Non-Gaussian unique factors}
  \end{subfigure}
  
  \caption{\textcolor{revision}{For one thousand trials, the size of the selected subset $\hat{S}$ (left), the size of the intersection of $\hat{S}$ and the population subset $S$ (middle), and the population sum of squared canonical correlations between the variables in $\hat{S}$ and $S$ (right) for the different simulation settings described in \Cref{sec:model_selection_sim}. Color indicates the average population $R^2$ from the regression of the variables not in $S$ on those in $S$.} }
  \label{fig:model_selec_sim}. 
\end{figure}

We provide the results from our simulations in \Cref{fig:model_selec_sim}. As expected, our method under selects when the signal size is too small. In these cases, the covariance looks reasonably diagonal to begin with, so it is difficult to gauge whether more variables are needed to achieve a diagonal residual covariance. For reasonable to large signal sizes the procedure typically suggests selecting 20 or 21 variables, and it seldom suggests selecting more than 22. In these settings, the selected subset $\hat{S}$ is almost always a superset of the population subset $S$. The results when the unique factors are non-Gaussian versus Gaussian look nearly identical, suggesting that our procedure has reasonable finite sample performance even when the unique factors are highly non-Gaussian \textcolor{revision2}{and $n$ is not much larger than $p$}.

\subsubsection{Reducing Survey Length via Our Procedure}
\label{sec:model_selection_survey}

In recent years, psychologists have been increasingly interested in creating shorter versions of existing surveys. \cite{Kruyen} found that, among the 2273 articles that appeared in six peer-reviewed psychological journals from 2005 to 2010, 170 articles contained abbreviated surveys. Developing a novel shortened survey was a main contribution in 84 of these articles.

Despite the increasing demand for shortened surveys, there is no standardized way of shortening them. Psychologists default to a number of arbitrary heuristic approaches that require extensive human labor and are susceptible to personal biases. Furthermore, reliability, the one concrete and standard measure of a shortened survey's quality, \textcolor{revision}{is defined in the restrictive setting of classical test theory \citep{Allen}. Estimates of reliability are highly sensitive to classical test theory's assumptions, and these assumptions are often violated in modern, multi-dimensional settings \citep{McNeish}.}

We propose using our procedure to shorten surveys. Essentially, we suggest selecting a subset of questions that, once accounted for, render the information in the remaining questions independent. To demonstrate our procedure's effectiveness, we consider the Big Five Inventory (BFI) \citep{John}, a $p = 44$ question personality survey. There has been much interest in reducing the BFI survey, as evidenced by articles proposing a 10-question version \citep{Rammstedt}, a 15-question version \citep{Gerlitz}, and multiple 20-question versions \citep{Engvik, Gouveia, Tucakovic}.  Each question in the survey is attributed to one of the five personality factors \citep{Digman}: extraversion, agreeableness, conscientiousness,  neuroticism, and openness. 

We consider the BFI dataset from \cite{Zhang}, which has responses from $n=228$ undergraduate students at a large US public university. It can be found on CRAN in the \texttt{EFAutilities} package. We apply our procedure at level $\alpha=0.05$ and search for the minimizing subset using our swapping algorithm with just one random initialization (using more than one random initialization gives identical results). Within less than a tenth of a second, our procedure selects a size-19 subset of the original 44 questions, cutting the survey in more than half.

\textcolor{revision}{
As is the case for all shortened surveys, we evaluate the quality of our survey by judging its validity and reliability \citep{Kruyen}. We point readers who are unfamiliar with validity and reliability to the relevant introductory sections of \cite{Kruyen}.} \newline

\noindent \textbf{Validity}: \textcolor{revision}{
 We evaluate the construct validity of our reduced survey, i.e., the degree to which it captures the relevant construct of interest. We cannot evaluate predictive validity (the degree to which our survey is predictive of an external criterion) because we are working with a stand-alone dataset. Examination of our selected questions suggest they indeed capture the relevant construct of interest. \Cref{table:factor_attribution} provides a breakdown of how many questions are attributed to each personality factor in both the original and the reduced survey. The selected questions are distributed fairly evenly across the five personality factors, and, in concordance with best practices \citep{Costello}, our method selects at least three questions per factor. Furthermore, performing a 5-factor exploratory factor analysis reveals that the reduced survey maintains the original survey's factor structure, another standard indicator of construct validity \citep{Kruyen}. \Cref{table:factor_loadings} displays the loadings from this factor analysis.}\newline 

\begin{figure}
\centering
\includegraphics[scale=0.58]{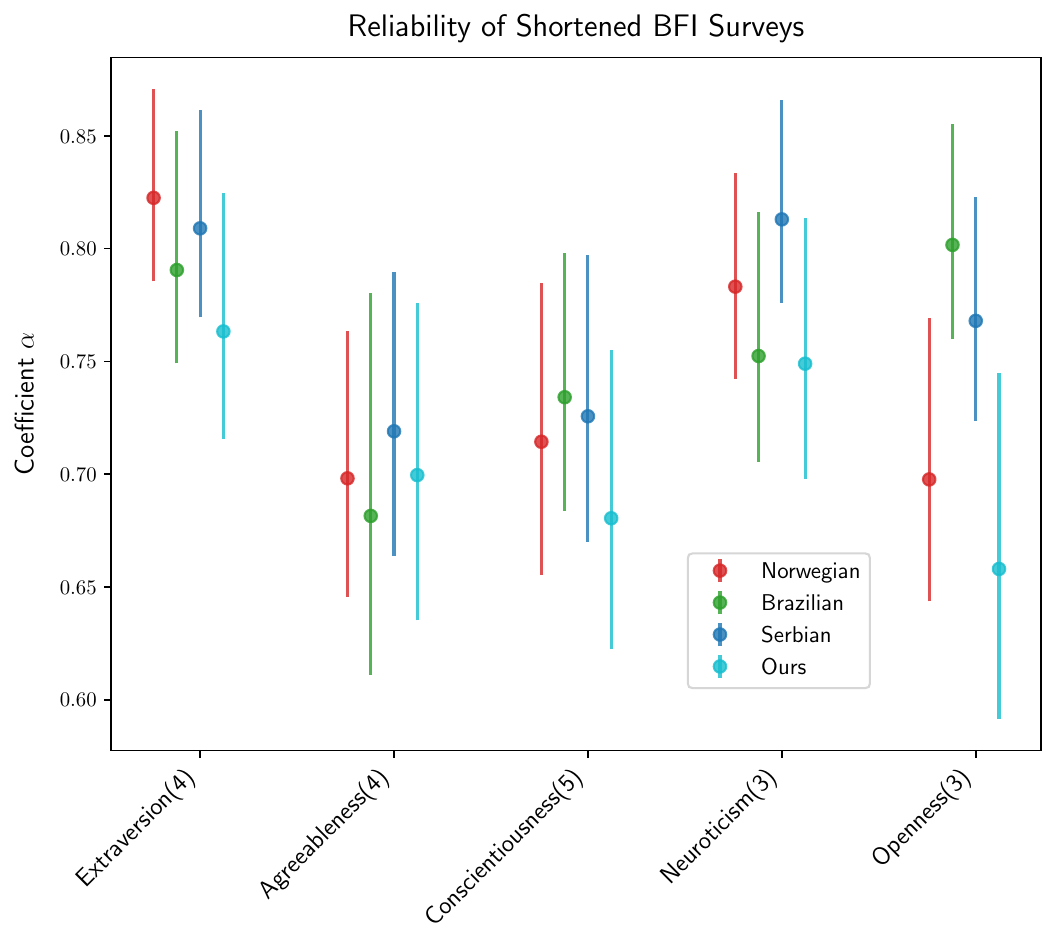}
\caption{\textcolor{revision}{For the five personality sub-surveys, we give Cronbach's coefficient $\alpha$ with 95\% bootstrap CIs ($B=1000$ bootstrap samples) for our survey and the Norwegian \citep{Engvik}, Brazilian  \citep{Gouveia}, and Serbian \citep{Tucakovic} 20-question surveys. The x-axis gives our number of selected questions per sub-survey. All other surveys chose four questions per sub-survey.}}
\label{fig:reliability}
\end{figure}

\noindent \textbf{Reliability}: \textcolor{revision}{ Reliability measures the consistency of a survey's responses across multiple re-administrations to the same individual. Although defining and estimating reliability relies heavily on the restrictive assumptions of classical test theory, it is still of much practical interest to pyschometricians. To measure reliability, we first split our reduced survey into five sub-surveys, each consisting of questions attributed to the same factor. Then we compute Cronbach's coefficient $\alpha$ \citep{Cronbach} for each of these five sub-surveys. For the sake of comparison, we do the same for the three pre-existing 20-question versions of the BFI survey. The results are given in \Cref{fig:reliability}. Along with being similar to the reliabilities of the three pre-existing surveys, our reliabilities are both (1) within the range of reliabilities one typically sees from shortened surveys \citep[Figure 3]{Kruyen} and (2) within the range of reliabilities contemporary pyschometricians consider adequate \citep{Clark}. Admittedly, our method aims to find the most parsimonious subset, and parsimony can be at odds with reliability. If the practitioner is particularly concerned about reliability, they can always add additional questions. } 

\begin{table}
\centering
\begin{tabular}{ |p{3cm}||p{3.8cm}|p{3.2cm}|}
 \hline
Factor & \# in Reduced Survey & \# in Original Survey   \\
 \hline
 Extraversion  & 4 &  8 \\
 Agreeableness  & 4  &  9 \\
 Conscientiousness  & 5 &  9 \\
 Neuroticism  &  3 &  8  \\
 Openness &  3 &  10 \\
 \hline
\end{tabular}
\caption{The number of questions attributed to each of the five personality factors for both the original and reduced surveys from \Cref{sec:model_selection_survey}.}
\label{table:factor_attribution}
\end{table}

\begin{table}[ht]
  \centering
  \begin{tabular}{|c|c|c|c|c|c|}
  \hline
   & Factor 1 & Factor 2 & Factor 3 & Factor 4 & Factor 5 \\
  \hline
  Extraversion 1 & 0.11 & \textbf{0.77} & 0.10 & -0.06 & -0.03 \\
  Extraversion 2 & -0.14 & \textbf{0.54} & 0.10 & 0.25 & 0.13 \\
  Extraversion 3 & 0.23 & \textbf{-0.54} & 0.07 & 0.19 & -0.05 \\
  Extraversion 4 & -0.02 & \textbf{0.81} & -0.01 & 0.10 & 0.02 \\
  Agreeableness 1 & 0.03 & 0.17 & -0.06 & \textbf{0.47 }& 0.15 \\
  Agreeableness 2 & 0.14 & -0.16 & 0.16 & \textbf{-0.66} & 0.02 \\
  Agreeableness 3 & 0.12 & -0.08 & 0.21 & \textbf{0.67} & 0.07 \\
  Agreeableness 4 & \textbf{0.35} & 0.09 & -0.06 & \textbf{-0.52} & -0.07 \\
  Conscientiousness 1 & 0.10 & 0.11 & -0.01 & 0.23 & \textbf{0.48} \\
  Conscientiousness 2 & 0.09 & 0.11 & 0.14 & -0.01 & \textbf{-0.60} \\
  Conscientiousness 3 & -0.03 & -0.09 & 0.24 & -0.03 & \textbf{0.55} \\
  Conscientiousness 4 & 0.06 & 0.11 & -0.02 & 0.05 & \textbf{0.66} \\
  Conscientiousness 5 & \textbf{0.50}& -0.03 & 0.01 & 0.12 & \textbf{-0.36} \\
  Neuroticism 1 & \textbf{-0.57 }& 0.02 & 0.03 & 0.05 & 0.10 \\
  Neuroticism 2 & \textbf{0.63} & -0.07 & -0.15 & 0.07 & -0.08 \\
  Neuroticism 3 & \textbf{0.79} & 0.03 & -0.02 & -0.13 & 0.09 \\
  Openness 1 & 0.01 & 0.11 & \textbf{0.68} & 0.01 & -0.06 \\
  Openness 2 & -0.10 & 0.04 & \textbf{0.76} & -0.01 & 0.03 \\
  Openness 3 & 0.05 & -0.14 & \textbf{0.50} & 0.07 & 0.00 \\
  \hline
  \end{tabular}
  \caption{Factor loadings (after rotation) from a exploratory factor analysis on the reduced 19-question survey from \Cref{sec:model_selection_survey}. As is common in the psychometrics literature, loadings with magnitudes larger than 0.30-0.32 are bolded \citep{Costello}.}
  \label{table:factor_loadings}
\end{table}

\section{Acknowledgements}
We thank Ron Kahn,  Ked Hogan, and  Bobby Luo from BlackRock for sharing ideas and their example with us.  We thank Stephen Boyd, John Cherian, Kevin Guo, Tim Morrison, Yash Nair, Asher Spector, Rob Tibshirani, and James Yang for helpful discussions. Anav Sood would like to especially acknowledge Kevin Guo for helpful discussions surrounding the probabilistic view and associated application. \textcolor{revision2}{We would also like to thank both rounds of anonymous referees for their valuable comments and suggestions that greatly helped improve the quality of this manuscript.} Trevor Hastie was partially supported by grant DMS-2013736 from the National Science Foundation, and grant 5R01 EB
001988-21 from the National Institutes of Health.

\bibliographystyle{plainnat}
\bibliography{bibliography.bib}

\begin{appendix}

\section{Additional Simulation Details and Results}
\label{sec:sims_appdx}

\subsection{Missing Data Simulation Setting}
\label{sec:missing_data_dist_appdx}

We describe the distribution used in \Cref{sec:missing_sim}. Recall the PCSS model:
\begin{align}
\tag{\ref{eq:pcss_model}, revisited}
\begin{split}
&X_S \sim F\\
&X_{-S} \mid X_S \sim N(\mu_{-S} + \W(X_S - E_{F}[X_{S}]), \sigma^2 \I_{p-k}). 
\end{split} 
\end{align}
We set the ground truth subset to be  $S = \{1, 2, 3, 4\}$ and let $X_{S}$ have a mean-zero multivariate Gaussian distribution with equicorrelated covariance $0.75\cdot \I_{4} + 0.25 \cdot 11^\top$. We set $\sigma^2 = 0.15$. The matrix $\W$ is given below:
\begin{equation*}
   \W = \begin{bmatrix}
        \sqrt{17/90} & \sqrt{17/90} & \sqrt{17/90} & 0 \\
        \sqrt{17/50} &  \sqrt{17/50} & -\sqrt{17/50} & 0 \\
        \sqrt{17/50} & -\sqrt{17/50} &  \sqrt{17/50} & 0 \\
        \sqrt{17/50} & -\sqrt{17/50} & -\sqrt{17/50} &  0 \\
       -\sqrt{17/50} &  \sqrt{17/50} &  \sqrt{17/50} & 0  \\
       -\sqrt{17/50} &  \sqrt{17/50} & -\sqrt{17/50} &  0 \\
       -\sqrt{17/50} & -\sqrt{17/50} & \sqrt{17/50} &  0  \\
       -\sqrt{17/90} & -\sqrt{17/90} & -\sqrt{17/90} &  0 \\
        0        &  \sqrt{17/90} &  \sqrt{17/90} &  \sqrt{17/90} \\
        0        &  \sqrt{17/50} &  \sqrt{17/50} & -\sqrt{17/50} \\
        0        &  \sqrt{17/50} & -\sqrt{17/50} & \sqrt{17/50} \\
        0        &  \sqrt{17/50} & -\sqrt{17/50} & -\sqrt{17/50} \\
        0        & -\sqrt{17/50} &  \sqrt{17/50} &  \sqrt{17/50} \\
        0        & -\sqrt{17/50} &  \sqrt{17/50} & -\sqrt{17/50} \\
        0        & -\sqrt{17/50} & -\sqrt{17/50} &  \sqrt{17/50}] \\
        0        & -\sqrt{17/90} & -\sqrt{17/90} & -\sqrt{17/90} 
    \end{bmatrix}
\end{equation*}

It is designed so the rows have non-zero entries of equal magnitude, rows with the same non-zero entries have a different configuration of signs, and the variance of each variable is one. 

\subsection{Simulation Setting for Selecting Subset Size}
\label{sec:model_selec_dist_appdx}

We describe the distribution used in \Cref{sec:model_selection_sim}. Recall the subset factor model:
\begin{align}
\tag{\ref{eq:subset_factor_model}, revisited}
\begin{split}
&X_S \sim F \qquad \epsilon \sim (0, \D) \qquad \epsilon_1 \perp , \dots, \perp \epsilon_{p-k} \qquad \Cov(X_S, \epsilon) = \0 \\
&X_{-S} = \W(X_{S} - E_{F}[X_S]) + \mu_{-S} + \epsilon.
\end{split} 
\end{align}
We set the ground truth subset to be  $S = \{1, \dots, 20\}$ and let $X_{S}$ have a mean-zero multivariate Gaussian distribution with block diagonal covariance.  Specifically there are five equally sized blocks, each of which is the equicorrelated matrix $0.5\cdot \I_{4} + 0.5 \cdot 11^\top$. The matrix $\W$ is somewhat sparse, and the non-zero entries are $1$ or $-1$ with probability $1/2$. We document the non-zero entries of $\W$ here:

\begin{equation*}
    \footnotesize{
    \W_{1:7, 1:8} = \begin{bmatrix} -1 &  1 &  1 & -1 &  1 &  1 &  1 &  1 \\  
       -1 &  1 &  1 & -1 & -1 &  1 &  1 &  1 \\
       -1 &  1 & -1 &  1 &  1 &  1 &  1 &  1 \\
        1 & -1 &  1 & -1 &  1 & -1 & -1 & -1 \\
       -1 &  1 & -1 &  1 &  1 &  1 &  1 &  1 \\
        1 & -1 & -1 &  1 & -1 & -1 & -1 &  1 \\
        1 &  1 &  1 &  1 &  1 & -1 &  1 &  1  \end{bmatrix}},
    \footnotesize{ 
    \W_{8:14, 5:12} = 
        \begin{bmatrix} 1 & -1 &  1 &  1 &  1 &  1 & -1 & -1 \\
        1 &  1 & -1 & -1 &  1 & -1 &  1 & -1 \\
        1 & -1 & -1 & -1 & -1 & -1 &  1 & -1 \\
       -1 &  1 &  1 & -1 & -1 & -1 &  1 &  1 \\
        1 &  1 & -1 & -1 &  1 &  1 &  1 &  1 \\
        1 &  1 &  1 & -1 & -1 & -1 &  1 &  1 \\
       -1 &  1 &  1 & -1 &  1 &  1 &  1 &  1 \end{bmatrix}}
\end{equation*}

\begin{equation*}
    \footnotesize{\W_{15:21, 9:16} = \begin{bmatrix} 1 &  1 &  1 &  1 & -1 &  1 & -1 & -1 \\
        1 &  1 &  1 & -1 & -1 &  1 &  1 &  1 \\
        1 &  1 & -1 &  1 &  1 &  1 &  1 &  1 \\
        1 & -1 &  1 & -1 &  1 & -1 & -1 & -1 \\
        1 &  1 &  1 & -1 &  1 & -1 & -1 &  1 \\
        1 &  1 & -1 &  1 &  1 &  1 &  1 &  1 \\
        1 & -1 &  1 &  1 &  1 & -1 & -1 &  1 \end{bmatrix}}, 
    \footnotesize{\W_{22:30, 13:20} =
    \begin{bmatrix} 1 &  1 & -1 & -1 &  1 &  1 &  1 & -1 \\
       -1 &  1 & -1 & -1 &  1 & -1 & -1 & -1 \\
        1 &  1 & -1 & -1 &  1 &  1 &  1 &  1 \\
        1 & -1 &  1 & -1 &  1 & -1 &  1 & -1 \\
       -1 &  1 & -1 & -1 & -1 &  1 & -1 &  1 \\
       -1 & -1 & -1 & -1 &  1 &  1 & -1 &  1 \\
       -1 & -1 & -1 & -1 &  1 &  1 & -1 & -1 \\
        1 & -1 & -1 &  1 &  1 &  1 &  1 & -1 \\
       -1 & -1 & -1 & -1 & -1 &  1 & -1 &  1 \end{bmatrix}}
\end{equation*}
To define $\D$, first we define the diagonal matrix $\widetilde{\D}$ where $\widetilde{\D}_{ii} = i \mod 6$. Then we set $\D = s \cdot \tilde{\D}$, where we use $s \in \R$ to control the amount of signal. The values of $s$ we use are $s = 0.254, 0.812, 2.71, 9.2, 30.0$. The unique factors $\epsilon \sim (0, \D)$ are generated independently of the principal variables. When the unique factors are non-Gaussian, ten of them are $\textup{Exp}(1) - 1$ random variables, ten are Rademacher random variables that take value $1$ or $-1$ with probability $1/2$, and ten are $t$-distributed random variables with $3$ degrees of freedom. In the non-Gaussian case we scale the unique factors so their variance matches $\D$. The indices of the exponential random variables are $1, 11, 13, 17, 19, 20, 23, 24, 26, 27$, of the Rademacher random variables are $3,  4,  8,  9, 12, 14, 21, 22, 29, 30$, and of the $t$-distributed random variables are $2,  5,  6,  7, 10, 15, 16, 18, 25, 28$. These indices were chosen from a random permutation. 

We also ran the same simulations in the case where $\X_{S} \sim N(0, \I_{k})$, $\epsilon \sim N(0, \eta \I_k)$ for some $\eta > 0$, $X_S \perp \epsilon$, and the rows of $\W$ are randomly generated unit vectors. The results (not shown) are notably better than the ones we chose to display. 

\subsection{Additional Simulations for Selecting Subset Size}
\label{sec:model_selec_results_appdx}

\begin{figure}
  \centering
  \begin{subfigure}{\textwidth}
    \centering
    \includegraphics[scale=0.3]{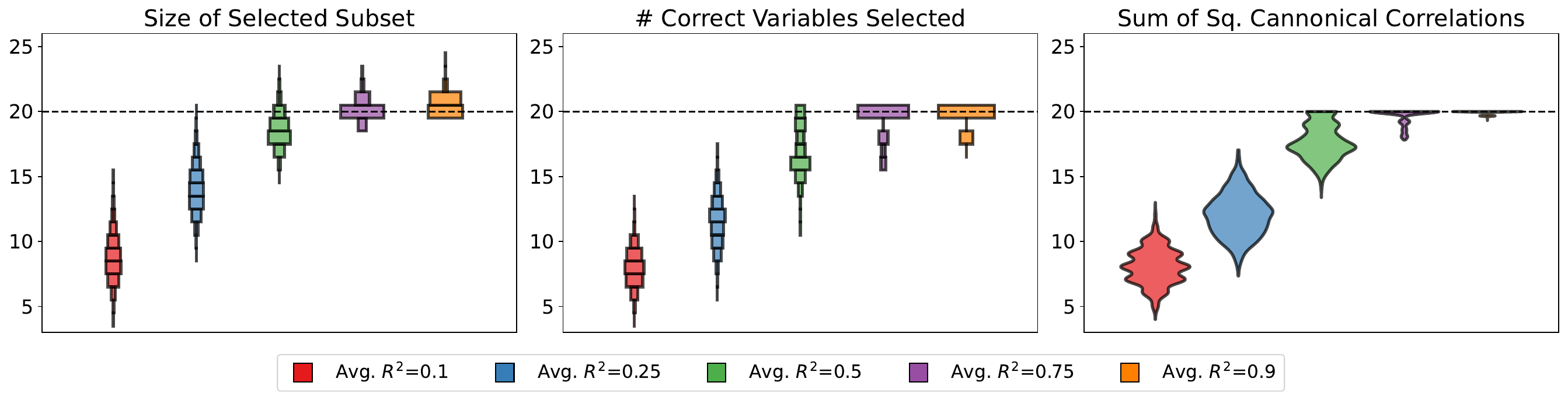}
    \caption{$n=100$}
  \end{subfigure}
  
  \begin{subfigure}{\textwidth}
    \centering
    \includegraphics[scale=0.3]{model_selec_results_n=200_gaussian.pdf}
    \caption{$n=200$}
  \end{subfigure}

  \begin{subfigure}{\textwidth}
    \centering
    \includegraphics[scale=0.3]{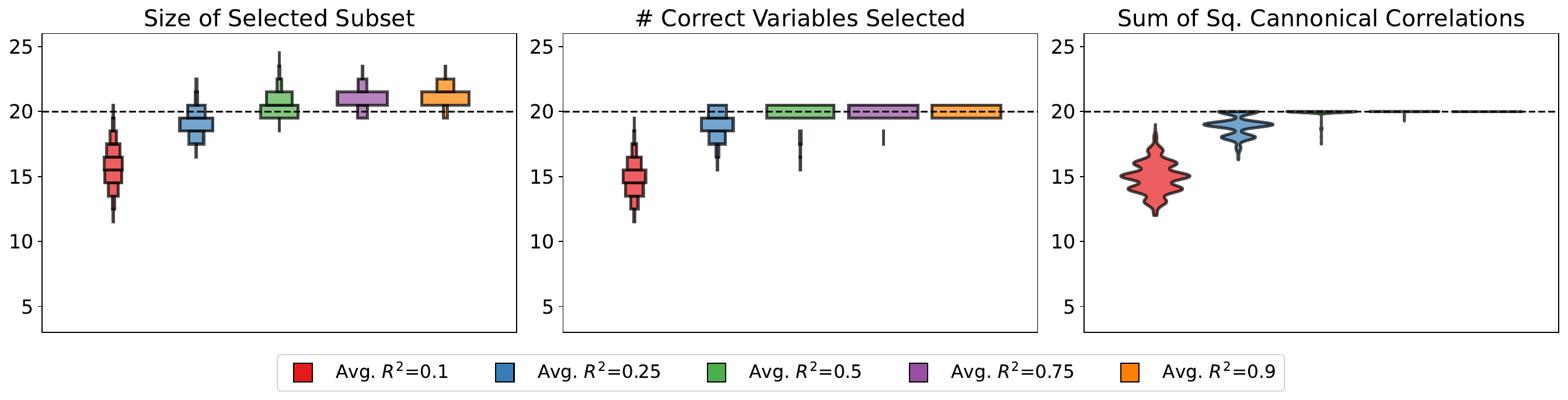}
    \caption{$n=300$}
  \end{subfigure}
  
  \begin{subfigure}{\textwidth}
    \centering
    \includegraphics[scale=0.3]{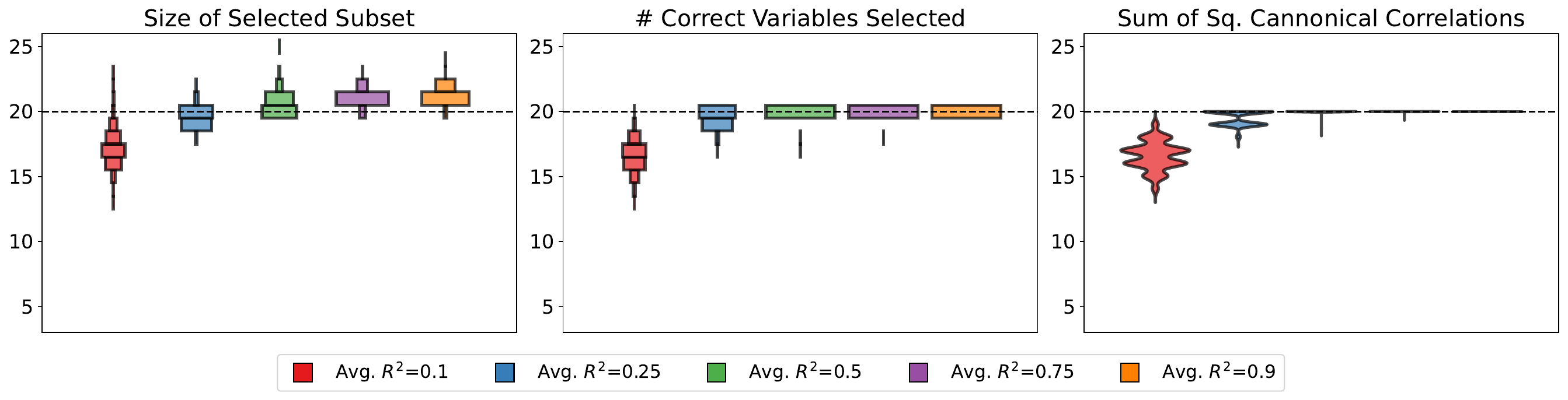}
    \caption{$n=400$}
  \end{subfigure}

  \begin{subfigure}{\textwidth}
    \centering
    \includegraphics[scale=0.3]{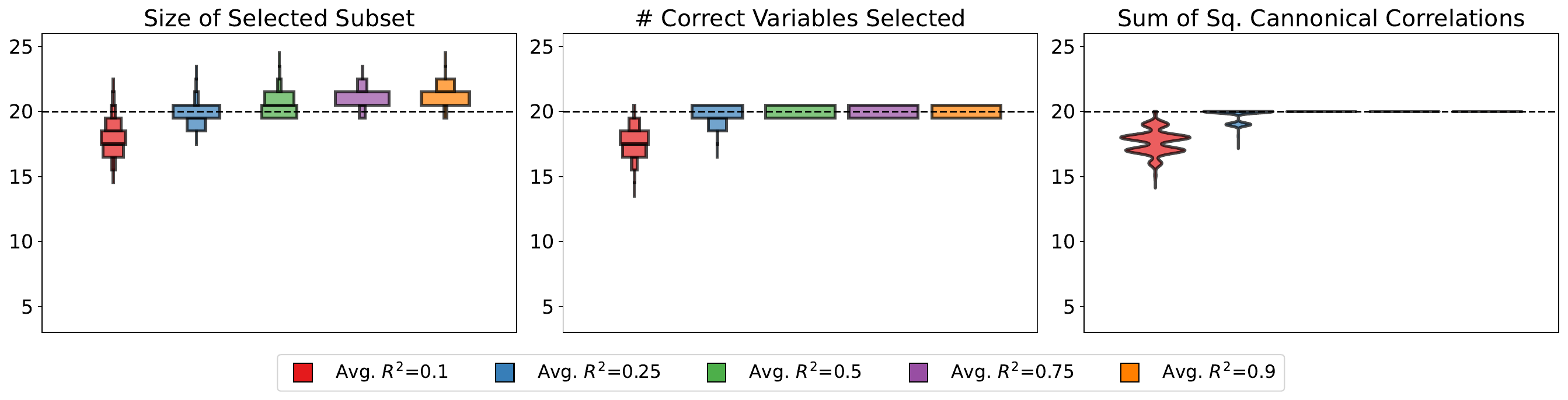}
    \caption{$n=500$}
  \end{subfigure}
  
  \caption{Results for the case of Gaussian unique factors. For description of the plots, see the caption of \Cref{fig:model_selec_sim}}
  \label{fig:gaussian_model_selec_sim}
\end{figure}

\begin{figure}
  \centering
  \begin{subfigure}{\textwidth}
    \centering
    \includegraphics[scale=0.3]{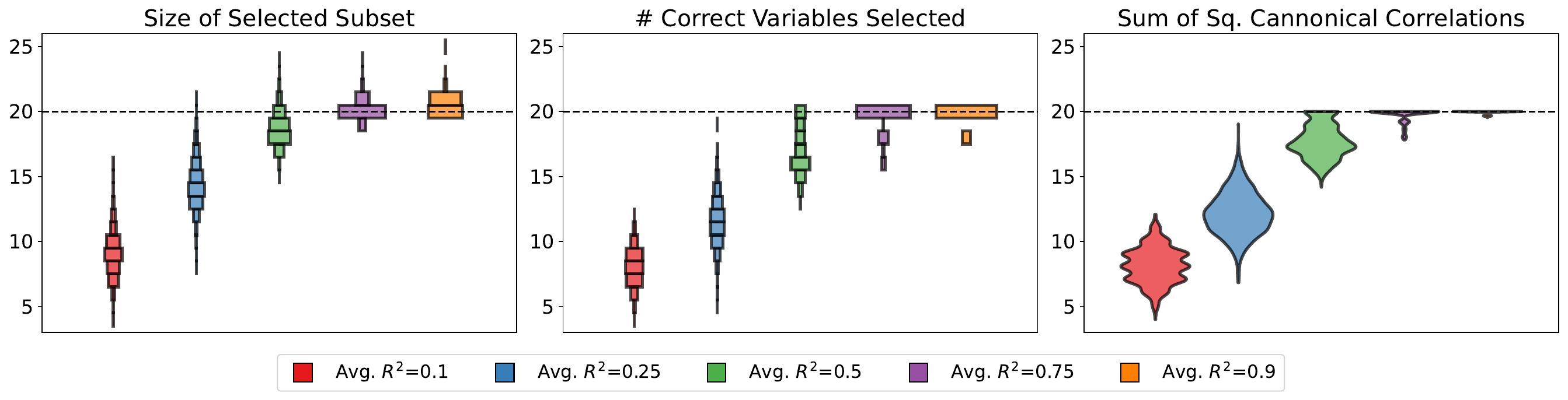}
    \caption{$n=100$}
  \end{subfigure}
  
  \begin{subfigure}{\textwidth}
    \centering
    \includegraphics[scale=0.3]{model_selec_results_n=200_non_gaussian.pdf}
    \caption{$n=200$}
  \end{subfigure}

  \begin{subfigure}{\textwidth}
    \centering
    \includegraphics[scale=0.3]{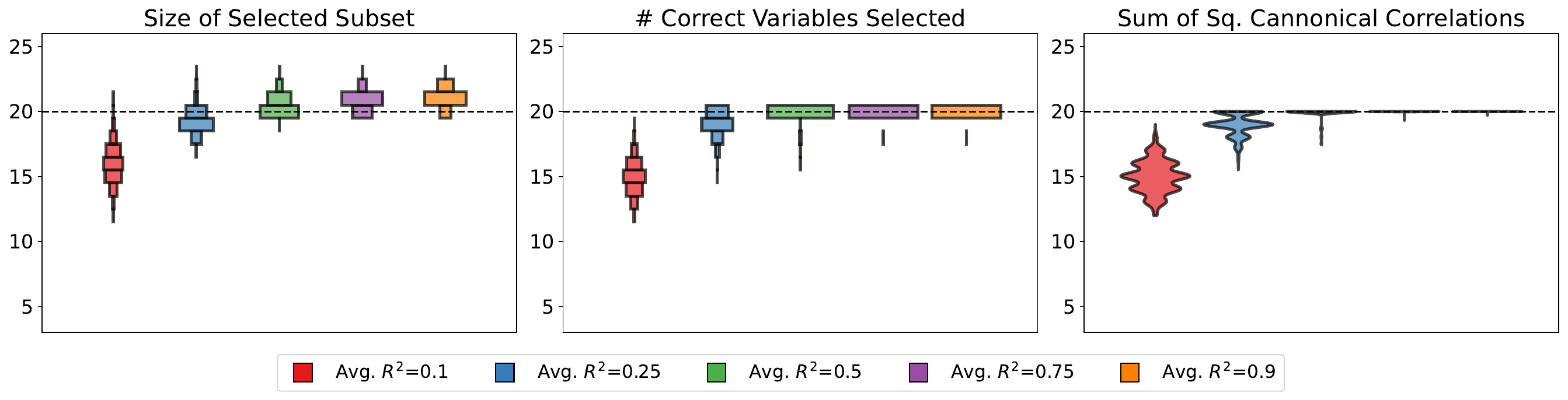}
    \caption{$n=300$}
  \end{subfigure}
  
  \begin{subfigure}{\textwidth}
    \centering
    \includegraphics[scale=0.3]{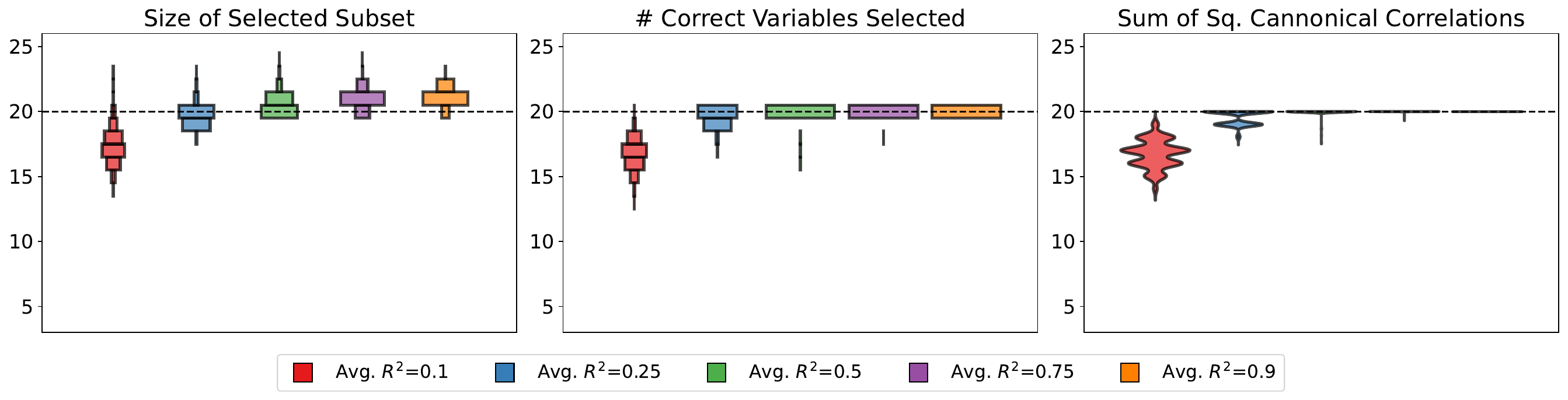}
    \caption{$n=400$}
  \end{subfigure}

  \begin{subfigure}{\textwidth}
    \centering
    \includegraphics[scale=0.3]{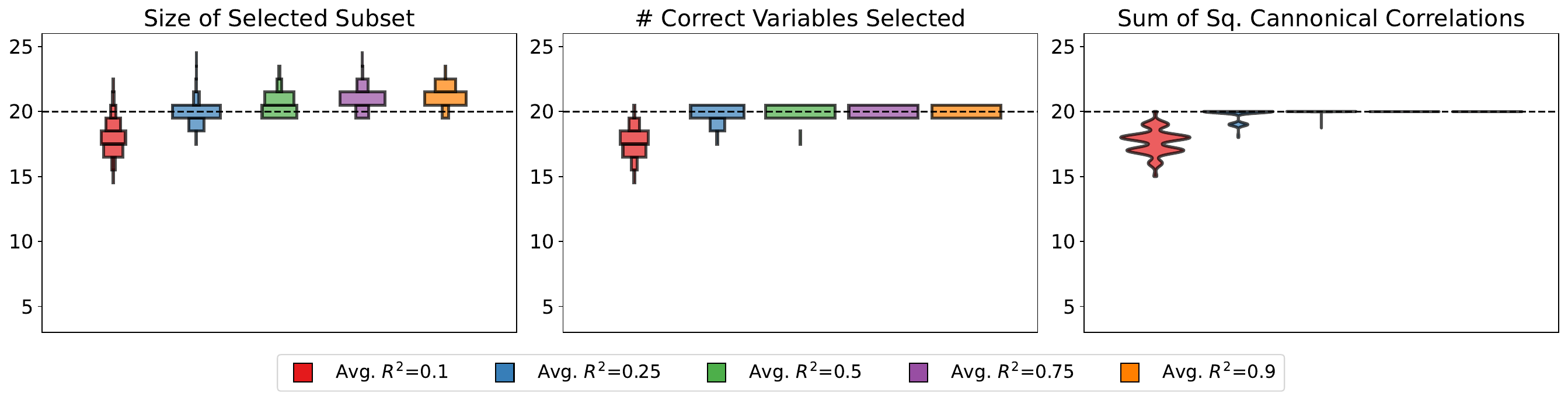}
    \caption{$n=500$}
  \end{subfigure}
  
  \caption{Results for the case of non-Gaussian unique factors. For description of the plots, see the caption of \Cref{fig:model_selec_sim}}
  \label{fig:non_gaussian_model_selec_sim}
\end{figure}

We give results from additional simulations, where we repeat the simulations from \Cref{sec:model_selection_sim} but for different sample sizes $n$. \Cref{fig:gaussian_model_selec_sim} depicts the results for the Gaussian unique factor case and \Cref{fig:non_gaussian_model_selec_sim} for the non-Gaussian unique factor case. 

\subsection{Selecting Subset Size with a Forward-Backward Algorithm}
\label{sec:forward_backward_appdx}
\textcolor{revision}{
We rerun our simulation from \Cref{sec:model_selection_sim} but now using a forward-backward algorithm to find the smallest subset size for which we fail to reject. This is a natural choice, as forward-backward algorithms are commonly used for model selection tasks \citep{Borboudakis}. Our algorithm completes a forward stage and then a backward stage. During each iteration of the forward stage, the algorithm greedily adds the variable that results in the largest decrease of the test statistic $T(\cdot)$ to the selected set. The forward stage continues until we find a subset for which we fail to reject. Then it begins the backward stage. On each iteration of the backward stage, the algorithm greedily removes the variable whose removal results in the smallest increase of $T(\cdot)$. The backward stage continues until removing any additional variable would result in a subset for which we reject. Conveniently, after one such forward and backward stage, our algorithm has converged, because removing any additional variables would result in rejecting, and there is no need to add more variables since we have already found a subset size for which we fail to reject.}

\begin{figure}
  \centering
  \begin{subfigure}{\textwidth}
    \centering
    \includegraphics[scale=0.315]{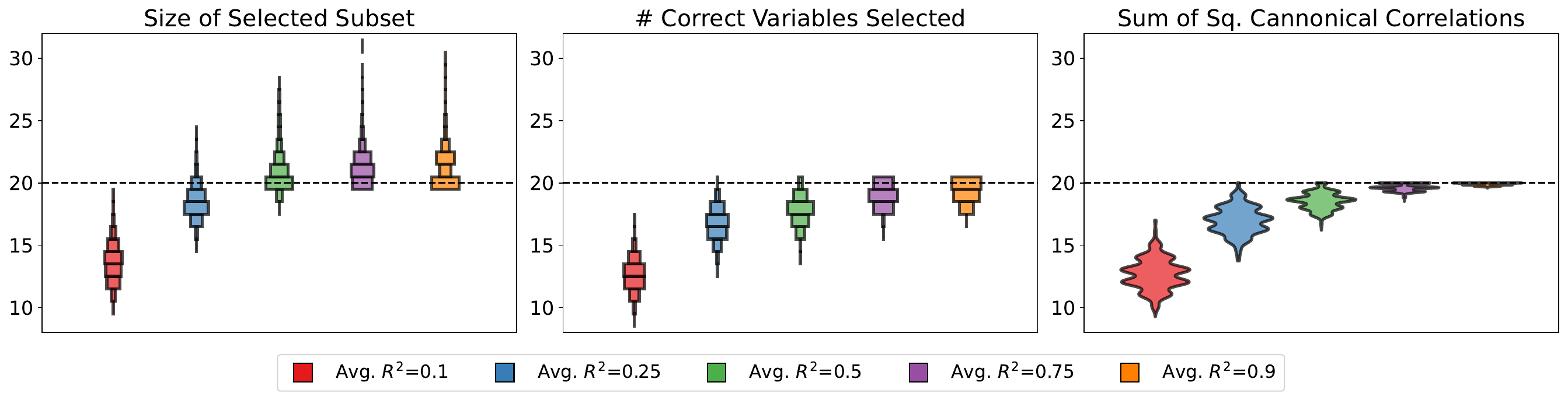}
    \caption{Gaussian unique factors}
  \end{subfigure}
  \bigskip 
  \begin{subfigure}{\textwidth}
    \centering
    \includegraphics[scale=0.315]{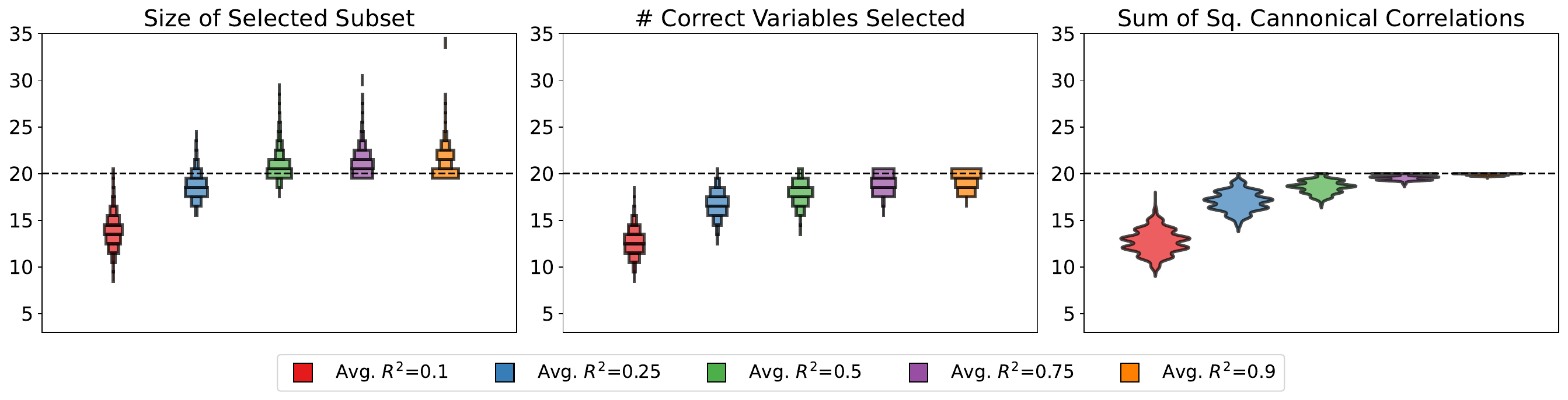}
    \caption{Non-Gaussian unique factors}
  \end{subfigure}
  
  \caption{\textcolor{revision}{Replication of the simulations whose results are in depicted \Cref{fig:model_selec_sim}, but now using our forward-backward algorithm to perform subset search. For description of the plots, see caption of \Cref{fig:model_selec_sim}.}}
  \label{fig:model_selec_sim_fb}
\end{figure}

\textcolor{revision}{
The results from replicating \Cref{sec:model_selection_sim}'s simulations with our forward-backward algorithm are given in \Cref{fig:model_selec_sim_fb}. Although the forward-backward algorithm is faster, it performs considerably worse than our swapping algorithm. Compared to the swapping algorithm, our forward-backward algorithm over selects more often and more drastically. Also, it often selects worse subsets, in the sense that the overlap between the variables it selects and the true population factor set is noticeably smaller.  When we replicate the experiment for sample sizes other than $n=200$ as we did for swapping in \Cref{sec:model_selec_results_appdx}, we see the same degradation in performance (results not shown).}

\section{More on Selecting the Appropriate Subset Size}
\label{sec:model_selection_appdx}
In this appendix, we provide more details surrounding the procedures presented in \Cref{sec:model_selection}. Recall that we want to identify $k^*$, the smallest $k$ for which the data generating distribution belongs to the subset factor model, and that the procedures we consider output $\hat{k}$, the smallest $k$ for which we are unable to reject the null that a size-$k$ subset is sufficient. In  \Cref{sec:model_selection_finite_sample_appdx} we consider the restricted case where the unique factors are Gaussian and independent of the principal variables, and establish a test for this null with finite sample validity. In \Cref{sec:model_selection_asymptotic_appdx}, we show that the same test maintains asymptotic validity in the unrestricted case. This is immediately sufficient to imply the correctness of \Cref{thm:error_control}, the proof of which is detailed in \Cref{thm:error_control:proof}. In \Cref{sec:model_selection_selected_subset_appdx}, we discuss some desirable properties of the (ideally) outputted subset $\hat{S}$ and some practical considerations surrounding searching for it. Finally, in \Cref{sec:model_selection_pcss_appdx} we give a discussion that establishes the analogous procedure for the PCSS model \eqref{eq:pcss_model}. 

Throughout we consider $n > p$ data points  $x^{(1)}, \dots, x^{(n)} \in \R^p$ that are sampled from some distribution and have sample mean $\hat{\mu}$ and sample covariance $\hat{\bSigma}$. 

\subsection{A Finite-Sample Test when the Unique Factors are Gaussian}
\label{sec:model_selection_finite_sample_appdx}

The goal of this section is to develop a test that is level $\alpha$ in finite samples for the null $\overline{H}_0: |S| \leq k$ that the data is drawn from from a distribution in the $k$-dimensional subset factor model \eqref{eq:subset_factor_model} with Gaussian unique factors $\epsilon$ that are independent of the principal variables $X_S$. This is the set of distributions that satisfy 
\begin{align}
\label{eq:diag_pcss_model}
\begin{split}
&X_S \sim F\\
&X_{-S} \mid X_S \sim N(\mu_{-S} + \W(X_S - E_{F}[X_{S}]), \D). 
\end{split} 
\end{align}
for some size-$k$ subset $S$, where $\D \succ 0$ is an arbitrary diagonal matrix. In other words, it is the PCSS model \eqref{eq:pcss_model} where the covariance of $X_{-S} \mid X_{S}$ is now allowed to be diagonal (rather than just isotropic). 

\subsubsection{Motivating a Test Statistic}
\label{sec:model_selection_test_stat_appdx}

Our test statistic more or less comes from a generalized likelihood ratio test (LRT). Specifically we compare the data's likelihood under a totally unrestricted multivariate Gaussian model to its likelihood under a restricted multivariate Gaussian model, where the restricted multivariate Gaussian model only considers distributions that are also in $k$-dimensional subset factor model. Suppose that $\hat{\bSigma} \succ \0$. It is well known that the minimized negative log-likelihood, scaled by $1/n$, under the unrestricted Gaussian model is given by 
\begin{equation*}
    \frac{1}{2}\log |\hat{\bSigma}| + \frac{p}{2}(1 + \log(2\pi))
\end{equation*}
From  \citep[9.1.2]{Petersen}, the above can be written as 
\begin{equation*}
     \frac{1}{2}\log |\hat{\bSigma}_{U}| + \frac{1}{2}\log |\hat{\bSigma}_{-U} - \hat{\bSigma}_{-U, U} \hat{\bSigma}^{-1}_{U}\hat{\bSigma}_{U, -U}| + \frac{p}{2}(1 + \log(2\pi)),
\end{equation*}
for any size-$k$ subset $U$. \Cref{sec:gaussian_pcss_mle_proof} in the proof of \Cref{thm:css_is_mle} tells us that the minimized negative log-likelihood, also scaled by $1/n$, under the restricted Gaussian model is 
\begin{equation*}
    \min_{U \subset [p] :|U| = k}\frac{1}{2} \log|\hat{\bSigma}_{U}| + \frac{1}{2}\log(|\Diag(\hat{\bSigma}_{-U} - \hat{\bSigma}_{-U, U}\hat{\bSigma}_{U}^{-1}\hat{\bSigma}_{U, -U})|) + \frac{p}{2}(1 + \log(2\pi)) 
\end{equation*}
The generalized LRT statistic is two times the difference in log likelihoods, and is thus given by  
\begin{equation*}
    \min_{U \subset [p] :|U| = k} n  \log \left( \frac{|\Diag(\hat{\bSigma}_{-U} - \hat{\bSigma}_{-U, U}\hat{\bSigma}_{U}^{-1}\hat{\bSigma}_{U, -U})|}{|\hat{\bSigma}_{-U} - \hat{\bSigma}_{-U, U} \hat{\bSigma}^{-1}_{U}\hat{\bSigma}_{U, -U}|} \right).
\end{equation*}

Our test statistic generalizes the LRT statistic by allowing for singular $\hat{\bSigma}$:
\begin{equation}
\tag{\ref{eq:test_stat}, revisited}
T_k = \min_{\substack{U \subset [p]: |U| = k}} T(U),  \qquad  T(U) = n \log \left(\frac{|\Diag(\hat{\bSigma}_{-U} - \hat{\bSigma}_{-U, U} \hat{\bSigma}_{U}^{+}\hat{\bSigma}_{U, -U})|}{|\hat{\bSigma}_{-U} - \hat{\bSigma}_{-U, U} \hat{\bSigma}_{U}^{+}\hat{\bSigma}_{U, -U}| } \right) 
\end{equation}
There may be size-$k$ subsets $U$ for which $|\Diag(\hat{\bSigma}_{-U} - \hat{\bSigma}_{-U, U}\hat{\bSigma}_{U}^{+}\hat{\bSigma}_{U, -U})| = |\hat{\bSigma}_{-U} - \hat{\bSigma}_{-U, U} \hat{\bSigma}^{+}_{U}\hat{\bSigma}_{U, -U}| = 0$, and we adopt the convention that $T(U) = 0$ for such $U$. In the case that $k = 0$, the same reasoning tells us that $T_0 = n\log\left( |\Diag(\hat{\bSigma})|/|\bSigma| \right)$. 

\subsubsection{Determining a Critical Value}
\label{sec:model_selection_crit_val_appdx}

We will come up with a critical value by studying the distribution of $T(S)$, where $S$ is a size-$k$ subset for which the data generating distribution satisfies \eqref{eq:diag_pcss_model}. Since $T_k = \min_{\substack{U \subset [p]: |U| = k}} T(U) \leq T(S)$, rejecting when $T_k$ is larger than the appropriate quantile of $T(S)$ will be sufficient to guarantee type I error control.   

To study the distribution of $T(S)$ we leverage techniques from fixed-X Ordinary Least Squares. It suffices to understand the distribution of $\hat{\bSigma}_{-S} - \hat{\bSigma}_{-S, S} \hat{\bSigma}_{S}^{+}\hat{\bSigma}_{S, -S}$, i.e, the distribution of the residual covariance from the regression of the $x_{-S}^{(i)}$ on the $x^{(i)}_S$. We are able to show that, conditional on the $x^{(i)}$,  this covariance has a Wishart distribution. Then, via the Wishart distribution's Bartlett decomposition \citep[Corollary 7.2.1]{anderson1958}, we establish that the quantiles of $T(S)$ are bounded above by those of 
\begin{equation}
\tag{\ref{eq:null_dist}, revisited}
    n \sum_{j=2}^{p-k} \log \left(1 + \frac{\tilde{\chi}^{2}_{j - 1}}{\chi^2_{n- k - j}} \right).
\end{equation}
In \eqref{eq:null_dist} $\{ \chi^2_{\ell} \}, \{ \tilde{\chi}^{2}_{\ell}\}$ are mutually independent chi-squared random variables with degrees of freedom specified by their subscript. When $k = p - 1$, \eqref{eq:null_dist} should be interpreted as a point mass at zero, and the same argument holds when $k = 0$. 

Denoting the $(1-\alpha)$ quantile of \eqref{eq:null_dist} as $Q_{n, p, k}(1-\alpha)$, \Cref{lem:finite_sample_validity} both summarizes this result and also establishes finite sample validity the test that rejects when $T_k > Q_{n, p, k}(1-\alpha)$. 
\begin{lemma}
\label{lem:finite_sample_validity}
    Consider $n > p$ samples  $x^{(1)}, \dots, x^{(n)} \in \R^p$ drawn from a distribution $P$ that satisfies  \eqref{eq:diag_pcss_model} for some size-$k$ set $S$ where $0 \leq k < p$. Then the quantiles of $T(S)$ are bounded above by the quantiles of \eqref{eq:null_dist}. As a consequence, $P(T_k > Q_{n, p, k}(1-\alpha)) \leq \alpha$. 
\end{lemma}

\subsection{An Asymptotic Test for the Subset Factor Model}
\label{sec:model_selection_asymptotic_appdx}

We now show that the test from \Cref{sec:model_selection_finite_sample_appdx} maintains asymptotic validity for the more general null $H_{0, k} : |S| \leq k$ that the data is drawn from a distribution in the $k$-dimensional subset factor model. Recall that the $k$-dimensional subset factor model is the set of distributions satisfying 
\begin{align}
\tag{\ref{eq:subset_factor_model}, revisited}
\begin{split}
&X_S \sim F \qquad \epsilon \sim (0, \D) \qquad \epsilon_1 \perp , \dots, \perp \epsilon_{p-k} \qquad \Cov(X_S, \epsilon) = \0 \\
&X_{-S} = \W(X_{S} - E_{F}[X_S]) + \mu_{-S} + \epsilon.
\end{split} 
\end{align}
for some size-$k$ subset $S$. 

\Cref{lem:asymptotic_dist} tells us that, should the data be drawn from a distribution satisfying \eqref{eq:subset_factor_model} for some size-$k$ set $S$, then $T(S)$ has a pivotal chi-squared limiting distribution. The degrees of freedom match what one would expect from Wilks' theorem \citep{Wilks}, although our setting is much more general. The proof is a careful application of the delta method and standard, but its result is surprising. Typically, we would expect the limiting distribution of a statistic like $T(S)$ to depend on fourth moments and therefore not be pivotal.

\begin{lemma}
    \label{lem:asymptotic_dist}
    Consider samples $x^{(1)}, \dots, x^{(n)} \in \R^p$ drawn from a distribution that satisfies \eqref{eq:subset_factor_model} for some size-$k$ set $S$ where $0 \leq k < p$. With $p$ and $k$ fixed, $T(S) \rightsquigarrow \chi^2_{(p-k)(p-k-1)/2}$ as the number of samples $n$ tends to infinity. 
\end{lemma}

In \Cref{lem:asymptotic_dist}, $\chi^2_{(p-k)(p-k-1)/2}$ should be interpreted as a point mass at zero when $k = p -1$. 

Using \Cref{lem:asymptotic_dist}, \Cref{lem:asymptotic_validity} establishes the asymptotic validity of the same test from \Cref{sec:model_selection_finite_sample_appdx} under the more general null $H_{0, k} : |S| \leq k$. Using \Cref{lem:asymptotic_dist}, we can easily show that the $(1-\alpha)$ quantile of \eqref{eq:null_dist} converges to the $(1-\alpha)$ quantile of the $\chi^2_{(p-k)(p-k-1)/2}$ distribution, which we will denote $Q_{p, k}(1-\alpha)$. Because $T(S)$ has a limiting distribution that admits a density, it is then immediate that the difference between $P(T(S) > Q_{n, p, k}(1-\alpha))$ and $P(T(S) > Q_{p, k}(1-\alpha))$ converges to zero, and this is sufficient to imply the result.  

\begin{lemma}
    \label{lem:asymptotic_validity}
    For samples $x^{(1)}, \dots, x^{(n)} \in \R^p$ drawn from a distribution $P$ that satisfies \eqref{eq:diag_pcss_model} for some size-$k$ set $S$ where $0 \leq k < p$, $\limsup_{n \rightarrow \infty} P(T_k > Q_{n, p, k}(1 - \alpha)) \leq \alpha$. 
\end{lemma}

Our decision to use the quantiles of \eqref{eq:null_dist} rather than those of the $\chi^2_{(p-k)(p-k-1)/2}$ distribution is intentional, even though the latter guarantee the same asymptotic validity. In many practical settings, $n$ is not sufficiently larger than $p$ for our asymptotic guarantees to be reflected in our finite sample results. Although rejecting according to the quantiles of \eqref{eq:null_dist} only guarantees finite sample validity when the unique factors are Gaussian (and independent of the principal variables), we may hope that this validity is somewhat robust to deviations from Gaussianity. If so, rejecting according to the quantiles of \eqref{eq:null_dist} should improve performance for smaller sample sizes while maintaining the same asymptotic guarantees. 

\begin{figure}
\centering
\includegraphics[scale=0.6]{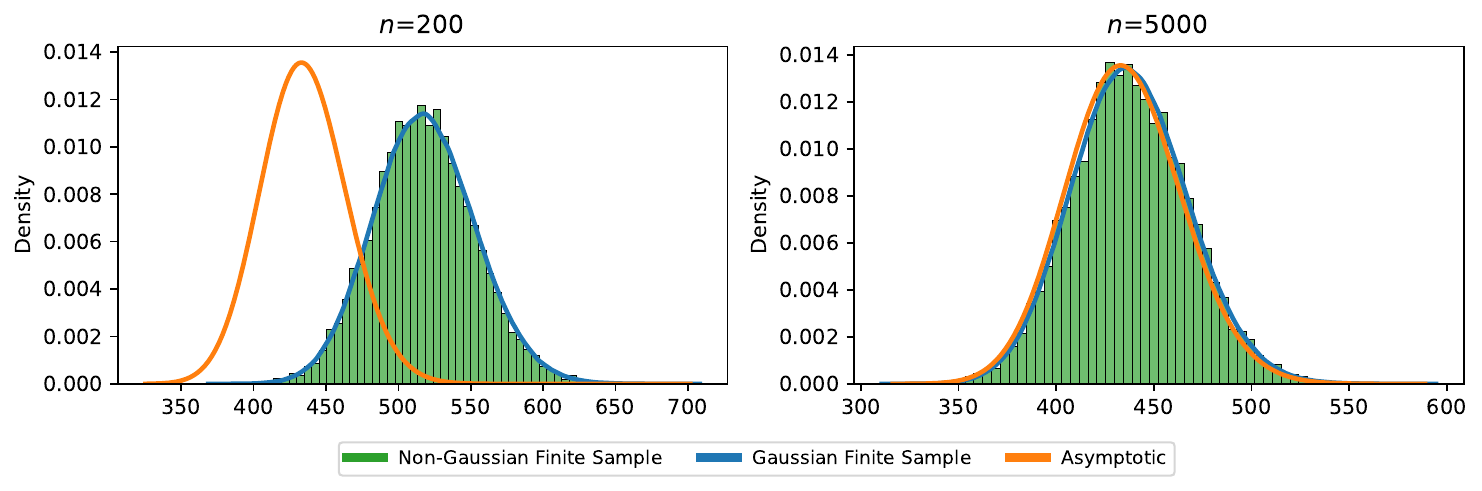}
\caption{In the case of $p=50$ variables and a size $k=20$ subset $S$, the exact finite sample distribution of $T(S)$ for $n=200$ (left) and $n=5000$ (right) samples in the cases of both Gaussian and non-Gaussian unique factors, plotted against its limiting $\chi^2_{(p-k)(p-k-1)/2}$ distribution.}
\label{fig:dist_comparison}
\end{figure}

\Cref{fig:dist_comparison} illustrates the benefits of rejecting according to the quantiles of \eqref{eq:null_dist}. In \Cref{fig:dist_comparison}, we consider settings with $p = 50$ variables, a size $k=20$ factor set $S$, and either $n=200$ or $n=5000$ samples, and plot three different distributions: (1) the limiting chi-squared distribution of $T(S)$; (2) the distribution \eqref{eq:null_dist}; and (3) the distribution of $T(S)$ in an example case where the unique factors are not Gaussian. In this non-Gaussian example, the thirty unique factors consist of ten centered $\text{Poi}(\lambda)$ random variables with $\lambda=1, \dots, 10$, ten centered chi-squared random variables with $1, \dots, 10$ degrees of freedom, and ten uniform random variables that range from $-a$ to $a$ for $a = 1, \dots, 10$. The principal variables are independent standard Gaussians and regression coefficient matrix $\W \in \R^{(p-k) \times k}$ has uniformly random unit vector rows. \Cref{fig:dist_comparison} has two important takeaways. First, at reasonable sample sizes, the limiting chi-squared distribution is a poor approximation for the actual distribution of $T(S)$, and rejecting according to its quantiles will lead to over-selection. Second, even when the unique factors are not Gaussian and the sample size is reasonably small, \eqref{eq:null_dist} is an excellent approximation for the actual distribution of $T(S)$. \textcolor{revision2}{This remained to be the case when we varied $n$, $p$, $k$, and the distributions of unique factors.}

\subsection{Examination of the Selected Subset}
\label{sec:model_selection_selected_subset_appdx}

Ultimately, having settled on selecting a size-$\hat{k}$ subset, we would ideally the size-$\hat{k}$ subset $\hat{S}$ that minimizes $T(U)$. Unlike in \Cref{sec:model_selection}, $\hat{S}$ in this section refers to any actual global minimizer of $T(U)$ that would be found via exhaustive search. In what follows, we give a useful re-characterization of $\hat{S}$. 

First, we re-characterize $\hat{S}$ as an MLE. Given that $T(U)$ is motivated by the LRT, this should not come as a surprise. Consider the case that every $(k + 1)$ by $(k +1)$ principal sub-matrix of $\hat{\bSigma}$ is full rank. Then we can use \citep[9.1.2]{Petersen} to compute for any size-$k$ subset $U$ that 

\begin{align*}
    n  \log \left( \frac{|\Diag(\hat{\bSigma}_{-U} - \hat{\bSigma}_{-U, U}\hat{\bSigma}_{U}^{-1}\hat{\bSigma}_{U, -U})|}{|\hat{\bSigma}_{-U} - \hat{\bSigma}_{-U, U} \hat{\bSigma}^{-1}_{U}\hat{\bSigma}_{U, -U}|} \right) &= n  \log \left( \frac{|\hat{\bSigma}_{U}||\Diag(\hat{\bSigma}_{-U} - \hat{\bSigma}_{-U, U}\hat{\bSigma}_{U}^{-1}\hat{\bSigma}_{U, -U})|}{|\hat{\bSigma}_{U}||\hat{\bSigma}_{-U} - \hat{\bSigma}_{-U, U} \hat{\bSigma}^{-1}_{U}\hat{\bSigma}_{U, -U}|} \right) \\
    &=  n  \log \left( \frac{|\hat{\bSigma}_{U}||\Diag(\hat{\bSigma}_{-U} - \hat{\bSigma}_{-U, U}\hat{\bSigma}_{U}^{-1}\hat{\bSigma}_{U, -U})|}{|\hat{\bSigma}|} \right)\\
    &= n \left( \log(|\hat{\bSigma}_U|) + \log(|\Diag(\hat{\bSigma}_{-U} -  \hat{\bSigma}_{-U, U}\hat{\bSigma}_{U}^{-1}\hat{\bSigma}_{U, -U} ) |) \right)\\
    &\qquad \qquad - n\log(|\hat{\bSigma}|) 
\end{align*}
So, in this case, every minimizer $\hat{S}$ of $T(U)$ belongs to  
\begin{equation}
    \label{eq:minimizing_subset}
    \argmin_{U \subset [p] : |U| = k} \log(|\hat{\bSigma}_U|) + \log(|\Diag(\hat{\bSigma}_{-U} -  \hat{\bSigma}_{-U, U}\hat{\bSigma}_{U}^{-1}\hat{\bSigma}_{U, -U} ) |)
\end{equation}
Correspondingly, in this case, the subsets belonging to \eqref{eq:minimizing_subset} are exactly the maximum likelihood estimators for $S$ in the restricted Gaussian model that only consists of distributions also in the  $\hat{k}$-dimensional subset factor model (see \Cref{sec:gaussian_pcss_mle_proof} in the proof of \Cref{thm:css_is_mle}). 

Solving \eqref{eq:minimizing_subset} also handles the case that every $(k + 1)$ by $(k +1)$ principal sub-matrix of $\hat{\bSigma}$ is not full rank. In this case, there must be a size-$k$ subset $U$ such that $T(U) = 0$, and we should fail to reject. However, every $(k + 1)$ by $(k +1)$ principal sub-matrix of $\hat{\bSigma}$ is not full rank if and only if the minimized value of \eqref{eq:minimizing_subset} is $-\infty$. Thus solving \eqref{eq:minimizing_subset} is sufficient to handle both cases. 

We can easily modify our algorithms from \Cref{sec:algorithms} to efficiently search for a subset that solves \eqref{eq:minimizing_subset}. We detail how to do this in \Cref{sec:other_algs_appdx}. 

Because $\hat{S}$ results in the most empirically diagonal residual covariance, it is immediate that $\hat{S}$ asymptotically (fixed $p$, $n$ goes to infinity) recovers a subset that results in an exactly diagonal population residual covariance, provided that $\hat{k} \geq k^*$. Unfortunately, this does not necessarily mean that $\hat{S}$ will recover a subset $S$ for which the subset factor model \eqref{eq:subset_factor_model} is satisfied, because the subset factor model requires that the unique factors are independent. We can at best guarantee that they are uncorrelated. Nonetheless, aside from some pathological cases, we can  guarantee recovery.  

\Cref{lem:subset_recovery} uses the following less restrictive version of the subset factor model to state our recovery result:

\begin{align}
\begin{split}
&X_S \sim F \qquad \epsilon \sim (0, \D) \qquad \Cov(X_S, \epsilon) = \0 \\
&X_{-S} = \W(X_{S} - E_{F}[X_S]) + \mu_{-S} + \epsilon.
\end{split} \label{eq:uncorrelated_subset_factor_model}
\end{align}
 
\begin{lemma}
    \label{lem:subset_recovery}
    Consider samples $x^{(1)}, \dots, x^{(n)} \in \R^p$ drawn from a distribution $P$ that satisfies \eqref{eq:uncorrelated_subset_factor_model} for some size-$k$ set $S$. Suppose that $P$ has population covariance $\bSigma$ such that $\bSigma_{S} \succ 0 $. Then, with probability one, $P$ eventually satisfies \eqref{eq:uncorrelated_subset_factor_model} with $\hat{S}$ as the number of samples $n$ tends to infinity. 
\end{lemma}

\subsection{The PCSS Model}
\label{sec:model_selection_pcss_appdx}

We provide an exactly analogous discussion for the PCSS model \eqref{eq:pcss_model}. Recall that the $k$-dimensional PCSS model is the set of distributions satisfying 
\begin{align}
\tag{\ref{eq:pcss_model}, revisited}
\begin{split}
&X_S \sim F\\
&X_{-S} \mid X_S \sim N(\mu_{-S} + \W(X_S - E_{F}[X_{S}]), \sigma^2 \I_{p-k}). 
\end{split} 
\end{align}
for some size-$k$ set $S$. Again, our procedure returns $\hat{k}$, the smallest $k$ for which we fail to reject the null $\widetilde{H}_0: |S| \leq k$ that the data generating distribution belongs to the $k$-dimensional PCSS model. Like before, we provide a test for this null that is valid in finite samples and we comment on some qualities of the selected subset $\hat{S}$. Because the PCSS model makes a Gaussian assumption, we do not need to discuss asymptotics. \newline 

\subsubsection{Motivating a Test Statistic}
Like in \Cref{sec:model_selection_test_stat_appdx}, we consider a generalized LRT. Specifically we compare the data's likelihood under a totally unrestricted multivariate Gaussian model to its likelihood under a restricted multivariate Gaussian model, where the restricted multivariate Gaussian model only considers distributions that are also in $k$-dimensional PCSS model. Again suppose that $\bSigma \succ \0$. Identical reasoning as in \Cref{sec:model_selection_test_stat_appdx} along with computations from the proof of \Cref{thm:css_is_mle} (see \Cref{sec:gaussian_pcss_mle_proof}) imply that the LRT test statistic is 
\begin{equation*}
    \min_{U \subset [p] : |U| = k} n \log \left( \frac{\left(\tr(\hat{\bSigma}_{-U} - \hat{\bSigma}_{-U, U}\hat{\bSigma}^{-1}_{U}\hat{\bSigma}_{U, -U})/ (p-k) \right)^{p-k}} {| \hat{\bSigma}_{-U} - \hat{\bSigma}_{-U, U}\hat{\bSigma}^{-1}_{U}\hat{\bSigma}_{U, -U}|} \right)
\end{equation*}
Again, we will consider a generalization of this statistic that allow $\hat{\bSigma}$ to be singular:
\begin{equation*}
    \widetilde{T}_k = \min_{\substack{U \subset [p]: |U| = k}} \widetilde{T}(U),  \qquad  \widetilde{T}(U) = n \log \left( \frac{\left(\tr(\hat{\bSigma}_{-U} - \hat{\bSigma}_{-U, U}\hat{\bSigma}^{+}_{U}\hat{\bSigma}_{U, -U})/ (p-k) \right)^{p-k}} {| \hat{\bSigma}_{-U} - \hat{\bSigma}_{-U, U}\hat{\bSigma}^{+}_{U}\hat{\bSigma}_{U, -U}|} \right)
\end{equation*}

\subsubsection{Determining a Critical Value}
Like in \Cref{sec:model_selection_crit_val_appdx}, we will consider a size-$k$ subset $S$ for which the data generating distribution satisfies \eqref{eq:pcss_model} and study the distribution of $\widetilde{T}(S)$. For simplicity we will assume that the principal variable distribution $F$ admits a density. Like before, the residual covariance still has a conditional Wishart distribution, and we can use the Wishart's Bartlett decomposition to show that $\widetilde{T}(S)$ has distribution 
\begin{equation}
    \label{eq:pcss_null_dist}
    n\log \left(\left(\frac{\left(\tilde{\chi}^2_{(p-k)(p-k-1)/2} + \sum_{j=1}^{p-k} \chi^2_{n - k - j}\right)}{p-k}\right)^{p-k}\bigg/\left(\prod_{j=1}^{p-k} \chi^2_{n - k - j}\right)\right)
\end{equation}
In \eqref{eq:pcss_null_dist}, $\{ \chi^2_{\ell} \}, \{ \tilde{\chi}^{2}_{\ell}\}$ are mutually independent chi-squared random variables with degrees of freedom specified by their subscript. Defining $\widetilde{Q}_{n, p, k}(1-\alpha)$ as the $(1-\alpha)$-quantile of \eqref{eq:pcss_null_dist}, \Cref{lem:pcss_finite_sample_validity} establishes a valid finite sample test for the null $\widetilde{H}_0 : |S| \leq k$. 

\begin{lemma}
    \label{lem:pcss_finite_sample_validity}
    Consider $n > p$ samples  $x^{(1)}, \dots, x^{(n)} \in \R^p$ drawn from a distribution $P$ that satisfies  \eqref{eq:pcss_model} for some size-$k$ set $S$ where $0 \leq k < p$. Suppose further that $P$ admits a density. Then $\widetilde{T}(S)$ has distribution \eqref{eq:pcss_null_dist} and, as a consequence, $P(\widetilde{T}_k > \widetilde{Q}_{n, p, k}(1-\alpha)) \leq \alpha$.
\end{lemma}

\Cref{prop:pcss_error_control} establishes error control of our procedure for the PCSS model. It is analogous to \Cref{thm:error_control}. 
\begin{proposition}
    \label{prop:pcss_error_control}
    Consider $n > p$ samples $x^{(1)}, \dots, x^{(n)} \in \R^p$ from a distribution $P$ that admits a density and let $k^*$ be the smallest $k$ for which $P$ belongs to the $k$-dimensional PCSS model \eqref{eq:pcss_model}. If $\hat{k}$ is the smallest $k$ for which the test statistic $\widetilde{T}_k$ is not strictly larger than the critical value $\widetilde{Q}_{n, p, k}(1 - \alpha)$, then 
    \begin{equation*}
        P(\hat{k} > k^*) \leq \alpha.
    \end{equation*}
\end{proposition}

\subsubsection{Examining the Selected Subset}

As in \Cref{sec:model_selection_selected_subset_appdx} we ultimately select the subset $\widetilde{S}$ that globally minimizes $\widetilde{T}(U)$. Like before, we can characterize this subset as an MLE. Supposing that every $k$ by $k$ principal sub-matrix of $\hat{\bSigma}$ is full rank and no size-$k$ subset empirically perfectly linearly reconstructs the remaining variables, similar computations to those in \Cref{sec:model_selection_selected_subset_appdx} shows that a size-$k$ subset $U$ minimizes $\widetilde{T}(U)$ if and only if it solves
\begin{equation}
    \label{eq:pcss_minimizing_subset}
    \argmin_{U \subset [p]: |U| = k} \log|\hat{\bSigma}_{U}| + (p-k)\log( \tr(\hat{\bSigma}_{-U} - \hat{\bSigma}_{-U, U}\hat{\bSigma}^{-1}_{U}\hat{\bSigma}_{U, -U})/(p-k))
\end{equation}
Under the same conditions, it is also true that $U$ solves \eqref{eq:pcss_minimizing_subset} if and only if it is the MLE for $S$ in the restricted multivariate Gaussian model that only contains distributions also in the $k$-dimensional PCSS model (see \Cref{sec:gaussian_pcss_mle_proof} in the proof of \Cref{thm:css_is_mle})

If the conditions from the previous paragraph are not met, it is unfortunately the case that the subset that minimizes $\widetilde{T}(U)$ may not solve \eqref{eq:pcss_minimizing_subset}. The conditions from the previous paragraph are not met, however, if and only if the minimized objective of \eqref{eq:pcss_minimizing_subset} is $-\infty$. Solving \eqref{eq:pcss_minimizing_subset} thus warns us if this is the case, but it is not totally sufficient as was true in \Cref{sec:model_selection_selected_subset_appdx}. 

We show how to search for the subset that solves \eqref{eq:pcss_minimizing_subset} using \Cref{sec:algorithms}'s algorithms in \Cref{sec:other_algs_appdx}. 

Analogous to \Cref{lem:subset_recovery}, \Cref{lem:pcss_subset_recovery} uses the following generalization of the PCSS model to state a subset recovery result:
\begin{align}
\begin{split}
&X_S \sim F \qquad \epsilon \sim (0, \sigma^2I) \qquad \Cov(X_S, \epsilon) = \0 \\
&X_{-S} = \W(X_{S} - E_{F}[X_S]) + \mu_{-S} + \epsilon.
\end{split} \label{eq:uncorrelated_pcss_model}
\end{align}
It says that $\widetilde{S}$ eventually will result in a isotropic population residual covariance.  

\begin{lemma}
    \label{lem:pcss_subset_recovery}
    Consider samples $x^{(1)}, \dots, x^{(n)} \in \R^p$ drawn from a distribution $P$ that satisfies \eqref{eq:uncorrelated_pcss_model} for some size-$k$ set $S$. Suppose further that, at the population level, some set of $k$ variables perfectly linearly reconstruct the remaining. Then, with probability one, $P$ eventually satisfies \eqref{eq:uncorrelated_pcss_model} with $\widetilde{S}$ as the number of samples $n$ tends to infinity. 
\end{lemma}

\section{Efficient Subset Search for other Objectives}
\label{sec:other_algs_appdx}

To modify Algorithm \ref{alg:greedy} and Algorithm \ref{alg:swap} to perform subset search under a new objective we simply need to alter $f$ in (\ref{eq:alg_function}) so that $i \not \in U \subset [p]$ that minimizes $f(i, \cdot)$ also minimizes the new objective over the subsets $U + i$. In this section, we explicitly give $f(i, \cdot)$ that allow us to search for subsets that minimize $T(\cdot)$ (see \Cref{sec:model_selection}),  $\widetilde{T}(\cdot)$ (see \Cref{sec:model_selection_pcss_appdx}), and McCabe's other three criteria. We generalize all of McCabe's criteria by changing inverses to pseudo-inverses, thereby allowing for singular $\bSigma$. Correctness of our proposed $f_{\cdot}(i, \cdot)$ are implied by derivations in \Cref{sec:alg_correctness_appdx}. 

For McCabe's criteria, Table \ref{table:runtimes} contrasts the time complexity of our algorithms with that of naive implementations. In the naive implementation, the practitioner recomputes the objective as written with no optimizations when adding/removing elements to/from the current subset. For both our algorithms and the naive implementation, the time complexity of searching for subsets that minimize $T(\cdot)$ and $\widetilde{T}(\cdot)$ matches that of McCabe's second criterion.  \newline

\begin{table}
\centering
\begin{tabular}{ |p{5cm}||p{3.5cm}|p{3.5cm}|  }
 \hline
 Criterion/Method & Algorithm \ref{alg:greedy} & Naive \\
 \hline
 McCabe's First Criterion  & $O(\min\{p^2k, pk^3\})$   & $O(p^4k)$ or $O(pk^4)$\\
 McCabe's Second Criterion &  $O(p^2k)$  & $O(p^3k^2)$\\
 McCabe's Third Criterion  &$O(p^3k)$ & $O(p^3k^2)$\\
 McCabe's Fourth Criterion &$O(p^3k)$ & $O(p^3k^2)$\\
 \hline
\end{tabular}
\caption{A time complexity comparison of our algorithms to the naive algorithms that, when features are added or removed from the current subset, recompute the objective as written with no optimizations. The reported time corresponds to one complete run of Algorithm \ref{alg:greedy} or one iteration of Algorithm \ref{alg:swap}. McCabe himself pointed out that his first criteria of minimizing $|\bSigma_{-S} - \bSigma_{-S, S}\bSigma_{S}^{-1}\bSigma_{S, -S}|$ is equivalent to maximizing $|\bSigma_{S}|$, so we report the naive time complexities for both versions of the objective.}
\label{table:runtimes}
\end{table}

\noindent \textbf{Minimizing $T(\cdot)$:} The discussion in \Cref{sec:model_selection_selected_subset_appdx} tells us that it suffices to find a subset that minimizes $\log |\bSigma_{S}| + \log(|\Diag(\bSigma_{R(X_{-S},X_{S})})|)$. To perform this search, use 
\begin{align*}
    &f(i, U, \bSigma_{R(X, X_{U})}) = \log((\bSigma_{R(X, X_{U})})_{ii})\\
    &\qquad \qquad + \sum_{j \not \in U + i} \log \left((\bSigma_{R(X, X_{U})})_{jj} - (\bSigma_{R(X, X_{U})})^2_{ij}/(\bSigma_{R(X, X_{U})})_{ii} \cdot I_{(\bSigma_{R(X, X_{U})})_{ii} > 0 } \right)
\end{align*}

\noindent \textbf{Minimizing $\widetilde{T}(\cdot)$:} The discussion in \Cref{sec:model_selection_selected_subset_appdx} suggests finding a subset that minimizes $\log |\bSigma_{S}| + (p-k)\log(\tr(\bSigma_{R(X_{-S},X_{S})})/(p-k) )$. If the objective value is not $-\infty$ then solving this problem is sufficient. Otherwise we cannot say anything concrete. To perform this search, use 
\begin{align*}
    &f(i, U, \bSigma_{R(X, X_{U})}) = \log((\bSigma_{R(X, X_{U})})_{ii}) \\
    &\qquad \qquad + (p -k) \cdot \log \left( \tr(\bSigma_{R(X, X_{U})}) - \|(\bSigma_{R(X, X_{U})})_{\bullet i}\|_2^2/(\bSigma_{R(X, X_{U})})_{ii} \cdot  I_{(\bSigma_{R(X, X_{U})})_{ii} > 0 } \right)
\end{align*}

\noindent \textbf{McCabe's First Criterion:} McCabe's first criterion suggests selecting a subset that minimizes $|\bSigma_{-S} - \bSigma_{-S, S}\bSigma_{S}^{-1}\bSigma_{S, -S}|$. We generalize this to selecting the subset which minimizes $|\bSigma_{-S} - \bSigma_{-S, S}\bSigma_{S}^{+}\bSigma_{S, -S}| = |\bSigma_{R(X, X_{S})}|$. For simplicity, we will assume that there exists a size-$k$ subset such that $|\bSigma_{S}| > 0 $. If this is not the case, then every size $k$ subset has redundant variables which can be perfectly linearly reconstructed by other variables in the subset, and the practitioner should use a smaller subset size.  To perform subset search according to McCabe's first criterion use:
\begin{equation*}
    f(i, \bSigma_{R(X, X_{U})}) = -(\bSigma_{R(X, X_{U})})_{ii}
\end{equation*}
Alternatively, as we add/remove variables to/from $U$, we can keep track of $\bSigma^{+}_{U}$ instead of  $\bSigma_{R(X, X_{U})}$ using \citep[3.2.7]{Petersen} and instead use
\begin{equation*}
    f(i, \bSigma, \bSigma^{+}_{U}) = -(\bSigma_{ii} - \bSigma_{iU}\bSigma_{U}^{+}\bSigma_{Ui})
\end{equation*}
\Cref{lem:residual_cov_update} implies that this is equivalent to our earlier suggestion, but it has smaller time complexity when $k \ll \sqrt{p}$. \newline 

\noindent \textbf{McCabe's Third Criterion:} McCabe's third criterion suggests selecting a subset that minimizes $\| \bSigma_{-S} - \bSigma_{-S, S}\bSigma_{S}^{-1} \bSigma_{S, -S} \|^2_F$. We generalize this to selecting the subset which minimizes $\|\bSigma_{-S} - \bSigma_{-S, S}\bSigma_{S}^{+} \bSigma_{S, -S} \|_F = \|\bSigma_{R(X, X_S)}\|^2_F$. To perform subset search according to McCabe's third criterion use:
\[f(i, \bSigma_{R(X, X_{U})}) = [(\|\beta\|_2^2/\beta_i)^2 - 2 \beta^\top\bSigma_{R(X, X_{U})}\beta/\beta_i] \cdot I_{\beta_i > 0}\] 
where $\beta = (\bSigma_{R(X, X_{U})})_{\bullet i}$.\newline 

\noindent \textbf{McCabe's Fourth Criterion:} McCabe's fourth criterion suggests selecting a subset that maximizes $\tr(\bSigma_{-S}^{-1} \bSigma_{-S, S}\bSigma_{S}^{-1} \bSigma_{S, -S})$. This is equivalent to maximizing the sum of squared cannonical correlations between $X_{S}$ and $X_{-S}$. We generalize this to selecting the subset which maximizes $\tr(\bSigma_{-S}^{+} \bSigma_{-S, S}\bSigma_{S}^{+} \bSigma_{S, -S})$ which has the same interpretation in the more general case where $\bSigma$ is singular. Consider a currently selected subset $U$. Fix an $i \not \in U$ and let $V = U + i$. Take $j$ and $h$ to be the rank of $i$ in $V$ and $-U$ respectively. So, for example, $(\bSigma_{V})_{jj} = \bSigma_{ii}$ and $(\bSigma_{-U})_{hh} = \bSigma_{ii}$ Then, to perform subset search in accordance with McCabe's fourth criterion use 
\begin{align*}
&f(i, U, \bSigma, \bSigma_{R(X_{-U}, X_{U})}, \bSigma_{R(X_{U}, X_{-U})}, \bSigma^{+}_{-U}, \bSigma^{+}_{U}) =\\
&\qquad \tr( (\bSigma^{+}_{V})_{-j,-j} \bSigma_{R(X_{U}, X_{-U})}) + \frac{\beta^\top \bSigma^{+}_{V} \beta}{\beta_j} I_{\beta_j > 0}  - I_{(\bSigma_{R(X_{-U}, X_{U})})_{hh} > 0}
\end{align*}
where $\beta =  \bSigma_{V, i} - \bSigma_{V, -V}\bSigma_{-V}^{+}\bSigma_{-V, i}$, and we compute $\bSigma_{V}^+$ from  $\bSigma_{U}^+$ and $\bSigma^+_{-V}$ from $\bSigma^+_{-U}$ using \citep[3.2.7]{Petersen}. We keep track of $\bSigma_{U}^+$ and $\bSigma^+_{-U}$ as we add/remove elements to/from $U$ using \citep[3.2.7]{Petersen} and keep track of $\bSigma_{R(X_{-U}, X_{U})}$ and $\bSigma_{R(X_{U}, X_{-U})}$ using \Cref{lem:residual_cov_update} as usual. 

If $\bSigma \succ \0$, then we can give an exact closed form for the replacement $f$ given above:
\[ f(i, U, \bSigma, \bSigma_{R(X_{-U}, X_{U})}, \bSigma^{-1}_{-U}, \bSigma^{-1}_{U}) =  \frac{\beta_{-j}^\top\bSigma^{-1}_{U}\beta_{-j}}{\beta_{j}} - \frac{(\alpha^\top\beta_{-j})^2}{\delta \beta_{j}} - \frac{\alpha^\top\bSigma_{R(X_U, X_{-U})}\alpha + 2\alpha^\top\beta_{-j} - \beta_{j}}{\delta } \]
where $\alpha = \bSigma_{U}^{-1}\bSigma_{Ui}$, $\delta = \bSigma_{ii} - \bSigma_{iU}\alpha$ and $\beta$ is as before, but now we can explicitly compute $\bSigma_{-V}^+ = \bSigma_{-V}^{-1} = (\bSigma_{-U}^{-1})_{-h, -h} - (\bSigma_{-U})_{-h, h}(\bSigma_{-U})_{h, -h}/(\bSigma_{-U})_{hh}$. Also, as we add elements to $U$, \citep{Khan} now allows us to keep track of $\bSigma_{U}^+ = \bSigma_{U}^{-1}$ and $\bSigma^+_{-U} = \bSigma^{-1}_{-U}$ with simpler closed form updates. 

% This whole appendix was added as part of second revision 
\section{The High-Dimensional Consistency of CSS}
\label{sec:high_dim_consistency_appdx}

In this section we characterize a high-dimensional setting where the size-$k$ CSS estimate \eqref{eq:css} one gets from the sample covariance is consistent for the size-$k$ CSS solution according to the population covariance \eqref{eq:pv}. In particular, we fix $k$ and imagine observing $n$ samples from a distribution $P$ over $\R^p$, where $p$ is possibly growing with $n$. We denote the population covariance under $P$ as $\bSigma \in \S_+^{p \times p}$. Our analysis applies in the proportional asymptotic regime, where $p/n$ converges to a constant. For the result to apply, $P$ must satisfy three assumptions as $n$ grows.

\begin{assumption}[Light-tailed data]
    \label{ass:sub_gaussian}
     The sub-Gaussian norm (see \citep[Definition 2.5.6]{Vershynin}) of each variable is bounded above by a constant independent of $p$ and $n$.
\end{assumption}

\begin{assumption}[Low-dimensional invertibility]
    \label{ass:invertibility}
    The minimum of the smallest eigenvalues of all the $k \times k$ principal sub-matrices of $\bSigma$ is bounded below by a positive constant independent of $p$ and $n$.
\end{assumption}

\begin{assumption}[Separation]
    \label{ass:seperation} 
    For some $\delta > 0$, the achieved CSS objective for the population best subset 
    \begin{equation*}
    S^{*} = \argmin_{S \subseteq [p], |S| = k} \tr(\bSigma - \bSigma_{\bullet S} \bSigma^{+}_S \bSigma_{\bullet S})
    \end{equation*}
    is well separated from the objective achieved by any other subset 
    \begin{equation*}
        \min_{S \neq S^*} \tr(\bSigma - \bSigma_{\bullet S} \bSigma^{+}_S \bSigma_{\bullet S}) - \tr(\bSigma - \bSigma_{\bullet S^*} \bSigma^{+}_{S^*} \bSigma_{\bullet S^*})  = \Omega(p^{1/2 + \delta})
    \end{equation*}
\end{assumption}

Note that \Cref{ass:invertibility} is much weaker than assuming a lower bound on the minimum eigenvalue of $\bSigma$. Under \Cref{ass:invertibility} ,  $\bSigma$ can still be singular. 

Under the above three assumptions, \Cref{thm:high_dim_consistency_appdx} guarantees that the in-sample CSS solution is consistent for the population CSS solution.

\begin{theorem}[High-dimensional setting where CSS is consistent]
    \label{thm:high_dim_consistency_appdx}
     Suppose we observe samples $x^{(1)}, \dots, x^{(n)} \in \mathbb{R}^{p}$ with sample covariance $\hat{\bSigma} \in \S_{+}^{p \times p}$ from a distribution $P$ with population covariance $\bSigma \in \S_{+}^{p_\times p}$. Fix $k$, and consider the proportional asymptotic regime where $p/n$ converges to a constant as $n, p \rightarrow \infty$. If $P$ satisfies \Cref{ass:sub_gaussian}, \Cref{ass:invertibility}, and \Cref{ass:seperation} as $n, p \rightarrow \infty$, then any in-sample CSS solution 
    \begin{equation*}
        \hat{S} \in \argmin_{S \subseteq [p], |S| = k} \tr(\hat{\bSigma} - \hat{\bSigma}_{\bullet S} \hat{\bSigma}^{+}_S \hat{\bSigma}_{\bullet S})
    \end{equation*}
    is consistent for the population CSS solution 
    \begin{equation*}
        S^{*} \in \argmin_{S \subseteq [p], |S| = k} \tr(\bSigma - \bSigma_{\bullet S} \bSigma^{+}_S \bSigma_{\bullet S}), 
    \end{equation*}
    i.e., 
    \begin{equation*}
        \lim_{n \rightarrow \infty} P(\hat{S} = S^*) = 1.
    \end{equation*}
\end{theorem}
Note that, by \Cref{ass:seperation}, the population CSS solution $S^*$ is eventually unique.

The rest of this appendix is a proof of \Cref{thm:high_dim_consistency_appdx}. We use $\| \cdot \|_2$ to denote the spectral norm of a matrices. We denote the sorted eigenvalues of a positive semi-definite matrix $\A \in \S^{q \times  q}_{+}$ as $\lambda_1(\A) \geq \dots \geq \lambda_q(\A) \geq 0$. For a matrix $\M \in \R^{q \times k}$ we denote its sorted top-$k$ singular values as $s_1(\M) \geq \dots \geq s_k(\M) \geq 0$. We use $c_i$ to denote positive constants that are independent of $n$ and $p$. If we introduce a new $c_i$, that means that such a $c_i$ exists. Also, we make use of the stochastic calculus notation of \citep[Chapter 2]{Vaart}. 

\subsection{Central Argument}

First we consider the case that $X \sim (0, \bSigma)$, $\bSigma \in \mathbb{S}_+^{p \times p}$ is mean-zero and let $\widetilde{\bSigma}_{ij} = n^{-1}\sum_{\ell = 1}^n X_i^{(\ell)}X_j^{(\ell)}$ be our estimate of $\bSigma = E[XX^{\top}]$. We will use the notation 
\begin{equation*}
    \text{Obj}(\bOmega, U) = p^{-1/2 - \delta}\tr(\bOmega  - \bOmega _{\bullet U} \bOmega _{U}^{+} \bOmega _{U \bullet}).
\end{equation*}
to denote the CSS objective for the covariance $\bOmega \in \S_{+}^{p \times p}$, scaled by $p^{-1/2 - \delta}$. Our goal is to show that $\tilde{S}$, which is a minimizer (break ties randomly) of $\text{Obj}(\widetilde{\bSigma}, S)$ for size-$k$ subsets $S$, is consistent for $S^*$, which is the minimizer of $\text{Obj}(\bSigma, S)$ for size-$k$ subsets $S$. By \Cref{ass:seperation}, $\text{Obj}(\bSigma, S)$ eventually has a unique minimizer.

\Cref{ass:seperation} tells us exactly that 
\begin{equation*}
\min_{S \neq S^*}\text{Obj}(\bSigma, S) - \text{Obj}(\bSigma, S^*) = \Omega(1),
\end{equation*}
If we show that 
\begin{equation*}
    \max_{S \subseteq [p]: |S|=k} \widetilde{\Delta}(S) = o_p(1), \qquad \widetilde{\Delta}(S) = |\text{Obj}(\bSigma, S) - \text{Obj}(\widetilde{\bSigma}, S)|
\end{equation*}
then $\tilde{S}$ will be consistent for $S^*$ via the following argument: 
\begin{align*}
P(\tilde{S} = S^*) &=  P(\text{Obj}(\widetilde{\bSigma}, S^*) < \min_{S \neq S^*} P(\text{Obj}(\widetilde{\bSigma}, S) )  \\
                        &\geq P(\max_{S \subseteq [p]: |S|=k} \widetilde{\Delta}(S) < \frac{1}{2}(\min_{S \neq S^*}\text{Obj}(\bSigma, S) -  \text{Obj}(\bSigma, S^*) )) \\
                        &= 1 - o(1)
\end{align*}
Considering an arbitrary size-$k$ subset $S$, we fix $t > 0$ and devote the rest of the argument to showing that
\begin{equation}
\label{eq:overall_tail_bound}
    P(\widetilde{\Delta}(S) > t) = o\left({{p}\choose{k}}^{-1}\right).
\end{equation}
Then, the fact that $\max_{S \subseteq [p]: |S|=k} \widetilde{\Delta}(S) = o_p(1)$ follows trivially from a union bound. \newline 

We start by re-writing $\widetilde{\Delta}(S)$. For two square positive semi-definite matrices $\A$ and $\B$, von Neumann's trace inequality \citep{Mirsky} implies that $|\tr(\A \B)| \leq |\tr(\A)| |\tr(\B)|$. Then, for four square positive semi-definite matrices $\widetilde{\A}$, $\widetilde{\B}$, $\A$, and $\B$, we have that 
\begin{align*}
    |\tr(\widetilde{\A} \widetilde{\B} - \A \B)| &= |\tr(\widetilde{\A} \widetilde{\B} - \widetilde{\A}\B + \widetilde{\A}\B - \A \B)| \\
                 &\leq |\tr(\widetilde{\A} (\widetilde{\B} - \B) )| + | \tr((\widetilde{\A} - \A)\B) |\\
                 &\leq |\tr(\widetilde{\A} )| \cdot |\tr(\widetilde{\B} - \B) | + |\tr(\widetilde{\A} - \A )| \cdot |\tr(\B) |.
\end{align*}
Using this fact we can reconfigure $\widetilde{\Delta}(S)$
\begin{align*}
    &\widetilde{\Delta}(S) \\
    &= p^{-1/2 - \delta}|\tr(\widetilde{\bSigma} - \widetilde{\bSigma}_{\bullet S} \widetilde{\bSigma}_S^{+} \widetilde{\bSigma}_{S \bullet}) -  \tr(\bSigma - \bSigma_{\bullet S} \bSigma_{S}^{+} \bSigma_{S \bullet}) |\\
        &\leq p^{-1/2 - \delta}| \tr(\widetilde{\bSigma} - \bSigma)| + p^{-1/2 - \delta}| \tr(\widetilde{\bSigma}_{S\bullet}\widetilde{\bSigma}_{\bullet S} \widetilde{\bSigma}^{+}_{S} - \bSigma_{S\bullet}\bSigma_{\bullet S} \bSigma^{+}_{S})|\\
        &\leq p^{-1/2 - \delta}| \tr(\widetilde{\bSigma} - \bSigma)| + p^{-1/2 - \delta} |\tr(\widetilde{\bSigma}_{S\bullet}\widetilde{\bSigma}_{\bullet S})| \cdot |\tr(\widetilde{\bSigma}_S^+ - \bSigma_S^+)| + p^{-1/2 - \delta}|\tr(\widetilde{\bSigma}_{S\bullet}\widetilde{\bSigma}_{\bullet S} - \bSigma_{S\bullet}\bSigma_{\bullet S} )| \cdot  |\tr(\bSigma^{+}_{S} )|\\
        &\leq p^{-1/2 - \delta}| \tr(\widetilde{\bSigma} - \bSigma)| + p^{-1/2 - \delta} |\tr(\widetilde{\bSigma}_{S\bullet}\widetilde{\bSigma}_{\bullet S} -\bSigma_{S\bullet}\bSigma_{\bullet S})| \cdot |\tr(\widetilde{\bSigma}_S^+ - \bSigma_S^+)| \\
        &\qquad  + p^{-1/2 - \delta} |\tr(\bSigma_{S\bullet}\bSigma_{\bullet S})| \cdot |\tr(\widetilde{\bSigma}_S^+ - \bSigma_S^+)| + p^{-1/2 - \delta}|\tr(\widetilde{\bSigma}_{S\bullet}\widetilde{\bSigma}_{\bullet S} - \bSigma_{S\bullet}\bSigma_{\bullet S} )| \cdot  |\tr(\bSigma^{+}_{S} )|\\
\end{align*}

We can further bound some of these terms. We know from \Cref{ass:sub_gaussian} that $\max_{i \in [p]} \bSigma_{ii} \leq c_1$ and therefore also $\max_{i \in [p], j \in [p]} |\bSigma_{ij}| \leq c_1$, so $|\tr(\bSigma_{S\bullet} \bSigma_{\bullet S}) | \leq kp \max_{i \in [p], j \in [p]} \bSigma^2_{ij} \leq c_2p$. We also know from \Cref{ass:invertibility} that $|\tr(\bSigma_S^+)| \leq k\|\bSigma_S^{-1} \|_2 \leq c_3$. Hence, we have 

\begin{align*}
\widetilde{\Delta}(S) &\leq  \underbrace{p^{-1/2 - \delta}| \tr(\widetilde{\bSigma} - \bSigma)|}_{\widetilde{\Delta}_1(S)} + p^{-1/2 - \delta} |\tr(\widetilde{\bSigma}_{S\bullet}\widetilde{\bSigma}_{\bullet S} -\bSigma_{S\bullet}\bSigma_{\bullet S})| \cdot |\tr(\widetilde{\bSigma}_S^+ - \bSigma_S^+)| \\
        &\qquad  + c_2 \underbrace{p^{1/2 - \delta}  |\tr(\widetilde{\bSigma}_S^+ - \bSigma_S^+)|}_{\widetilde{\Delta}_3(S)} + c_3 \underbrace{p^{-1/2 - \delta} |\tr(\widetilde{\bSigma}_{S\bullet}\widetilde{\bSigma}_{\bullet S} - \bSigma_{S\bullet}\bSigma_{\bullet S} )|}_{\widetilde{\Delta}_2(S)} \\
\end{align*}

If we provide appropriate tail bounds for the $\widetilde{\Delta}_i(S)$, i.e. show that for any $t > 0$ that 
\begin{equation*}
    P(\widetilde{\Delta}_i(S) > t) =  o\left({{p}\choose{k}}^{-1}\right),
\end{equation*}
where the right-hand-side bound is independent of $S$. Then it is straightforward to establish the tail bound \eqref{eq:overall_tail_bound} for $\widetilde{\Delta}(S)$. We devote the rest of the argument to establishing this tail bound for $\widetilde{\Delta}_1(S)$, $\widetilde{\Delta}_2(S)$, and $\widetilde{\Delta}_3(S)$.

\subsection{Establishing Tail Bounds}

\subsubsection{Tail bound for $\widetilde{\Delta}_1(S)$} 

We start by getting a concentration bound for the entries of $\widetilde{\bSigma}_{ij}$. Using the Orlicz norm notation from Section 2.7.1 of \cite{Vershynin}, \Cref{ass:sub_gaussian} implies that $\| X_i\|_{\psi_2} \leq c_4$. We can use \citep[Lemma 2.7.7]{Vershynin} to see that  $\| X_iX_j \|_{\psi_1} \leq \| X_i\|_{\psi_2}\| X_j\|_{\psi_2} \leq c_5$ implying that all the $X_iX_j$ are sub-exponential. For any $x \geq 0$, \citep[Corollary 2.8.3 ]{Vershynin} gives the following concentration bound for the entries $\widetilde{\bSigma}_{ij}$:
\begin{equation}
    \label{eq:sub_exp_concentration}
    P(|\widetilde{\bSigma}_{ij} - \bSigma_{ij}| \geq x ) \leq 2 \exp \left( -c_6 \min\left(x^2, x\right) n \right)
\end{equation}

Now, we use the union bound along with our sub-exponential concentration result
\begin{align*}
    P(\Delta_1(S) > t) &\leq  P( p^{-1/2 - \delta} \sum_{i=1}^p |\widetilde{\bSigma}_{ii} - \bSigma_{ii}| > t) & \text{ (triangle inequality) } \\
        &\leq \sum_{i=1}^p P\left( |\widetilde{\bSigma}_{ii} - \bSigma_{ii}| > p^{-1/2 + \delta} t  \right) & \text{ (union bound) } \\
        &\leq 2p \exp \left( -c_6 \min\left(np^{-1 + 2\delta} t^2, np^{-1/2 + \delta}t \right)  \right) & \text{ (sub-exponential concentration) }\\
        &= o\left({{p}\choose{k}}^{-1}\right)
\end{align*}

\subsubsection{Tail bound for $\widetilde{\Delta}_2(S)$} 

To get a tail bound for $\Delta_2(S)$ we need a concentration result for $\widetilde{\bSigma}^2_{ij}$. Fix some constant $c_7 > 0$ and define $\widetilde{B}_{ij} = \max\{|2\bSigma_{ij} + c_7|,  |2\bSigma_{i j} - c_7| \}$. Two things are true. First, if $|\widetilde{\bSigma}_{ij} + \bSigma_{ij}| > \widetilde{B}_{ij}$, then it must be the case that $|\widetilde{\bSigma}_{ij} - \bSigma_{ij}| > c_7$. Second $\widetilde{B}_{ij} \leq c_8$ for some constant. Considering some $x \geq 0$, we can now get a concentration result for $\widetilde{\bSigma}^2_{ij}$.   

\begin{align*}
    &P\left( |\widetilde{\bSigma}_{ij}^2 - \bSigma_{ij}^2  | > x \right) & \\
    &= P\left( |\widetilde{\bSigma}_{ij}^2 -  \bSigma_{ij}^2  | > x, |\widetilde{\bSigma}_{ij} + \bSigma_{ij}| \leq \widetilde{B}_{ij}  \right) + P\left( |\widetilde{\bSigma}_{ij}^2 - \bSigma_{ij}^2  | > x, |\widetilde{\bSigma}_{ij} + \bSigma_{ij}| > \widetilde{B}_{ij}  \right) & \\
    &\leq  P\left( |\widetilde{\bSigma}_{ij} -\bSigma_{i j }  | > x/\widetilde{B}_{ij}\right) + P\left(|\widetilde{\bSigma}_{ij} +  \bSigma_{ij}| > \widetilde{B}_{ij}  \right) & \\
    &\leq P\left( |\widetilde{\bSigma}_{ij} -\bSigma_{i j }  | > x/c_8 \right) + P\left(|\widetilde{\bSigma}_{ij} -  \bSigma_{ij}| > c_7 \right) &  \text{ (definition of $\widetilde{B}_{ij}$)}\\
    &\leq 2\exp ( -c_9 \min\left(x^2, x\right) n ) + 2\exp(-c_{10} n) & \text{ (sub-exponential concentration) }
\end{align*}

With this result under our belt, getting a tail bound for  $\widetilde{\Delta}_2(S)$ is straight-forward.
 
\begin{align*}
    &P(\widetilde{\Delta}_2(S) > t) \\
    &\leq  P( p^{-1/2-\delta}\sum_{i \in S} \sum_{j = 1}^p  |\widetilde{\bSigma}_{i j}^2 - \bSigma_{i j }^2 | > t) & \text{(triangle inequality)}\\
    &=\sum_{i \in S} \sum_{j = 1}^p   P\left( |\widetilde{\bSigma}_{i j}^2 - \bSigma_{i j }^2  | > k^{-1}p^{-1/2 + \delta} t \right) & \text{(union bound)}\\
    &\leq 2pk \exp \left( -c_{11} \min\left(np^{-1 + 2\delta}t^2, np^{-1/2 + \delta}\right)  \right) + 2pk\exp(-c_{10} n)  & \text{(concentration for $\widetilde{\bSigma}^2_{ij}$)}\\
    &=  o\left({{p}\choose{k}}^{-1}\right) & 
\end{align*}

\subsubsection{Tail bound for $\widetilde{\Delta}_3(S)$}

We will get a tail bound for $\Delta_3(S)$ via a concentration inequality for the singular values of $  \X_{\bullet S} \bSigma^{-1/2}_{S}\in \R^{n \times k}$, where $\boldsymbol{X}_{\bullet S} \in \R^{n \times k}$ is random matrix with i.i.d samples of $X_S$ as rows. Essentially, concentration of these singular values this will imply concentration of the eigenvalues of $\widetilde{\bSigma}_S^{-1}$.

Using \Cref{ass:sub_gaussian} and \Cref{ass:invertibility}, we can show that the rows of $\X_{\bullet S}\bSigma^{-1/2}_{S} \in \R^{n \times k}$ are isotropic sub-gaussian random vectors with bounded sub-gaussian norm: 

\begin{align*}
    \| \bSigma_S^{-1/2} X_S  \|_{\psi_2} &= \sup_{\| v\|_2 = 1} \|(\bSigma_S^{-1/2} X_S)^{\top}v \|_{\psi_2} \\
    &= \sup_{\| v\|_2 = 1} \left( \left \| X_S^{\top} \left(\frac{\bSigma_S^{-1/2}v}{\| \bSigma_S^{-1/2}v\|_2 } \right) \right\|_{\psi_2} \|\bSigma_S^{-1/2}v\|_2  \right)\\
    &\leq \|\bSigma_S^{-1/2} \|_2 \sup_{\|\tilde{v}\|_2 = 1} \|\tilde{v}^{\top}X_S \|_{\psi_2}\\
    &\leq \|\bSigma_S^{-1} \|^{1/2}_2 \sup_{\|\tilde{v}\|_2 = 1} \sum_{i=1}^k |v_i| \|X_S\|_{\psi_2}\\
    &\leq \|\bSigma_S^{-1} \|^{1/2}_2 \max_{i \in [p]} \|X_i\|_{\psi_2} \sup_{\|\tilde{v}\|_2 = 1} \| \tilde{v}\|_1 \\
    &\leq c_{12}.
\end{align*}
Note that in the above $v, \tilde{v} \in \R^k$ have fixed (non-growing) dimension because $k$ is fixed. \citep[Theorem 4.6.1]{Vershynin}, then tells us that there is a constant $c_{13}$ such that for any $x \geq c_{13}$, 
\begin{equation}
    \label{eq:singular_value_concentration}
    P\left(\max_{i \in [k]}| s_i(\X_{\bullet S}\bSigma_S^{-1/2}) -  n^{1/2} | \geq x \right) \leq 2\exp(-c_{14} (x - c_{13})^2),
\end{equation}

To apply this result we need to note two facts. First, whenever the first $k$ singular values of $\X_{\bullet S}\bSigma_S^{-1/2}$ are positive, that means that $\bSigma_S^{-1/2} \widetilde{\bSigma}_{S} \bSigma_S^{1/2}$ is invertible, so $\widetilde{\bSigma}_{S}$ must be invertible. Second, whenever $\widetilde{\bSigma}_S$ is invertible,
\begin{align*}
    |\tr(\widetilde{\bSigma}_S^+ - \bSigma_S^{+})| &=  |\tr(\widetilde{\bSigma}_S^{-1} - \bSigma_S^{-1})|\\
                                                   &= | \tr((\widetilde{\bSigma}_S^{-1}  \bSigma_{S} - \I_k) \bSigma_{S}^{-1})|\\
                                                   &\leq |\tr(\widetilde{\bSigma}_S^{-1}  \bSigma_{S} - \I_k)| |\tr(\bSigma_{S}^{-1})|\\
                                                   &\leq k\|\bSigma_{S}^{-1} \|_2 | \tr(\bSigma_S^{1/2} \widetilde{\bSigma}_S^{-1} \bSigma^{1/2}_S - \I_k) |\\
                                                   &\leq k\|\bSigma_{S}^{-1} \|_2 | \tr((\bSigma_S^{-1/2} \widetilde{\bSigma}_S \bSigma^{-1/2}_S)^{-1} - \I_k) |\\
                                                   &\leq k^2\|\bSigma_{S}^{-1} \|_2 | \max_{i \in [k]} | s^{-2}_i(n^{-1/2}\X_{\bullet S}\bSigma_S^{-1/2}) - 1| \\
                                                   &\leq c_{15}\max_{i \in [k]} | s^{-2}_i(n^{-1/2}\X_{\bullet S}\bSigma_S^{-1/2}) - 1|
\end{align*}
where we have again made use of \Cref{ass:invertibility}. To finalize our argument, we will need to make use of the following lemma. 

\begin{lemma}
    \label{lem:inverse_local_lip}
    Fix some $a \in (0, 1)$. There exists some $b>0$ such that for all  $x > 0$ 
    \begin{equation}
        |x - 1| < a \implies b|x - 1| \geq |x^{-2} - 1|
    \end{equation}
\end{lemma}
\begin{proof}
If $x = 1$ the right hand inequality is always true, so consider $x \neq 1$.  If $|x - 1| < a$, then 
\begin{align*}
    \frac{|x^{-2} - 1| }{|x - 1|} &= \frac{|1- x^2| }{x^2|x- 1|}\\
                                 &= \frac{1 + x}{x^2}\\
                                 &\leq \frac{2 + a}{(1 - a)^2}             
\end{align*}
So we can take $b = (2 + a)/{(1-a)^2}$. 
\end{proof}

Now we can prove the tail bound for $\widetilde{\Delta}_3(S)$. Fix some $c_{16} \in (0, 1)$. Then for large enough $n$, 
\begin{align*}
 &P\left(\widetilde{\Delta}_3 > t  \right)  \\
 &\leq P\left(p^{1/2 - \delta} |\tr(\widetilde{\bSigma}_S^+ - \bSigma_S^{+})| > t, \max_{i \in [k]} | s_i(n^{-1/2}\X_{\bullet S}\bSigma_S^{-1/2}) - 1| < c_{16} \right)\\
 &\qquad + P\left(\max_{i \in [k]} | s_i(n^{-1/2}\X_{\bullet S}\bSigma_S^{-1/2}) - 1| \geq c_{16}\right)  \\
 &\leq P\left(\max_{i \in [k]} | s^{-2}_i(n^{-1/2}\X_{\bullet S}\bSigma_S^{-1/2}) - 1| > c_{17}p^{-1/2 + \delta}t, \max_{i \in [k]}| s_i(n^{-1/2}\X_{\bullet S}\bSigma_S^{-1/2}) - 1| < c_{16} \right)\\
 &\qquad + P\left(\max_{i \in [k]} | s_i(n^{-1/2}\X_{\bullet S}\bSigma_S^{-1/2}) - 1| \geq c_{16}\right)  \\
 & & \hspace{-380pt} \text{($\widetilde{\bSigma}_{S}$ is invertible on the event)}\\
 &\leq P\left(\max_{i \in [k]} | s_i(n^{-1/2}\X_{\bullet S}\bSigma_S^{-1/2}) - 1| > c_{18}p^{-1/2 + \delta}t, \max_{i \in [k]}| s_i(n^{-1/2}\X_{\bullet S}\bSigma_S^{-1/2}) - 1| < c_{16} \right)\\
 &\qquad + P\left(\max_{i \in [k]} | s_i(n^{-1/2}\X_{\bullet S}\bSigma_S^{-1/2}) - 1| \geq c_{16}\right)  \\
 & & \hspace{-300pt} \text{(\Cref{lem:inverse_local_lip} with $a=c_{16}$)}\\
 &\leq P\left(\max_{i \in [k]} | s_i(\X_{\bullet S}\bSigma_S^{-1/2}) - n^{1/2}| >  c_{18}n^{1/2}p^{-1/2 + \delta}t\right) + P\left(\max_{i \in [k]} | s_i(\X_{\bullet S}\bSigma_S^{-1/2}) - n^{1/2}| \geq n^{1/2}  c_{16}\right) \\
 & \leq 2 \exp(-c_{19} (n^{1/2}p^{-1/2 + \delta}t - c_{20})^2) + 2\exp(-c_{21} (n^{1/2} - c_{22})^2) \\
 & & \hspace{-150pt}\text{(apply \Cref{eq:singular_value_concentration} once $n$ large enough) }\\
 &= o_p\left({{p}\choose{k}}^{-1}\right).
\end{align*}

\subsection{Estimation with Mean}
\label{sec:high_dim_consistency_mean_appdx}

Now we allow $X \sim (\mu, \bSigma), \bSigma \in \S_{+}^{p \times p}$ to have a non-zero mean. Now with

\begin{equation*}
\widetilde{\bSigma} = n^{-1} \sum_{\ell=1}^n (X_i^{(\ell)} - \mu)(X_i^{(\ell)} - \mu)^T,
\end{equation*}
we still let $\widetilde{S}$ be a minimizer of $\text{Obj}(\widetilde{\bSigma}, S)$ (with ties broken randomly). Also still we let $S^*$ be the (eventually unique) minimizer of $\text{Obj}(\bSigma, S)$ for size-$k$ subsets $S$. Now we define $\hat{S}$ to be the minimizer of $\text{Obj}(\hat{\bSigma}, S)$ (with ties broken randomly) where
\begin{equation*}
\widetilde{\bSigma} = n^{-1} \sum_{\ell=1}^n (X_i^{(\ell)} - \hat{\mu})(X_i^{(\ell)} - \hat{\mu})^T, \qquad  \hat{\mu} = n^{-1}\sum_{\ell=1}^n X_i^{(\ell)},
\end{equation*}
and we want to show that $\hat{S}$ is consistent for $S^*$. 

By our earlier arguments, we know that 
\begin{equation*}
    \max_{S \subseteq [p]: |S| = k} \widetilde{\Delta}(S) = o_p(1), \qquad \widetilde{\Delta}(S) = p^{-1/2 + \delta}|\text{Obj}(\widetilde{\bSigma}, S) - \text{Obj}(\bSigma, S)|
\end{equation*}
If we can show that also
\begin{equation*}
    \max_{S \subseteq [p]: |S| = k} \widehat{\Delta}(S) = o_p(1), \qquad \widehat{\Delta}(S) = p^{-1/2 + \delta}|\text{Obj}(\hat{\bSigma}, S) - \text{Obj}(\widetilde{\bSigma}, S)|.
\end{equation*}
then we have consistency, because 
\begin{align*}
&P(\hat{S} = S^*) \\
&=   P(\text{Obj}(\hat{\bSigma}, S^*) < \min_{S \neq S^*} P(\text{Obj}(\hat{\bSigma}, S) )  \\
&\geq P\left(\max_{S \subseteq [p]: |S|=k} \widehat{\Delta}(S) < \frac{1}{4}(\min_{S \neq S^*}\text{Obj}(\bSigma, S) -  \text{Obj}(\bSigma, S^*) ) , \max_{S \subseteq [p]: |S|=k} \widetilde{\Delta}(S) < \frac{1}{4}(\min_{S \neq S^*}\text{Obj}(\bSigma, S) -  \text{Obj}(\bSigma, S^*) )\right) \\
&= 1 - o(1)
\end{align*}

Again, we consider an arbitrary size-$k$ subset $S$, fix $t > 0$, and devote the rest of the argument to showing that
\begin{equation*}
    P(\widehat{\Delta}(S) > t) = o\left({{p}\choose{k}}^{-1}\right).
\end{equation*}
Then, the fact that $\max_{S \subseteq [p] : |S| =k} \widehat{\Delta}(S) = o_p(1)$ follows trivially from a union bound.  \newline 

Again we reconfigure $\widehat{\Delta}(S)$ and bound it, similar to how we did before.

\begin{align*}
    &\widehat{\Delta}(S) \\
    &= p^{-1/2 - \delta}|\tr(\hat{\bSigma} - \hat{\bSigma}_{\bullet S} \hat{\bSigma}_S^{+} \hat{\bSigma}_{S \bullet}) -  \tr(\widetilde{\bSigma} - \widetilde{\bSigma}_{\bullet S} \widetilde{\bSigma}_{S}^{+} \widetilde{\bSigma}_{S \bullet}) |\\
        &\leq p^{-1/2 - \delta}| \tr(\hat{\bSigma} - \widetilde{\bSigma})| + p^{-1/2 - \delta}| \tr(\hat{\bSigma}_{S\bullet}\hat{\bSigma}_{\bullet S} \hat{\bSigma}^{+}_{S} - \widetilde{\bSigma}_{S\bullet}\widetilde{\bSigma}_{\bullet S} \widetilde{\bSigma}^{+}_{S})|\\
        &\leq p^{-1/2 - \delta}| \tr(\hat{\bSigma} - \widetilde{\bSigma})| + p^{-1/2 - \delta} |\tr(\hat{\bSigma}_{S\bullet}\hat{\bSigma}_{\bullet S})| \cdot |\tr(\hat{\bSigma}_S^+ - \widetilde{\bSigma}_S^+)| + p^{-1/2 - \delta}|\tr(\hat{\bSigma}_{S\bullet}\hat{\bSigma}_{\bullet S} - \widetilde{\bSigma}_{S\bullet}\widetilde{\bSigma}_{\bullet S} )| \cdot  |\tr(\widetilde{\bSigma}^{+}_{S} )|\\
        &\leq p^{-1/2 - \delta}| \tr(\hat{\bSigma} - \widetilde{\bSigma})| + p^{-1/2 - \delta} |\tr(\hat{\bSigma}_{S\bullet}\hat{\bSigma}_{\bullet S} -\widetilde{\bSigma}_{S\bullet}\widetilde{\bSigma}_{\bullet S})| \cdot |\tr(\hat{\bSigma}_S^+ - \widetilde{\bSigma}_S^+)|\\
        &\qquad  + p^{-1/2+\delta} |\tr(\widetilde{\bSigma}_{S\bullet}\widetilde{\bSigma}_{\bullet S} - \bSigma_{S\bullet}\bSigma_{\bullet S})| \cdot |\tr(\hat{\bSigma}_S^+ - \widetilde{\bSigma}_S^+)| +  p^{-1/2 - \delta} |\tr( \bSigma_{S\bullet}\bSigma_{\bullet S})| \cdot |\tr(\hat{\bSigma}_S^+ - \widetilde{\bSigma}_S^+)| \\
        &\qquad + p^{-1/2 - \delta}|\tr(\hat{\bSigma}_{S\bullet}\hat{\bSigma}_{\bullet S} - \widetilde{\bSigma}_{S\bullet}\widetilde{\bSigma}_{\bullet S} )| \cdot  |\tr(\widetilde{\bSigma}^{+}_{S} - \bSigma_S^+) | + p^{-1/2 - \delta}|\tr(\hat{\bSigma}_{S\bullet}\hat{\bSigma}_{\bullet S} - \widetilde{\bSigma}_{S\bullet}\widetilde{\bSigma}_{\bullet S} )| \cdot  |\tr(\bSigma^{+}_{S} )| \\
        &\leq \underbrace{p^{-1/2 - \delta}| \tr(\hat{\bSigma} - \widetilde{\bSigma})|}_{\widehat{\Delta}_1(S)} + \underbrace{p^{-1/2 - \delta} |\tr(\hat{\bSigma}_{S\bullet}\hat{\bSigma}_{\bullet S} -\widetilde{\bSigma}_{S\bullet}\widetilde{\bSigma}_{\bullet S})|}_{\widetilde{\Delta}_2(S)} \cdot |\tr(\hat{\bSigma}_S^+ - \widetilde{\bSigma}_S^+)|\\
        &\qquad  + p^{-1/2+\delta} |\tr(\widetilde{\bSigma}_{S\bullet}\widetilde{\bSigma}_{\bullet S} - \bSigma_{S\bullet}\bSigma_{\bullet S})| \cdot |\tr(\hat{\bSigma}_S^+ - \widetilde{\bSigma}_S^+)| +  c_2 \underbrace{p^{1/2 + \delta} |\tr(\hat{\bSigma}_S^+ - \widetilde{\bSigma}_S^+)|}_{\widehat{\Delta}_3(S)} \\
        &\qquad + p^{-1/2 - \delta}|\tr(\hat{\bSigma}_{S\bullet}\hat{\bSigma}_{\bullet S} - \widetilde{\bSigma}_{S\bullet}\widetilde{\bSigma}_{\bullet S} )| \cdot  \underbrace{|\tr(\widetilde{\bSigma}^{+}_{S} - \bSigma_S^+) |}_{\widetilde{\Delta}_3(S)} + c_3 \underbrace{p^{-1/2 - \delta}|\tr(\hat{\bSigma}_{S\bullet}\hat{\bSigma}_{\bullet S} - \widetilde{\bSigma}_{S\bullet}\widetilde{\bSigma}_{\bullet S} )|}_{\widehat{\Delta}_2(S)} \\
\end{align*}

Our above results already imply tail bounds for the $\widetilde{\Delta}_i(S)$, i.e., we know for any $t > 0$ that 
\begin{equation*}
    P(\widetilde{\Delta}_i(S) > t) =  o\left({{p}\choose{k}}^{-1}\right),
\end{equation*}
We know show the same tail bounds for the $\widehat{\Delta}_i(S)$, i.e., we show for any $t > 0$ that 
\begin{equation*}
    P(\widehat{\Delta}_i(S) > t) =  o\left({{p}\choose{k}}^{-1}\right),
\end{equation*}
It is then straightforward to establish the required tail bound \eqref{eq:overall_tail_bound} for $\widehat{\Delta}(S)$.

Prior to proceeding, we note it is easy to verify that 

\begin{align*}
    \widetilde{\bSigma} - \hat{\bSigma} = \mu(\mu - \hat{\mu} )^T + \hat{\mu}( \hat{\mu} - \mu)^T \implies \widetilde{\bSigma}_{ij} - \hat{\bSigma}_{ij} = (\hat{\mu}_i - \mu_i)(\hat{\mu}_j - \mu_j).
\end{align*} 

\subsubsection{Tail bound for $\widehat{\Delta}_1(S)$} 

By \Cref{ass:sub_gaussian}, the $X_i$ are sub-Guassian and  
$\max_i \|X_i\|_{\psi_2}$ is bounded by some constant $c_{23}$. Therefore, \cite[Theorem 2.6.3]{Vershynin} gives us sub-Gaussian concentration, i.e.,  for any $x > 0$:
\begin{equation*}
    P(|\hat{\mu}_i - \mu_i| \geq x) \leq 2 \exp\left( -c_{24} t^2 n \right). 
\end{equation*}
We can then get our tail bound by applying a union bound.
\begin{align*}
      P(\widehat{\Delta}_1(S) > t) &= P( p^{-1/2 - \delta} \sum_{i=1}^p (\hat{\mu}_i - \mu_i)^2 > t) & \\
        &\leq \sum_{i=1}^p P\left( |\hat{\mu}_i - \mu| > (tp^{-1/2 + \delta} )^{\frac{1}{2}}  \right) & \text{ (union bound) } \\
        &\leq 2p \exp \left( -c_{24} tnp^{-1/2 + \delta}  \right) & \text{ (sub-Gaussian concentration) }\\
        &= o\left({{p}\choose{k}}^{-1}\right). 
\end{align*}

\subsubsection{Tail bound for $\widehat{\Delta}_2(S)$}

As was the case for $\widetilde{\Delta}_2(S)$, it suffices to examine the different $\hat{\bSigma}^2_{ij} - \widetilde{\bSigma}^2_{ij}$. Fix some constant $c_{25} > 0$ and define $\widehat{B}_{ij} = \max\{ 2\bSigma_{ij} + c_{25},   2\bSigma_{ij} - c_{25}\}$. Two things are true. First, rewriting 
\begin{equation*}
    \hat{\bSigma}_{ij} + \widetilde{\bSigma}_{ij} = 2\bSigma_{ij} + 2(\widetilde{\bSigma}_{ij} - \bSigma_{ij}) + (\hat{\bSigma}_{ij} - \widetilde{\bSigma}_{ij})
\end{equation*}
makes it clear that  $|\hat{\bSigma}_{ij} + \widetilde{\bSigma}_{ij} | > \widehat{B}_{ij}$ implies that either $2|\widetilde{\bSigma}_{ij} - \bSigma_{ij}| > c_{25}/2$ or $|\hat{\bSigma}_{ij} - \widetilde{\bSigma}_{ij}| > c_{25}/2$. Second, $\widehat{B}_{ij} \leq c_{26}$ for some constant. Thus for any $x \geq 0$,

\begin{align*}
    &P\left( |\hat{\bSigma}_{ij}^2 - \widetilde{\bSigma}_{ij}^2  | > x \right)  \\
    &= P\left( |\hat{\bSigma}_{ij}^2 -  \widetilde{\bSigma}_{ij}^2  | > x, |\hat{\bSigma}_{ij} + \widetilde{\bSigma}_{ij}| \leq \widehat{B}_{ij}  \right) + P\left( |\hat{\bSigma}_{ij}^2 - \widetilde{\bSigma}_{ij}^2  | > x, |\hat{\bSigma}_{ij} + \widetilde{\bSigma}_{ij}| > \widehat{B}_{ij}  \right)  \\
    &\leq  P\left( |\hat{\bSigma}_{ij} -\widetilde{\bSigma}_{i j }  | > x/\widehat{B}_{ij}\right) + P\left(|\hat{\bSigma}_{ij} +  \widetilde{\bSigma}_{ij}| > \widehat{B}_{ij}  \right)  \\
    &\leq P\left( |\hat{\bSigma}_{ij} - \widetilde{\bSigma}_{i j }  | > x/c_{26} \right) + P\left(|\widetilde{\bSigma}_{ij} - \bSigma_{ij}| > c_{25}/4 \right) + P\left(|\hat{\bSigma}_{ij} - \widetilde{\bSigma}_{ij}| > c_{25}/2 \right) \\
    & & \hspace{-150pt}\text{ (definition of $\widehat{B}_{ij}$ and union bound)}\\
    &= P\left( |\hat{\mu}_i - \mu_i|\cdot|\hat{\mu}_j - \mu_j| > x/c_{26} \right) + P\left(|\widetilde{\bSigma}_{ij} - \bSigma_{ij}| > c_{25}/4 \right) + P\left(|\hat{\mu}_i -\mu_i| \cdot |\hat{\mu}_j - \mu_j|  > c_{25}/2 \right) \\
    &\leq  P( |\hat{\mu}_i - \mu_i| > (x/c_{26})^{1/2})  + P( |\hat{\mu}_j - \mu_j|> (x/c_{26})^{1/2})  + P\left(|\widetilde{\bSigma}_{ij} - \bSigma_{ij}| > c_{25}/4 \right)\\
    &\qquad + P\left(|\hat{\mu}_i -\mu_i|  > (c_{25}/2)^{1/2}\right) + P\left( |\hat{\mu}_j - \mu_j|  > (c_{25}/2)^{1/2} \right) \\
    & & \hspace{-150pt}\text{(union bound)}\\
    &\leq 4\exp(-c_{27}xn ) + 5\exp(-c_{28}n)  \\
    & & \hspace{-300pt} \text{ (sub-Gaussian and sub-exponential concentration) }
\end{align*}

We can finish the argument the same way as we did for $\widetilde{\Delta}_2(S)$.

\begin{align*}
    &P(\widehat{\Delta}_2(S) > t) \\
    &\leq  P( p^{-1/2-\delta}\sum_{i \in S} \sum_{j = 1}^p  |\hat{\bSigma}_{i j}^2 - \widetilde{\bSigma}_{i j }^2 | > t) & \text{(triangle inequality)}\\
    &=\sum_{i \in S} \sum_{j = 1}^p   P\left( |\hat{\bSigma}_{i j}^2 - \widetilde{\bSigma}_{i j }^2  | > k^{-1}p^{-1/2 + \delta} t \right) & \text{(union bound)}\\
    &\leq 4pk\exp(-c_{29}t np^{-1/2 + \delta}) + 5pk\exp(-c_{30}n)   & \text{(concentration for $\hat{\bSigma}^2_{ij}$)}\\
    &=  o\left({{p}\choose{k}}^{-1}\right) & 
\end{align*}

\subsubsection{Tail bound for $\widehat{\Delta}_3(S)$}

We first develop the following bound, which holds whenever both $\hat{\bSigma}_S$ and $\widetilde{\bSigma}_S$ are invertible.

\begin{align*}
    |\tr(\hat{\bSigma}_S^+ - \widetilde{\bSigma}_S^+ )|  &= |\tr(\hat{\bSigma}_S^{-1} - \widetilde{\bSigma}_S^{-1} )|\\
    &= |\tr(\bSigma_S^{-1}\bSigma_S^{1/2} ( \hat{\bSigma}_S^{-1} - \widetilde{\bSigma}_S^{-1} )  \bSigma_S^{1/2} )| \\
    &\leq k \|\bSigma_S^{-1} \|_2 \cdot \left|\tr\left((\bSigma_S^{-1/2} \widetilde{\bSigma}_S\bSigma_S^{-1/2} + \bSigma_S^{-1/2} (\hat{\bSigma}_S - \widetilde{\bSigma}_S)\bSigma_S^{-1/2} )^{-1} 
   - (\bSigma_S^{-1/2} \hat{\bSigma}_S\bSigma_S^{-1/2})^{-1} \right)\right| \\  
   &\leq c_{31}\left|\tr\left((\bSigma_S^{-1/2} \widetilde{\bSigma}_S\bSigma_S^{-1/2} + \bSigma_S^{-1/2} (\hat{\bSigma}_S - \widetilde{\bSigma}_S)\bSigma_S^{-1/2} )^{-1} 
   - (\bSigma_S^{-1/2} \hat{\bSigma}_S\bSigma_S^{-1/2})^{-1} \right)\right| 
\end{align*}

Crucially, we can show that $\|\bSigma_S^{-1/2}(\hat{\bSigma}_{S} - \widetilde{\bSigma}_{S})\bSigma_S^{-1/2}\|_2$ concentrates to zero. For any $x \geq 0$, we have 
\begin{align*}
    &P(\|\bSigma_S^{-1/2}(\hat{\bSigma}_{S} - \widetilde{\bSigma}_{S}) \bSigma_S^{-1/2} \|_2 \geq x) &\\
    &\leq   P(\|\hat{\bSigma}_{S} - \widetilde{\bSigma}_{S}\|_2 \geq  \|\bSigma_S^{-1}\|_2^{-1}x) & \\
    &\leq P(\|\hat{\bSigma}_{S} - \widetilde{\bSigma}_{S}\|_F \geq \|\bSigma_S^{-1}\|_2^{-1} x) &\\
    &\leq \sum_{i \in S} \sum_{j \in S} P(|\hat{\mu_i} - \mu_i| \cdot |\hat{\mu}_j - \mu_j| \geq k^{-2}\|\bSigma_{S}^{-1} \|^{-2}_2 x^2) & \text{(triangle inequality + union bound)}\\
    &\leq \sum_{i \in S} \sum_{j \in S} P(|\hat{\mu_i} - \mu_i| \geq k^{-1}\|\bSigma_{S}^{-1}\|^{-1}_2 x) +  P(|\hat{\mu_j} - \mu_j| \geq k^{-1}\|\bSigma_{S}^{-1}\|^{-1}_2 x) & \text{(union bound)}\\
    &\leq 4 \exp(-c_{32}x^2n )  & \text{(sub-Gaussian concentration)}  
\end{align*}
Henceforth we refer to $\|\bSigma_S^{-1/2}(\hat{\bSigma}_{S} - \widetilde{\bSigma}_{S})\bSigma_S^{-1/2}\|_2$ as $\E$. We will need the following lemma. 

\begin{lemma}
    \label{lem:inverse_local_lip2}
    Fix some $a > 0$. There exists $b > 0$ such that for all $x, y \geq c$
    \begin{equation}
        |x^{-1} - y^{-1}| \leq b |x - y|
    \end{equation}
\end{lemma}
\begin{proof}
\begin{align*}
    |x^{-1} - y^{-1}| = \frac{|x-y|}{xy} \leq \frac{1}{c^2}|x-y|
\end{align*}
So one can take $b= 1/c^2$. 
\end{proof}

The Bauer-Fike theorem \citep{Bauer} tells us that every eigenvalue of $\bSigma_S^{-1/2} \widetilde{\bSigma}_S\bSigma_S^{-1/2} + \E$ is within $\|\E\|_2$ of an eigenvalue of  $\bSigma_S^{-1/2} \widetilde{\bSigma}_S\bSigma_S^{-1/2}$. Note that because $\bSigma_{S}$ is always full-rank, whenever $\bSigma_S^{-1/2} \widetilde{\bSigma}_S\bSigma_S^{-1/2} + \E$ is invertible, so is $\hat{\bSigma}_S$. Using Bauer-Fike, we find whenever $\widetilde{\bSigma}_{S}$ and $\hat{\bSigma}_{S}$ are both invertible that
\begin{align*}
    &\left|\tr\left((\bSigma_S^{-1/2} \widetilde{\bSigma}_S\bSigma_S^{-1/2} + \bSigma_S^{-1/2} (\hat{\bSigma}_S - \widetilde{\bSigma}_S)\bSigma_S^{-1/2} )^{-1} 
   - (\bSigma_S^{-1/2} \hat{\bSigma}_S\bSigma_S^{-1/2})^{-1} \right)\right| &\\
   &\leq k \left( \frac{1}{\lambda_k(\bSigma_S^{-1/2} \hat{\bSigma}_S\bSigma_S^{-1/2}) -  \|\E\|_2} -  \frac{1}{\lambda_1(\bSigma_S^{-1/2} \hat{\bSigma}_S\bSigma_S^{-1/2}) +  \| \E\|_2} \right) & \text{(Bauer-Fike)}\\
   &= k \left( \frac{1}{s^2_k(\widetilde{\X}_{\bullet S} \bSigma_{S}^{-1/2}) -  \|\E\|_2} -  \frac{1}{s_1^2(\widetilde{\X}_{\bullet S} \bSigma_{S}^{-1/2}) +  \| \E\|_2} \right) & \\
\end{align*}

where $\widetilde{\X}_{\bullet S} = \X_{\bullet S} - 1\mu_S^T$.

Now we can get our tail bound. Fix some $c_{33} \in (0, 1)$ and also some $c_{34} \in (0, (1 - c_{33})^2)$. We see for large enough $n$ that 

\begin{align*}
 &P\left(\widehat{\Delta}_3 > t  \right)  \\
 &\leq P\left(p^{1/2 - \delta} |\tr(\widehat{\bSigma}_S^+ - \widehat{\bSigma}_S^{+})| > t, \max_{i \in [k]} | s_i(n^{-1/2}\widetilde{\X}_{\bullet S}\bSigma_S^{-1/2}) - 1| \leq c_{33}, \|\E\|_2 \leq c_{34} \right) \\
 &\qquad + P(\max_{i \in [k]} | s_i(n^{-1/2}\widetilde{\X}_{\bullet S}\bSigma_S^{-1/2}) - 1| > c_{33}) + P(|\E\|_2 > c_{34}) \\
 & & \hspace{-300pt} \text{(union bound)}\\
&\leq P\left( |s_1^2(\widetilde{\X}_{\bullet S} \bSigma_{S}^{-1/2}) - s_k^{2}(\widetilde{\X}_{\bullet S} \bSigma_{S}^{-1/2})| + 2\|\E\|_2 > c_{35}k^{-1} p^{-1/2 + \delta}t, \max_{i \in [k]} | s_i(n^{-1/2}\widetilde{\X}_{\bullet S}\bSigma_S^{-1/2}) - 1| \leq c_{33}, \|\E\|_2 \leq c_{34} \right) \\
 &\qquad + P(\max_{i \in [k]} | s_i(\widetilde{\X}_{\bullet S}\bSigma_S^{-1/2}) - n^{1/2}| > c_{33}n^{1/2}) + P(|\E\|_2 > c_{34}) \\
 & & \hspace{-300pt} \text{(both $\hat{\bSigma}_S$ and $\widetilde{\bSigma}_S$ invertible on event, apply \Cref{lem:inverse_local_lip2})}\\
&\leq P\left( |s_1^2(\widetilde{\X}_{\bullet S} \bSigma_{S}^{-1/2}) - s_k^{2}(\widetilde{\X}_{\bullet S} \bSigma_{S}^{-1/2})| > c_{36} p^{-1/2 + \delta}t\right) \\
 &\qquad + P(\|\E\|_2 > c_{37} p^{-1/2 + \delta}t) + P(\max_{i \in [k]} | s_i(\widetilde{\X}_{\bullet S}\bSigma_S^{-1/2}) - n^{1/2}| > n^{1/2}c_{33}) + P(|\E\|_2 > c_{34}) \\
 & & \hspace{-300pt} \text{(union bound) }\\
 &\leq P\left( |s_1^2(\widetilde{\X}_{\bullet S} \bSigma_{S}^{-1/2}) - s_k^{2}(\widetilde{\X}_{\bullet S} \bSigma_{S}^{-1/2})| > c_{36} p^{-1/2 + \delta}t\right)\\
 & + 4\exp(-c_{38} p^{-1 + \delta}n t^2) + 2 \exp(-c_{39}(n^{1/2} - c_{40})^2)  + 4\exp(-c_{41}n)\\
 & & \hspace{-300pt} \text{(earlier concentration bounds once $n$ is large enough)}
\end{align*}

The last three terms are $o\left({{p}\choose{k}}^{-1}\right)$, so all that remains to show is that the first term is as well. By a union bound 
\begin{align*}
    &P\left( |s_1^2(\widetilde{\X}_{\bullet S} \bSigma_{S}^{-1/2}) - s_k^{2}(\widetilde{\X}_{\bullet S} \bSigma_{S}^{-1/2})| > c_{36} p^{-1/2 + \delta}t\right) \\
    &\leq P\left( |s_1^2(\widetilde{\X}_{\bullet S}\bSigma_S^{-1/2}) -1|  > c_{42} p^{-1/2 + \delta}t\right) + P\left( |s_k^2(\widetilde{\X}_{\bullet S} \bSigma_S^{-1/2}) -1|) > c_{42} p^{-1/2 + \delta}t\right)
\end{align*}

Thus, letting $c_{43} = \max\{|2 - c_{44}|, |2 + c_{44}|\}$, it suffices to see that 
\begin{align*}
  &P\left(\max_{i \in [k]}| s^2_i(n^{-1/2}\widetilde{\X}_{\bullet S}\bSigma_{S}^{-1/2}) - 1| >c_{42} p^{-1/2 + \delta}t \right)\\
    &= P\left(\max_{i \in [k]}| s^2_i(n^{-1/2}\widetilde{\X}_{\bullet S}\bSigma_{S}^{-1/2}) - 1| > c_{42} p^{-1/2 + \delta}t , \max_{i \in [k]}| s_i(n^{-1/2}\widetilde{\X}_{\bullet S} \bSigma_{S}^{-1/2}) + 1| \leq c_{43} \right)\\
    &\qquad + P\left(\max_{i \in [k]}| s_i(n^{-1/2}\widetilde{\X}_{\bullet S} \bSigma_{S}^{-1/2}) + 1| > c_{43} \right)\\
    &\leq P\left(\max_{i \in [k]}| s_i(n^{-1/2}\widetilde{\X}_{\bullet S} \bSigma_{S}^{-1/2}) - 1| > c_{45}p^{-1/2 + \delta}t \right) + P\left(\max_{i \in [k]}| s_i(n^{-1/2}\widetilde{\X}_{\bullet S} \bSigma_{S}^{-1/2}) - 1| > c_{44} \right)\\
    &\leq P\left(\max_{i \in [k]}| s_i(\widetilde{\X}_{\bullet S} \bSigma_{S}^{-1/2}) - n^{1/2}| > c_{45}n^{1/2}p^{-1/2 + \delta}t \right) + P\left(\max_{i \in [k]}| s_i(|\widetilde{\X}_{\bullet S} \bSigma_{S}^{-1/2}) - n^{1/2}| > c_{44}n^{1/2} \right)\\
    &\leq 2\exp(-c_{46} (n^{1/2}p^{-1/2 + \delta}t  - c_{47})^2) + 2\exp(-c_{48} (n^{1/2} - c_{49})^2)\\
    & & \hspace{-300pt} \text{(Apply \Cref{eq:singular_value_concentration} once $n$ large enough)}\\
    &= o\left({{p}\choose{k}}^{-1}\right)
\end{align*}
where we have applied the same technique as we did for $\widetilde{\Delta}_2(S)$. 

% The whole first half of this was changed for the second revision. 

\section{Additional Theoretical Results}
\label{sec:add_theory}

\subsection{Estimation Error of Regression Coefficients}
\label{sec:add_theory_error}

In this section, we use the finite sample bounds we developed in \Cref{sec:high_dim_consistency_appdx} to study the regression coefficients from the regression of $X_{-U}$ on $X_U$, where $U$ is some size-$k$ subset. More concretely, we aim to measure the quality of the observed regression coefficients $\hat{\bSigma}_{-U, U} \hat{\bSigma}_U^{-1}$ as an estimate of the population regression coefficients $\bSigma_{-U, U} \bSigma^{-1}_U$. We will use the $2$-to-$\infty$ norm to measure our estimation error,
\begin{equation*}
    \|\hat{\bSigma}_{-U, U} \hat{\bSigma}_U^{-1} - \bSigma_{-U, U} \bSigma_U^{-1} \|_{2 \rightarrow \infty} = \max_{j \not \in U} \|\hat{\bSigma}_U^{-1}\hat{\bSigma}_{U j} -  \bSigma_U^{-1}\bSigma_{Uj}  \|_2. 
\end{equation*}
Considering variables $j \not \in U$, the $2$-to-$\infty$ norm captures the worst-case $L^2$ estimation error for the regression coefficients from the regression of $X_j$ on $X_U$. Because our analysis comes from finite sample bounds, it applies even in the high-dimensional where the number of variables $p$ and the sample size $n$ are comparable in size. Along with providing a bound that holds for a specific subset $U$, we also provide a bound that holds uniformly over all size-$k$ subsets. Because it is uniform, it is informative even for a subset that is chosen by a possibly complicated and non-deterministic CSS algorithm. 

\Cref{prop:coefficient_estimation_error} states our error control result. In \Cref{prop:coefficient_estimation_error}, we adopt two of the assumptions from \Cref{sec:high_dim_consistency_appdx}, this time also requiring that constants be independent of $k$: 
\begin{assumption}[Light-tailed data]
    \label{ass:sub_gaussian_coeff}
     The sub-Gaussian norm (see \citep[Definition 2.5.6]{Vershynin}) of each variable is bounded above by a constant independent of $n$, $p$, and $k$.
\end{assumption}
\begin{assumption}[Low-dimensional invertibility]
    \label{ass:invertibility_coeff}
    The minimum of the smallest eigenvalues of all the $k \times k$ principal sub-matrices of $\bSigma$ is bounded below by a positive constant independent of $n$, $p$, and $k$.
\end{assumption}

\begin{proposition}
    \label{prop:coefficient_estimation_error} 
    Consider observing samples $x^{(1)}, \dots, x^{(n)} \in \R^p$ from a mean-zero distribution $X \sim P$ satisfying \Cref{ass:sub_gaussian_coeff} and \Cref{ass:invertibility_coeff}, and let $\hat{\bSigma} = n^{-1}\sum_{i=1}^n x^{(i)} (x^{(i)})^{\top}$ denote the sample covariance. For a size-$k$ subset $U$, let $M$ be the sub-Gaussian norm of $X_U$ and set $B = \max_{j \not \in U} \|\bSigma_{Uj} \|_2$. Then for any $t \in (0, 1)$,
    \begin{equation*}
        P(\|\hat{\bSigma}_{-U, U} \hat{\bSigma}_U^{-1} - \bSigma_{-U, U} \bSigma_U^{-1} \|_{2 \rightarrow \infty} >t ) \leq c_1 pk \exp\left( -\left( \frac{c_{2}t n^{1/2}}{k^{1/2} \max(BM^2, M^2, 1)} - 1 \right)_+^2  \right )
    \end{equation*}
    for some constants $c_1, c_2 > 0$ that are independent of $n, p, k$. As a consequence,
    \begin{equation*}
       P\left( \max_{U \subseteq [p]: |U| = k}\|\hat{\bSigma}_{-U, U} \hat{\bSigma}_U^{-1} - \bSigma_{-U, U} \bSigma_U^{-1} \|_{2 \rightarrow \infty} >t \right) \leq c_1 p^{k+1}k \exp\left( -\left( \frac{c_3 t n^{1/2}}{k^{2}} - 1 \right)_+^2  \right ),
    \end{equation*}
    where $c_3 > 0$ is another constant independent of $n, p, k$.
\end{proposition}
In \Cref{prop:coefficient_estimation_error}, the light-tail assumption is necessary to ensure that our estimates concentrate, and the invertibility assumption is necessary to ensure that the estimate $\hat{\bSigma}^{-1}_U$ of $\bSigma^{-1}_U$ is well-behaved.   We assume mean-zero data for simplicity. By applying identical arguments to those in \Cref{sec:high_dim_consistency_mean_appdx}, we can modify \Cref{prop:coefficient_estimation_error} to accommodate data with non-zero mean. 

While \Cref{prop:coefficient_estimation_error} may appear hard to parse at first, what it tells us is quite simple:  
excluding log factors, the 2-to-$\infty$ regression coefficient estimation error for a subset $U$ scales (up to logarithmic factors) at worst as $\sqrt{k/n} \max(BM^2, M^2, 1)$, where $M$ is the sub-Gaussian norm of $X_U$ and $B= \max_{j \not \in U} \|\Sigma_{U j}\|_2$ is the maximum $L^2$ norm of the covariance between the variables $U$ and variables $j \not \in U$. If the $B$ and $M^2$ can be bounded by constants that have no dependence on $k, n, $ or $p$, then the estimation error scales (up to logarithmic factors) at worst as $\sqrt{k/n}$. 

A situation where we may reasonably believe that the estimation error scales at worst (up to logarithmic factors) as $\sqrt{k/n}$ is if our data comes from the $k$-dimensional PCSS model \eqref{eq:pcss_model} and we consider coefficient estimates for population subset $S$. It is not unreasonable to imagine that the variables in $S$ have little dependence, as if the principal variables are highly correlated and explain redundant variation, a smaller and less dependent principal variable set may better model the data. Hence, we can reasonably restrict to the case where the sub-Gaussian norm of $X_S$ and the maximum eigenvalue of $\bSigma_S$ are bounded by a constant independent of $n$, $p$, and $k$. One simple case where this is true is when the principal variables are exactly independent (because of \Cref{ass:sub_gaussian_coeff}) . Henceforth, when we say bounded by a constant, we mean a constant independent of $n$, $p $ and $k$. In this case, we have immediately that $M$ is bounded by a constant. To deal with $B$, we note that by \Cref{ass:sub_gaussian_coeff}, 
the maximum variance of the variables in $-S$ is bounded above by a constant. Because of the covariance structure induced by the PCSS model, this means that $\max_{j \not \in S} \bSigma_{jS}\bSigma_S^{-1}\bSigma_{Sj} = \max_{j \not \in S} 
 \|\bSigma_{S}^{-1/2} \bSigma_{Sj}\|_2^2$ and therefore $\max_{j \not \in S}  \|\bSigma_{S}^{-1/2} \bSigma_{Sj}\|_2$ are also bounded above by a constants. Since
 \begin{equation*}
     B = \max_{j \not \in S}  \| \bSigma_{Sj}\|_2^2 \leq \|\bSigma_{S}^{1/2} \|_2  \max_{j \not \in S}  \| \bSigma_{S}^{-1/2}\bSigma_{Sj}\|_2 = \|\bSigma_{S} \|^{1/2}_2 \max_{j \not \in S}  \| \bSigma_{S}^{-1/2}\bSigma_{Sj}\|_2, 
 \end{equation*}
and $\|\bSigma_{S} \|^{1/2}_2$ is bounded above by a constant, so is $B$. Hence, if the principal variables are sufficiently non-dependent, we can reasonably imagine that the regression coefficient error for the principal variable set $S$ (up to logarithmic factors) scales at worst as $\sqrt{k/n}$.

In general, however, for a set of more correlated variables $U$, the sub-Gaussian norm of $X_U$ or the maximum eigenvalue of $\bSigma_U$ can grow with $k$. Still, even in the most general case, it is straightforward to use \Cref{ass:sub_gaussian_coeff} to bound $M \leq c_4 \sqrt{k}$ and $B \leq c_4 \sqrt{k}$ for some constant $c_4 > 0$. Hence we still find that the estimation error (up to logarithmic factors) scales at worst as $k^2/\sqrt{n}$. In fact, \Cref{prop:coefficient_estimation_error} tells us that this error bound holds uniformly over all size-$k$ subsets, so it applies even for a subset selected via a complex and possibly non-deterministic CSS algorithm.

Since $k$ is typically much smaller than $n$ and $p$, the estimation error is quite small in both cases, even when the number of variables $p$ and sample size $n$ are comparable in magnitude.  

\subsection{Selecting the Best Subset}
\label{sec:add_theory_subset}

\textcolor{revision}{
We identify a handful of settings where our greedy (\Cref{alg:greedy}) and/or swapping (\Cref{alg:swap}) algorithms find the best subset according to the CSS objective \eqref{eq:pv}. In the below, we consider performing CSS on a covariance matrix $\bSigma$ and suppose we are aiming to select a size-$k$ subset. Proofs of the claims can be found in \Cref{sec:subset_correctness_appdx} . }

\textcolor{revision}{
\begin{enumerate}
    \item  \textbf{Independence}: Supposing that $\bSigma$ is a diagonal matrix, both our greedy and swapping algorithms find an optimal solution.   
    \item \textbf{Perfect reconstruction}: Suppose $\bSigma$ is such that there exists a set of $k$ variables that perfectly linearly reconstruct the remaining. Then both our greedy and swapping algorithms will find such a set of variables, and therefore find an optimal solution. 
    \item \textbf{Equicorrelation}: If $\bSigma$ is a equicorrelation matrix, then all subsets have the same objective value by symmetry, and thus our algorithms trivially find an optimal solution. Note that our covariance view from \Cref{sec:equivalence} greatly simplifies our analysis of this setting.
    \item \textbf{Block Diagonal}: Suppose $\bSigma$ is a block diagonal correlation with $k$ blocks, i.e., there are $k$ groups of variables. Let $M_1 \in \R^{q \times q}, \dots, M_k \in \R^{q \times q}$ be the different blocks. Supposing that each group has a strong enough representative, i.e., there exists a row/column $i_j$ in each of the $j = 1, \dots, k$ blocks such that $\|M_{\bullet i_j} \|_2^2 > q/2 $, then our greedy algorithm will find the optimal solution. 
\end{enumerate}}

\textcolor{revision}{Empirically we verified that when $\bSigma_{ij} = \rho^{|i-j|}$ (covariance of an AR(1) process) our greedy and swapping algorithms are not guaranteed to find the optimal solution.}

\section{\texorpdfstring{$L^2$}{L2} Random Variables as a Hilbert Space}
\label{sec:hilbert_space_appdx}

The results of this section are generally known but hard to find covered in full generality. We give detailed account of them as they are particularly useful for proving the results in this article. Consider a probability space $(\Omega, \mathcal{F}, P)$. Upon identifying random variables which are equal almost surely in the same equivalence class, the random variables in $L^2(\Omega, \mathcal{F}, P)$ form a complete Hilbert space. Equality between random variables denotes membership to the same equivalence class, i.e., almost sure equality. For two real-valued random variables $X, Y \in L^2$, the inner product $\langle X, Y \rangle_{\mathcal{H}}$, given by $E[XY]$, induces a norm $\|X\|_{\mathcal{H}}^2 = E[X^2]$. As abuse of notation, when dealing with random vectors $Z \in \R^k$ with entries $Z_i$ in $L^2$, we let $\|Z\|^2_{\mathcal{H}} = \sum_{i=1}^k \|Z_i\|_{\mathcal{H}}^2$, so $\|Z\|^2_{\mathcal{H}} = \tr(E[ZZ^\top])$. 

Since we're working in a Hilbert space, we can project random variables onto linear subspaces of our Hilbert space. We only project onto finite dimensional subspaces, so we don't concern ourselves with the subtleties of infinite dimensional subspaces. If $\mathcal{G}$ is a subspace of $\mathcal{H}$, we denote the projection of $X$ onto $\mathcal{G}$ as $\Proj_{\mathcal{G}}(X)$. For random vectors, $\Proj_{\mathcal{G}}(Z)$ is the random vector with $i$th entry $\Proj_{\mathcal{G}}(Z_i)$. We denote the subspace spanned by the entries of $Z$ as $\Span(Z)$, but as an abuse of notation we write $\Proj_{Z}(\cdot)$ instead of $\Proj_{\Span(Z)}(\cdot)$. The following Lemma helps us compute projections:

\begin{lemma}
\label{lem:coefficient_minimiziation} 
For random vectors $X \in \R^p$, $Y \in \R^q$,
\[ \{ E[YX^\top]E[XX^\top]^{+} + \E: \E \in \R^{q \times p} \text{ such that } \E X = 0\} = \argmin_{\B \in \R^{q \times p}} E[\|Y - \B X\|_2^2].  \]
\end{lemma}
\Cref{thm:proj_props} provides some useful facts about these projections. 
\begin{theorem}
\label{thm:proj_props}
Consider random vectors $X \in \R^{p}$, $Y \in \R^{q}$ and define $P = E[YX^\top]E[XX^\top]^+ X$ and $R = Y - P$. Then,
\begin{enumerate}[label=(\roman*)]
    \item $P = \Proj_{X}(Y)$,
    \label{thm:proj_props_i}
    \item $E[PR^\top] = \0$ and $E[XR^\top] = \0$, i.e., each entry of $X$ and $P$ is orthogonal to each entry of $R$,
    \label{thm:proj_props_ii}
    \item $\|Y\|^2_{\mathcal{H}} = \|P\|^2_{\mathcal{H}} + \|R\|^2_{\mathcal{H}} $,
    \label{thm:proj_props_iii}
    \item  $E[PP^\top] = E[YX^\top]E[XX^\top]^{+}E[XY^\top]$ and $E[RR^\top] = E[YY^\top] - E[YX^\top]E[XX^\top]^+E[XY^\top]$. 
    \label{thm:proj_props_iv}
\end{enumerate}
\end{theorem}

The following is a useful corollary of \Cref{thm:proj_props}.

\begin{corollary}
\label{cor:uncorrelated_coeffs}
Considering random vectors $X \in \R^{p}$, $Y \in \R^{q}$, $Y = \W X + \epsilon$ with $E[ X \epsilon^\top] = 0$ if and only if $\W = E[YX^\top]E[XX^\top]^+ + \E$ where $\E \in \R^{q \times p}$ such that $\E X = 0$.
\end{corollary}

\section{Proofs and Derivations}
\label{sec:proofs_appdx}

\subsection{Proof of \texorpdfstring{\Cref{lem:coefficient_minimiziation}}{Lemma}}

Minimizing $E[\|Y - \B X\|_2^2] = E[\tr( (Y - \B X)(Y - \B X)^\top ] = \tr(E[YY^\top] - 2\B E[XY^\top] + \B E[XX^\top]\B^\top)$ with respect to $\B$ is an unconstrained convex minimization problem. Thus $\B$ is a minimizer if and only if it sets the first derivative to $0$. Using \citep[2.5.2]{Petersen} to compute the first derivative, we get $\B$ is a solution if and only if  $\B E[XX^\top] - E[YX^\top] = 0$. 

We check that $\B = E[YX^\top] E[XX^\top]^{+}$ is a solution. If $v \in \R^{p}$ is in the null space of $E[XX^\top]$, then $0 = v^\top E[XX^\top]v = E[(X^\top v)^2]$, so $X^\top v = 0$. Then also $E[YX^\top]v = E[Y(X^\top v)]= 0$. Then, for our choice of $v$, $(E[YX^\top] E[XX^\top]^{+}E[XX^\top] - E[YX^\top])v =0$. If $v \in \R^p$ is not in the null space of $E[XX^\top]$, then $E[XX^\top]^+E[XX^\top]v = v$, so $(E[YX^\top] E[XX^\top]^{+}E[XX^\top] - E[YX^\top])v =0$. Thus $(E[YX^\top]E[XX^\top]^{+}E[XX^\top] - E[YX^\top])v = 0$ for all $v \in \R^p$, so $E[YX^\top]E[XX^\top]^{+}E[XX^\top] - E[YX^\top] = 0$. 

For any $ \E \in \R^{q \times p}$ such that $\E X = 0$ it is obvious that $E[YX^\top]E[XX^\top]^{+} + \E$ is a solution. If $\tilde{\B}$ is also a solution, then  $\tilde{\B} E[XX^\top] - E[YX^\top] = 0$ and thus $(E[YX^\top] E[XX^\top]^{+} - \tilde{\B})E[XX^\top] = 0$. Thus $\tilde{\B} = E[YX^\top] E[XX^\top]^{+} + \E$ for some $ \E \in \R^{q \times p}$ such that $\E E[XX^\top] = \0$. Since also $\E E[XX^\top]\E^\top  = \0$, we have that $\tr(E[\E X (\E X)^\top])) = E[\|\E X \|^2_2]= 0$ so $\E X = 0$.

\subsection{Proof of \texorpdfstring{\Cref{thm:proj_props}}{Theorem}}
 We prove each result in the order the are presented. 
\begin{enumerate}[label=(\roman*)]
    \item The $i$th entry of $\Proj_{X}(Y)$ is given by $Y_i - (b^*)^\top X$, where $b^*$ minimizes $\|Y_i - b^\top X\|_{H}^2 = E[(Y_i - b^\top X)^2]$ over $b$. By \Cref{lem:coefficient_minimiziation}, $b^*$ must be given by the $i$th row of $E[YX^\top]E[XX^\top]^{+} + \E$ for some $ \E \in \R^{q \times p}$ such that $\E X = 0$. Thus, up to almost sure equality, that the $i$th entry of $\Proj_{X}(Y)$ is given by $Y_i - (E[YX^\top]E[XX^\top]^{+})_{i \bullet} X$. 
    \item From \ref{thm:proj_props_i} we know that $P_j$ is $\Proj_X(Y_j)$ and $R_i = Y_i - \Proj_X(Y_i)$. From standard properties of projections in Hilbert spaces, we know that $P_j \in \Span(X)$ and $R_i$ is orthogonal to anything in $\Span(X)$. thus $P_j$ and $R_i$ are orthogonal, meaning exactly that $E[PR^\top] = \0$. Same argument applies for $X$. 
    \item From \ref{thm:proj_props_ii} we know that $E[P_iR_i] = 0$. Thus, since $Y_i = P_i + R_i$ we have $ \|Y_i\|^2_{\mathcal{H}} = E[Y_i^2] = E[R_i^2] + E[P_i^2] =  \|P_i\|^2_{\mathcal{H}}+ \|R_i\|^2_{\mathcal{H}}$.
    \item We can compute $E[PP^\top] = E[YX^\top]E[XX^\top]^{+}E[XX^\top]E[XX^\top]^{+}E[XY^\top]  = E[YX^\top]E[XX^\top]^{+}E[XY^\top]  $. Since $Y = P + R$ and we know $E[PR^\top] = \0$ from \ref{thm:proj_props_ii}, we immediately get $E[YY^\top] = E[PP^\top] + E[RR^\top]$. So,  $E[RR^\top]= E[YY^\top] - E[PP^\top] = E[YY^\top] - E[YX^\top]E[XX^\top]^+E[XY^\top] $.
\end{enumerate}

\subsection{Proof of \texorpdfstring{\Cref{cor:uncorrelated_coeffs}}{Corollary}}
\label{cor:uncorrelated_coeffs:proof}
The reverse direction follows as if $\W = E[YX^\top]E[XX^\top]^+ + \E$, then $\W X = P$ and $\epsilon = R$ as in the statement of \Cref{thm:proj_props}. Thus by \Cref{thm:proj_props}\ref{thm:proj_props_ii},  $E[X \epsilon^\top] = 0$. For the forward direction, by the properties of projections in Hilbert spaces, it is only possible for $\epsilon_i$ to be uncorrelated with $X$ if $\W_{i \bullet}X$ equals $\Proj_{X}(Y_i) = P_i$.  Therefore we must have $\W =  E[YX^\top]E[XX^\top]^+ + \E$ where $\E \in \R^{q \times p}$ such that $\E X = 0$ by \Cref{thm:proj_props}\ref{thm:proj_props_i}. 

\subsection{Proof of \texorpdfstring{\Cref{prop:css_equiv_pv}}{Proposition} and \texorpdfstring{\Cref{prop:pop_css_equiv_pv}}{Proposition}}
By \Cref{lem:coefficient_minimiziation} and \Cref{thm:proj_props}\ref{thm:proj_props_iv}, 
\begin{equation*}
    \min_{\B \in  \R^{p \times k}} E[\|X - \B X_{S}\|_2^2] = \tr(E[XX^\top] - E[XX^\top]_{\bullet S} E[XX^\top]_S^{+}E[XX^\top]_{S \bullet}).
\end{equation*}
Taking $X$ to be mean-zero, this immediately implies \Cref{prop:pop_css_equiv_pv}. To prove \Cref{prop:css_equiv_pv}, let $X$ be uniformly random over the rows of the matrix $\X$ and evaluate the expectations on both sides. 

\subsection{Proof of \texorpdfstring{\Cref{thm:css_is_mle}}{Theorem}} 
We provide a proof of \Cref{thm:css_is_mle}. We then provide the analogous result when we allow the covariance of $X_{-S} | X_{S}$  to be diagonal (see \eqref{eq:diag_pcss_model}). Lastly we provide the analogous results for both these cases when we assume that the principal variable distribution $F$ is Gaussian.  

\subsubsection{PCSS Model}

Suppose we restrict $F$ to have a probability density bounded by $K < \infty$. Our model is parameterized by $\theta = (S, f, \mu_{-S}, \W, \sigma^2)$ where $f$ is the density of $F$. Let $\hat{\mu}$ and $\hat{\bSigma}$ be the sample mean and sample covariance of our data. Fix a size-$k$ subset $S$. The negative log-likelihood, scaled by $1/n$, under our model is given by 
\begin{align*}
    &-\frac{1}{n} \sum_{i=1}^n \ell(x^{(i)}; \theta) \\
    &=-\frac{1}{n} \sum_{i=1}^n \left(\ell(x_{S}^{(i)}; \theta) + \ell(x_{-S}^{(i)} | x_{S}^{(i)};\theta) \right)\\
    &= \frac{1}{n} \sum_{i=1}^n  \log(f(x_{S}^{(i)})) +  \frac{1}{2}\cdot\frac{1}{n}\sum_{i=1}^n \left( (p-k) \log(2\pi) + (p-k)\log \sigma^2 + \sigma^{-2}\|x_{-S}^{(i)} - \mu_{S} - \W(x_{S}^{(i)}  - \mu_{-S})\|_2^2)  \right)\\
    &= -\frac{1}{n} \sum_{i=1}^n f(x_{S}^{(i)}) + \frac{p-k}{2} \log \sigma^2 +  \frac{1}{2\sigma^2} \cdot \sum_{j=1}^{p-k} \left( \frac{1}{n}\sum_{i=1}^n ((x_{-S}^{(i)})_j - (\mu_{S})_j - \W_{j \bullet}(x_{S}^{(i)}  - \mu_{-S}))^2 \right)+ \frac{(p-k) \log(2\pi)}{2}.
\end{align*}
We know that $-\frac{1}{n} \sum_{i=1}^n \log(f(x_{S}^{(i)}))$ is bounded below by $-\log(K)$. By adding together $n$ bump functions are narrow enough and attain their maximum height of $K$ at each $x^{(i)}$ and placing any remaining mass in the tails of $f$, we can find an $f$ that attains this bound. Although $\mu_{S} = E_F[X_S]$ is not explicitly known, we can absorb $(\mu_{S})_j$ into $\W_{j \bullet} \mu_{-S}$ for each $j$, so it is an inconsequential parameter. So, without loss of generality we set $\mu_{S} = \hat{\mu}_{S}$ in the above. For $c > 0$, the function $\log x + \frac{c}{x}$ is minimized at $x = c$ and has minimized value $\log(c) + 1$. Thus we should pick $(\mu_{-S})_{j}$ and $\W_{j \bullet}$  to minimize $\frac{1}{n}\sum_{i=1}^n((x_{-S}^{(i)})_{j} - (\mu_{-S})_{j} - \W_{j \bullet}(x_{S}^{(i)} - \hat{\mu}_{S}) )^2$. This is a convex problem. From the first order condition one can compute the minimizing value for $(\mu_{-S})_j$ is $(\hat{\mu}_{-S})_j$, and then apply \Cref{lem:coefficient_minimiziation} to compute that the minimizing value of $\W_{j \bullet}$ is given by $(\hat{\bSigma}_{-S,S})_{j \bullet} \hat{\bSigma}_{S}^{+}$. Plugging in these minimizing values gives
\begin{equation*}
\frac{1}{n}\sum_{i=1}^n((x_{-S}^{(i)})_{j} - (\mu_{-S})_{j} - \W_{j \bullet}(x_{S}^{(i)} - \hat{\mu}_{S}) )^2 = (\hat{\bSigma}_{-S})_{jj} - (\hat{\bSigma}_{-S, S})_{j\bullet}\hat{\bSigma}^{+}_{S}(\hat{\bSigma}_{S, -S})_{\bullet j}. 
\end{equation*}
Since no $k$ variables perfectly linearly reconstruct the remaining, we know at least one of $(\hat{\bSigma}_{-S})_{jj} - (\hat{\bSigma}_{-S, S})_{j\bullet}\hat{\bSigma}^{+}_{S}(\hat{\bSigma}_{S, -S})_{\bullet j}$ is positive. The minimizing value of $\sigma^2$ is thus attained at
\begin{equation*}
  \sigma^2 = \frac{1}{p-k} \sum_{j=1}^{p-k}(\hat{\bSigma}_{-S})_{jj} - (\hat{\bSigma}_{-S, S})_{j\bullet}\hat{\bSigma}^{-1}_{S}(\hat{\bSigma}_{S, -S})_{\bullet j} = \tr(\hat{\bSigma}_{-S} - \hat{\bSigma}_{-S, S}\hat{\bSigma}^{+}_{S}\hat{\bSigma}_{S, -S})/(p-k).
\end{equation*}
Plugging in all our minimizing values, the negative log likelihood is 
\begin{equation*}
-\log(K) + \frac{p-k}{2} \log\left( \tr(\hat{\bSigma}_{-S} - \hat{\bSigma}_{-S, S}\hat{\bSigma}^{+}_{S}\hat{\bSigma}_{S, -S})/(p-k)\right) + \frac{p-k}{2}(1 + \log(2\pi)).
\end{equation*}
Thus the $S$ which minimizes the negative log-likelihood is the one which minimizes $\tr(\hat{\bSigma}_{-S} - \hat{\bSigma}_{-S, S}\hat{\bSigma}^{+}_{S}\hat{\bSigma}_{S, -S}) = \tr(\hat{\bSigma} - \hat{\bSigma}_{\bullet S}\hat{\bSigma}^{+}_{S}\hat{\bSigma}_{S \bullet})$, completing the claim. 

In the case that $F$ is completely unrestricted, our model is parameterized by $\theta = (S, F, \mu_{-S}, \W, \sigma^2)$. Fix a size-$k$ subset $S$. The negative log hybrid likelihood is given by
\begin{align*}
    &-\frac{1}{n} \sum_{i=1}^n \ell(x^{(i)}; \theta) \\
    &=-\frac{1}{n} \left(\sum_{i=1}^n \ell(x_{S}^{(i)}; \theta) - \ell(x_{-S}^{(i)} | x_{S}^{(i)};\theta)\right)\\
    &=-\frac{1}{n} \sum_{i=1}^n \log(F(x^{(i)}_{S})) - \frac{1}{n} \sum_{i=1}^n \ell(x_{-S}^{(i)} | x_{S}^{(i)};\theta).
\end{align*}
By \citep[Theorem 2.1]{Owen}, the optimal choice of $F$ is that which puts weight $1/n$ on each $x^{(i)}$. For this minimizing $F$, the negative log hybrid likelihood is $  \log(n) - \frac{1}{n} \sum_{i=1}^n \ell(x_{-S}^{(i)} | x_{S}^{(i)};\theta)$. The argument is then identical to the one above. 

\subsubsection{Diagonal Conditional Covariance}
We show the analogous result for the when we allow the covariance of $X_{-S} | X_{S}$  to be diagonal (see \eqref{eq:diag_pcss_model}). We will assume that every $(k + 1)$ by $(k + 1)$ principal sub-matrix of $\hat{\bSigma}$ is full rank. 

In place of $\sigma^2 > 0$ we have the diagonal matrix $\D \succ \0$ as a parameter. Fix a subset $S$ and again decompose the likelihood:
\begin{equation*}
   -\frac{1}{n} \sum_{i=1}^n \ell(x^{(i)}; \theta) = -\frac{1}{n} \sum_{i=1}^n \left(\ell(x_{S}^{(i)}; \theta) + \ell(x_{-S}^{(i)} | x_{S}^{(i)};\theta) \right). 
\end{equation*}
The maximum likelihood estimate of the density $f$ is the same as before, and finding the remaining maximizers is similar to before:
\begin{align*}
&-\frac{1}{n} \sum_{i=1}^n \ell(x_{-S}^{(i)} | x_{S}^{(i)}; \theta) \\
&=  \frac{1}{2} \sum_{j=1}^{p-k} \left( \log \D_{jj} +  \frac{1}{\D_{jj}} \cdot  \left( \frac{1}{n}\sum_{i=1}^n ((x_{-S}^{(i)})_j - (\mu_{S})_j - \W_{j \bullet}(x_{S}^{(i)}  - \mu_{-S}))^2 \right) \right) + \frac{(p-k) \log(2\pi)}{2}.
\end{align*}
Following the reasoning above, we set $\W_{j \bullet}$  to minimize $\frac{1}{n}\sum_{i=1}^n((x_{-S}^{(i)})_{j} - (\mu_{-S})_{j} - \W_{j \bullet}(x_{S}^{(i)} - \hat{\mu}_{S}) )^2$ resulting in a minimized value of $ (\hat{\bSigma}_{-S})_{jj} - (\hat{\bSigma}_{-S, S})_{j\bullet}\hat{\bSigma}^{-1}_{S}(\hat{\bSigma}_{S, -S})_{\bullet j} > 0$ where the inequality and invertability of $\hat{\bSigma}^{-1}_{S}$ come from our new assumption. Like before, this minimized value is the minimizing choice for $\D_{jj}$, implying that the minimizing $\D$ is given by $\Diag(\hat{\bSigma}_{-S} - \hat{\bSigma}_{-S, S}\hat{\bSigma}_{S}^{-1}\hat{\bSigma}_{S, -S})$.
The minimized negative log-likelihood is then given by  
\begin{equation*}
-\log(K) + \frac{1}{2}\log(|\Diag(\hat{\bSigma}_{-S} - \hat{\bSigma}_{-S, S}\hat{\bSigma}_{S}^{-1}\hat{\bSigma}_{S, -S})|) + \frac{p-k}{2}(1 + \log(2\pi)).
\end{equation*}
So a subset minimizes the negative log-likelihood if and only if it minimizes $\log(|\Diag(\hat{\bSigma}_{-S} - \hat{\bSigma}_{-S, S}\hat{\bSigma}_{S}^{-1}\hat{\bSigma}_{S, -S})|)$.

The argument for minimizing the negative hybrid log-likelihood is then identical to the one above. 

\subsubsection{Gaussian Principal Variable Distribution \texorpdfstring{$F$}{F}}
\label{sec:gaussian_pcss_mle_proof}
Now we derive the analogous results when we require that the principal variable distribution $F$ is a Gaussian, so $X_{S} \sim N(\mu_{S}, \C)$ where $\C \succ \0$. First consider the PCSS model with this restriction. Along with the assumption that no $k$ variables perfectly reconstruct the rest, we require that every $k$ by $k$ principal sub-matrix of $\bSigma$ is full rank. The model is parameterized by $(S, \mu, \C, \W, \sigma^2)$. Fix a subset $S$. The negative log-likelihood, scaled by $\frac{1}{n}$, is given by 
\begin{equation*}
    -\frac{1}{n} \sum_{i=1}^n \ell(x^{(i)}; \theta) = -\frac{1}{n} \sum_{i=1}^n \left(\ell(x_{S}^{(i)}; \theta) + \ell(x_{-S}^{(i)} | x_{S}^{(i)};\theta) \right).
\end{equation*}
Fix a size-$k$ subset $S$. We can ignore the dependence of $-\frac{1}{n} \sum_{i=1}^n \ell(x_{-S}^{(i)} | x_{S}^{(i)};\theta)$ on $\mu_{S}$ for the same reason in the proof of \Cref{thm:css_is_mle}. Minimizing $-\frac{1}{n} \sum_{i=1}^n (\ell(x_{S}^{(i)}; \theta)$ with respect to $\C$ and $\mu_{S}$ is equivalent finding the unrestricted MLE for a multivariate Gaussian. Since $\hat{\bSigma}_{S}$ is invertible by our new assumption, the minimizing values of $C$ and $\mu_{S}$ are $\hat{\bSigma}_{S}$ and $\hat{\mu}_S$. The minimizing values of $W, \sigma^2$ are then identical to those in the proof of \Cref{thm:css_is_mle} by the same computations, and the minimized value of the negative log-likelihood is 
\begin{equation*}
    \frac{1}{2} \log|\hat{\bSigma}_{S}| + \frac{p-k}{2} \log\left( \tr(\hat{\bSigma}_{-S} - \hat{\bSigma}_{-S, S}\hat{\bSigma}^{-1}_{S}\hat{\bSigma}_{S, -S})/(p-k)\right) + \frac{p}{2}(1 + \log(2\pi)). 
\end{equation*}
So the subset which minimizes the negative log-likelihood is one which minimizes $\log|\hat{\bSigma}_{S}| + (p-k)\log( \tr(\hat{\bSigma}_{-S} - \hat{\bSigma}_{-S, S}\hat{\bSigma}^{-1}_{S}\hat{\bSigma}_{S, -S})/(p-k)$. This is sufficient to prove the claim.  

In the case that the principal variable distribution $F$ is a Gaussian and the covariance of $X_{-S}\mid X_{S}$ is allowed to be diagonal, we again assume that every $(k + 1)$ by $(k + 1)$ principal sub-matrix of $\hat{\bSigma}$ is full rank. Then proof is identical to above. The subset which minimizes the negative log-likelihood is one which minimizes $\log |\bSigma_{S}| + \log(|\Diag(\hat{\bSigma}_{-S} - \hat{\bSigma}_{-S, S}\hat{\bSigma}_{S}^{-1}\hat{\bSigma}_{S, -S})|)$.

% This whole subsection was added as part of second revision 
\subsection{Proof of \texorpdfstring{\Cref{thm:high_dim_consistency}}{Theorem}}

Throughout this proof, we denote the sorted eigenvalues of a positive semi-definite matrix $\A \in \S^{q \times  q}_{+}$ as $\lambda_1(\A) \geq \dots \geq \lambda_q(\A) \geq 0$. For a matrix $\M \in \R^{q \times k}$ we denote its sorted top-$k$ singular values as $s_1(\M) \geq \dots \geq s_k(M) \geq 0$. We also let $\| \cdot \|_2$ denote the spectral norm of a matrix. 

In \Cref{sec:high_dim_consistency_appdx}, we identified a high-dimensional setting where the  in-sample CSS solution (which is exactly the MLE $\hat{S}$) is consistent for the population CSS solution. This is the subject of \Cref{thm:high_dim_consistency_appdx}, whose proof is the bulk of \Cref{sec:high_dim_consistency_appdx}. If we can show that (1) the assumptions of \Cref{thm:high_dim_consistency_appdx} are satisfied and (2) that the population subset $S$ is  also the population CSS solution once $n$ is large enough, then \Cref{thm:high_dim_consistency} is immediately implied by \Cref{thm:high_dim_consistency_appdx}. We recall the assumptions of \Cref{thm:high_dim_consistency_appdx} below. 

\setcounter{assumption}{0}

\begin{assumption}[Light-tailed data]
     The sub-Gaussian norm (see \citep[Definition 2.5.6]{Vershynin}) of each variable is bounded above by a constant independent of $p$ and $n$.
\end{assumption}

\begin{assumption}[Low-dimensional invertibility]
    The minimum of the smallest eigenvalues of all the $k \times k$ principal sub-matrices of $\bSigma$ is bounded below by a positive constant independent of $p$ and $n$.
\end{assumption}

\begin{assumption}[Separation]
    For some $\delta > 0$, the achieved CSS objective for the population best subset 
    \begin{equation*}
    S^{*} = \argmin_{U \subseteq [p], |U| = k} \tr(\bSigma - \bSigma_{\bullet U} \bSigma^{+}_U \bSigma_{\bullet U})
    \end{equation*}
    is well separated from the objective achieved by any other subset 
    \begin{equation*}
        \min_{U \neq S^*} \tr(\bSigma - \bSigma_{\bullet U} \bSigma^{+}_U \bSigma_{\bullet U}) - \tr(\bSigma - \bSigma_{\bullet S^*} \bSigma^{+}_{S^*} \bSigma_{\bullet S^*})  = \Omega(p^{1/2 + \delta})
    \end{equation*}
\end{assumption}

To show that the conditions of \Cref{thm:high_dim_consistency} imply these assumptions and also imply that the population subset $S$ is eventually the same as the population CSS minimizer $S^*$, we will need two linear algebra lemmas whose results are well-known \citep{Joriki,  Algebraic}. 

\begin{lemma}
\label{lem:block_eig}
    For any positive semi-definite matrix $\A$ written in block matrix notation
    \begin{equation*}
        \A = \begin{bmatrix}
            \A_{11} & \A_{12}\\
            \A_{21} & \A_{22},
        \end{bmatrix}
    \end{equation*}
    we have $\lambda_1(\A) \leq \lambda_1(\A_{11}) + \lambda_1(\A_{22})$.
\end{lemma}

\begin{proof}
    We recall the proof from \cite{Joriki}. Let $v$ be a unit vector. We can write $v = \begin{bmatrix} v_1 & v_2 \end{bmatrix}^T$ as $v = \|v_1\|\begin{bmatrix} v_1/\|v_1\| & 0 \end{bmatrix}^T + \|v_2\|_2\begin{bmatrix} 0  & v_2/\|v_2\|_2 \end{bmatrix}^T$. Thus we have written $v$ as $c_1 u_2 + c_2u_2$ where $u_1$, $u_2$, and $c = \begin{bmatrix} c_1 , c_2\end{bmatrix}^T \in \R^2$ are all unit vectors. Defining the matrix 
    \begin{equation*}
        \B = \begin{bmatrix} u_1^T \A_{11} u_1 & u_1^T \A_{12}u_2 \\
         u_2^T \A_{11} u_1 & u_2^T \A_{22}u_2 .
    \end{bmatrix}
    \end{equation*}
    it is the case that $v^T\A v = c^T\B c$ by construction. Therefore, 
    \begin{align*}
        v^T \A v &= c^T \B c\\
                 &\leq \lambda_1(\B) \\
                 &\leq \tr(\B)\\
                 &= u_1^T \A_{11} u_1 + u_2^T \A_{22}u_2 \\
                 &\leq \lambda_1(\A_{11}) + \lambda_1(\A_{22}) 
    \end{align*}
    Because the inequality holds for all unit vectors $v$, it must be the case that $\lambda_1(\A) \leq \lambda_1(\A_11) + \lambda_1(\A_22)$ as desired. 
\end{proof}

\begin{lemma}
\label{lem:schur_complement}
    For any positive definite matrix $\A \in \S^{k \times k }_{++}$ written in block matrix notation
    \begin{equation*}
        \A = \begin{bmatrix}
            \A_{11} & \A_{12}\\
            \A_{21} & \A_{22},
        \end{bmatrix}
    \end{equation*}
    the minimum eigenvalue of the Schur compelment $\A_{22} - \A_{21}\A_{11}^{-1}\A_{12}$ is at least as large as the minimum eigenvalue of $\A$. 
\end{lemma}
\begin{proof}
We recall the proof from \cite{Algebraic}. 
\[
\begin{bmatrix}
\A_{11} & \A_{12} \\
\A_{21} & \A_{22} 
\end{bmatrix} = 
\begin{bmatrix} 
\I_{k_1} & 0 \\
\A_{21}\A^{-1}_{11} & \I_{k_2} 
\end{bmatrix}
\begin{bmatrix}
\A_{11} & 0 \\
0 & \A_{22} - \A_{21}\A^{-1}_{11}\A_{12} 
\end{bmatrix}
\begin{bmatrix}
\I_{k_1} & \A_{11}^{-1}\A_{12} \\
0 & \I_{k_2} 
\end{bmatrix},
\]
Let 
\[
\mathcal{V} := \left\{ 
\begin{bmatrix}
x \\ y 
\end{bmatrix} \in \R^k : x + \A_{11}^{-1} \A_{12} y = 0
\right\}.
\]

Note that if \( z = \begin{bmatrix} x^T & y^T \end{bmatrix}^T \in \mathcal{V} \), then 
\[
\begin{bmatrix} 
\I_{k_1} & \A_{11}^{-1} \A_{12} \\
0 & \I_{k_2} 
\end{bmatrix}
\begin{bmatrix} 
x \\ y 
\end{bmatrix} = 
\begin{bmatrix} 
0 \\ y 
\end{bmatrix}.
\]
so 
\begin{equation}
    z^T \A z = y^T (\A_{22} - \A_{21}\A^{-1}_{11}\A_{12})y
\end{equation}
Also $\|z\|_2^2 \geq \|y \|_2^2$. This is sufficient to imply that  
\[
\min_{z\in \R^k, z \neq 0} \frac{z^T \A z}{z^T z} 
\leq \min_{z \in \mathcal{V}, z \neq 0} \frac{z^T \A z}{z^T z} \leq \min_{y \in \R^{k_{2}} \neq 0} \frac{y^T (\A_{22} - \A_{21}\A^{-1}_{11}\A_{12}) y}{y^T y} 
\]
which establishes the claim 
\end{proof}

Now we catalogue what the conditions of \Cref{thm:high_dim_consistency} tell us:
\begin{enumerate}
    \item Because all the sub-gaussian norm of each variable is bounded above by a constant, there must be some constant $C > 0$ such that all the entries of $\bSigma$ have absolute value bounded above by $C$. 
    \item There are constants $m > 0$ and $v> 0$ such that $\lambda_k(\bSigma_S) > m$ and $\sigma^2 \geq v$. 
    \item There are constants $\eta > 0$ and $c_1, \dots, c_k > 0$ such that, if $N_i$ is the number of entries of $\W_{\bullet i}$ whose absolute value is strictly less than $\eta$, then eventually $N_i \geq c_i p^{1/2 + \delta}$. As a consequence, there is some  constant $c > 0$ such that eventually $\min_{i \in [k]} N_i \geq c p^{1/2 + \delta}$. 
\end{enumerate}

Of course, all these constants are independent of $n$ and $p$. With this information in hand, we can show both that the assumptions hold and that $S$ will eventually be the same as $S^*$. \newline 

\noindent \textbf{Sub-Gaussianity}: \Cref{ass:sub_gaussian} is itself assumed in the satement of \Cref{thm:high_dim_consistency}. \newline 

\noindent \textbf{Low-dimensional invertibility}: Consider any size-$k$ subset $U$ that is entirely disjoint from $S$. We can write the joint covariance of all the variables in $U$ and $S$ in block matrix form as  
\begin{equation*}
    \bSigma_{S \cup U} = \begin{bmatrix} \bSigma_S & \bSigma_{SU}  \\
    \bSigma_{US} & \bSigma_{US}\bSigma_{S}^{-1}\bSigma_{SU} + \sigma^2 \I_k \end{bmatrix}
\end{equation*}
One can confirm by direct computation that 
\begin{equation*}
    \bSigma^{-1}_{S \cup U} = \begin{bmatrix} \bSigma_S^{-1} + \sigma^{-2} \bSigma_{S}^{-1}\bSigma_{SU}\bSigma_{US}\bSigma_S^{-1} & -\sigma^{-2}\bSigma_{S}^{-1} \bSigma_{SU} \\ - \sigma^{-2}\bSigma_{US} \bSigma_{S}^{-1}& \sigma^{-2}\I_k \end{bmatrix}.
\end{equation*}
We can bound $\|\bSigma_{SU}\bSigma_{US}\|_2 \leq \tr(\bSigma_{SU}\bSigma_{US}) \leq k^2C^2$. Then, applying \Cref{lem:block_eig}, we have 
\begin{align*}
    \lambda_1(\bSigma^{-1}_{S \cup U}) &\leq  \lambda_1(\bSigma_S^{-1} + \sigma^{-2} \bSigma_{S}^{-1}\bSigma_{SU}\bSigma_{US}\bSigma_S^{-1}) + \lambda_1(\sigma^{-2}\I_k )\\
    &= \|\bSigma_S^{-1} + \sigma^{-2} \bSigma_{S}^{-1}\bSigma_{SU}\bSigma_{US}\bSigma_S^{-1}\|_2 + \sigma^{-2}\\
    &\leq \|\bSigma_S^{-1}\|_2 + \sigma^{-2} \|\bSigma_{S}^{-1}\|^2_2 \|\bSigma_{SU}\bSigma_{US}\|_2 + \sigma^{-2} \\
    &\leq m^{-1} + v^{-1}m^{-2}k^2C^2 + v^{-1} . 
\end{align*}
As a consequence, we have that 
\begin{equation*}
    \lambda_k(\bSigma_{S \cup U}) \geq  (m^{-1} + v^{-1}m^{-2}k^2C^2 + v^{-1})^{-1} > 0.
\end{equation*}
Now if we consider any size-$k$ subset $\widetilde{U}$, the matrix $\bSigma_{\widetilde{U}}$ is a principal sub-matrix of $\bSigma_{S \cup U}$ for some choice of size-$k$ subset $U$. Therefore by Cauchy's eigenvalue interlacing theorem \cite{Bhatia}, 
\begin{equation*}
    \lambda_k(\bSigma_{\widetilde{U}}) \geq (m^{-1} + v^{-1}m^{-2}k^2C^2 + v^{-1})^{-1} > 0.
\end{equation*}
and we have established that \Cref{ass:invertibility} holds. \newline 

\noindent \textbf{Separation}: Now we establish both that \Cref{ass:seperation} holds and that $S^*$ and $S$ are eventually the same. Recall that $X_{\ell} = \W_{\ell \bullet} X_{S} + \epsilon$ for $\ell \not \in S$, where for ease of notation, we index the $p-k$ rows of $\W$ by elements of $-S$. Fix any size-$k$ subset $U$ that is not equal to $S$, so that there is some $i \in S$ that is not in $U$. To establish \Cref{ass:seperation} and that $S$ and $S^*$ are eventually equal, it will suffice to argue the following 
\begin{itemize}
    \item If $\ell \not \in S \cup U$ then 
    \begin{equation*}
    (\bSigma_{\ell \ell } - \bSigma_{\ell U
    }\bSigma_{U}^{-1}\bSigma_{U \ell}) -  (\bSigma_{\ell \ell } - \bSigma_{\ell S
    }\bSigma_{S}^{-1}\bSigma_{S \ell}) \geq 0.
    \end{equation*}
    \item If $\ell \not \in S \cup U$ and $\W_{\ell i} \geq \eta$, then 
    \begin{equation*}
    (\bSigma_{\ell \ell } - \bSigma_{\ell U
    }\bSigma_{U}^{-1}\bSigma_{U \ell}) -  (\bSigma_{\ell \ell } - \bSigma_{\ell S
    }\bSigma_{S}^{-1}\bSigma_{S \ell}) \geq \gamma
    \end{equation*}
    where $\gamma> 0$ is some constant. 
\end{itemize}
To see why this is sufficient, note that 
\begin{align*}
&\left(\tr(\bSigma -\bSigma_{\bullet U}\bSigma_{U}^{-1} \bSigma_{ U\bullet})  \right) -  \left(\tr(\bSigma- \bSigma_{\bullet S}\bSigma_{S}^{-1}\bSigma_{ S\bullet})  \right)\\
&\geq \sum_{\ell \not \in  S \cup U, \W_{\ell i} \geq w} \left[ (\bSigma_{\ell \ell } - \bSigma_{\ell U
}\bSigma_{U}^{-1}\bSigma_{U \ell}) -  (\bSigma_{\ell \ell } - \bSigma_{\ell S
}\bSigma_{S}^{-1}\bSigma_{S \ell}) \right]\\
& \qquad + \sum_{\ell \not \in  S \cup U, \W_{\ell i} < w} \left[ (\bSigma_{\ell \ell } - \bSigma_{\ell U
}\bSigma_{U}^{-1}\bSigma_{U \ell}) - (\bSigma_{\ell \ell } - \bSigma_{\ell S
}\bSigma_{S}^{-1}\bSigma_{S \ell}) \right]\\
&\qquad  - \left| \sum_{\ell \in S \cup U} (\bSigma_{\ell \ell } - \bSigma_{\ell U
}\bSigma_{U}^{-1}\bSigma_{U \ell}) -  (\bSigma_{\ell \ell } - \bSigma_{\ell S
}\bSigma_{S}^{-1}\bSigma_{S \ell}) \right|\\
&\geq \gamma (N_i - k)  - 4kC\\
&\geq \gamma \min_{j \in [k]}N_j - \gamma k - 4kC\\
&\geq \gamma c p^{\frac{1}{2} + \delta} - \gamma k - 4kC
\end{align*}
where the last inequality holds for large enough $n$ and we have also  bounded
\begin{align*}
    \left|\sum_{\ell \in S \cup U} \tr(\bSigma_{\ell \ell } - \bSigma_{\ell U }\bSigma_{U}^{-1}\bSigma_{U \ell}) -  \tr(\bSigma_{\ell \ell } - \bSigma_{\ell S }\bSigma_{S}^{-1}\bSigma_{S \ell}) \right| \leq \sum_{\ell \in S \cup U} 2|\tr(\bSigma_{\ell \ell })| \leq 4kC. 
\end{align*}
This bound has no dependence on $U$, and therefore it holds that for large enough $n$ that 
\begin{equation*}
    \min_{U \neq S} \tr(\bSigma - \bSigma_{\bullet U}\bSigma_{U}^{-1} \bSigma_{ U\bullet}) -  \tr(\bSigma - \bSigma_{\bullet S}\bSigma_{S}^{-1}\bSigma_{ S\bullet}) \geq \gamma c p^{\frac{1}{2} + \delta} - \gamma k - 4kC
\end{equation*}
Since $p$ is growing by assumption, this is sufficient to imply both \Cref{ass:seperation} and that $S$ is the eventually the same as the population CSS minimizer $S^*$.

We devote the rest of the proof to showing the claim. Let $\L$ be the Cholesky decomposition of $\bSigma_S$, so $\bSigma = \L^T\L$. If we write the population covariance in block matrix form with the subset $S$ as the first block and the remaining variables as the second block, then it is straightforward to check it admits the following Cholesky decomposition 
\begin{equation*}
    \bSigma = \begin{bmatrix} \L^T & \L^T \W^T \\ \0 & \sigma \I_{p-k}\end{bmatrix}^{T} \begin{bmatrix} \L^T & \L^T \W^T \\ \0 & \sigma \I_{p-k}\end{bmatrix} 
\end{equation*}
Let $v_1, \dots, v_p \in \R^p$ be the columns of this Cholesky decomposition. For a subset $T \subseteq [p]$, denote the span of 
$\{v_j: j \in T \}$ as $\Span(T)$, the projection of $x \in \R^p$ onto the span as $\Proj_{T}(x)$, and the component of $x$ that is orthogonal to this subspace as $R_T(x) = x - \Proj_T(x)$. By construction, the columns satisfy $v_s^T v_t = \bSigma_{st}$. Also they satisfy for $\ell \not \in S$ that 
\begin{equation}
    v_{\ell} = n_{\ell} + \sum_{j \in S} \W_{\ell j} v_j = n_{\ell} + \tilde{v}_{\ell}
\end{equation}
where
\begin{enumerate}
    \item the $n_{\ell}$ are orthogonal to each other and the subspace $\Span(S)$,
    \item the $n_{\ell}$ all have norm $\|n_{\ell}\|_2^2 = \sigma^2$, 
    \item the $\tilde{v}_{\ell} =  \sum_{j \in S} \W_{\ell j} v_j$ are in $\Span(S)$. 
\end{enumerate}
It is also easy to confirm that, for a subset $T \subseteq [p]$, the norm $\|R_{T}(v_j)\|_2^2$ is exactly the Schur complement $\bSigma_{jj} - \bSigma_{jT}\bSigma_{T}^{-1}\bSigma_{Tj}$. As a consequence, recalling the universal lower bound $m > 0$ for $\lambda_k(\bSigma_S)$, \Cref{lem:schur_complement} tells us that 
\begin{equation*}
    \|R_{S - j}(v_j)\|_2^2 \geq \lambda_{k}(\bSigma_{S}) \geq m,
\end{equation*}
where $S - j$ is the subset $S$ with $j$ removed.  \newline 

\noindent Now fix a size-k subset $U$ that is not equal to $S$. There must be at least some $i \in S$ that is not in $U$. Considering $\ell \not \in S \cup U$, our goal is to lower bound the difference between
\begin{align*}
     \bSigma_{\ell \ell} - \bSigma_{\ell U}\bSigma_U^{-1}\bSigma_{U\ell} = \min_{\beta \in \R^k} \|v_{\ell} - \sum_{j \in U} \beta_j v_j \|^2_2,
\end{align*}
where for notational simplicity we index the entries of $\beta$ by the $j \in U$, and $\bSigma_{\ell \ell} - \bSigma_{\ell S}\bSigma_S^{-1}\bSigma_{S\ell} = \sigma^2 $.  We can rewrite the objective
\begin{align*}
     &\|v_{\ell} - \sum_{j \in U} \beta_j v_j \|^2_2\\
     &= \|\tilde{v}_{\ell} - \sum_{j \in U \cap S} \beta_j v_j - \sum_{j \in U\setminus S} \beta_j \tilde{v}_j \|^2_2 +  \sigma^2 + \sigma^2 \sum_{j \in U \setminus S} \beta_j^2  \\
    &=\| \Proj_{U \cap S}( \tilde{v}_{\ell}) - \sum_{j \in U \setminus S} \beta_j \Proj_{U \cap S}(\tilde{v}_j)  - \sum_{j \in U \cap S} \beta_j v_j \|^2_2 +  \|  R_{U \cap S}(\tilde{v}_{\ell}) - \sum_{j \in U \setminus S} \beta_j R_{U \cap S}(\tilde{v}_j)  \|_2^2 +  \sigma^2 + \sigma^2 \sum_{j \in U \setminus S} \beta_j^2  \\ 
\end{align*}
We see immediately that $\bSigma_{\ell \ell} - \bSigma_{\ell U}\bSigma_U^{-1}\bSigma_{U\ell} \geq \sigma^2$. No matter what values of $\beta_j$ we pick for $j \in U\setminus S$, the quantity 
\begin{equation*}
    \Proj_{U \cap S}( \tilde{v}_{\ell}) - \sum_{j \in U\setminus S} \beta_j \Proj_{U \cap S}(\tilde{v}_j)
\end{equation*}
will always been in $\Span(S)$, so we will always set $\beta_j$ for $j \in U \setminus S$ to make the left-most term equal to zero. Thus, letting $q = |U \cap S|$ we are left with 
\begin{equation*}
   \bSigma_{\ell \ell} - \bSigma_{\ell U}\bSigma_U^{-1}\bSigma_{U\ell} = \argmin_{\beta \in \R^{q} } \|  R_{U \cap S}(\tilde{v}_{\ell}) - \sum_{j \in U \setminus S} \beta_j R_{U \cap S}(\tilde{v}_j)  \|_2^2 +  \sigma^2 \sum_{j \in U \setminus S} \beta_j^2 +  \sigma^2
\end{equation*}
The right hand side is a ridge regression problem with a unique solution $\beta^*$ that has a closed form. Let we let $\A \in \R^{p\times q}$ be the matrix whose columns are given by the vectors $R_{U \cap S}(\tilde{v}_i)$. Because 
\begin{equation}
    \|R_{U \cap S}(\tilde{v}_i)\|^2_2 \leq \|\tilde{v}_i \|_2^2 \leq \| v_i\|_2^2 \leq C
\end{equation}
We know that the trace and therefore largest eigenvalue of $\A^T\A$ is at most $kC$. This implies the largest singular value of $\A$ is at most $k^{1/2}C^{1/2}$. Let $\U \in \R^{p \times p}$ be the orthogonal matrix with columns $u_i$, whose first $q$ columns are the top $q$ left singular vectors of $\A$, \citep[Equation 3.47]{Hastie} implies for the optimal $\beta^*$ that 
\begin{align*}
    \|R_{U \cap S}(\tilde{v}_{\ell}) -  \sum_{j \in U \setminus S} \beta^*_j R_{U \cap S}(\tilde{v}_i) \|^2_2 &= \sum_{i = 1}^q
    \left(1 - \frac{s_i(\A)^2 }{s_i(\A)^2 + \sigma^2}\right)^2 (u_i^T R_{U \cap S}(\tilde{v}_{\ell}))^2 + \sum_{i=q+1}^p (u_i^T R_{U \cap S}(\tilde{v}_{\ell}))^2\\
    &\geq   \left(1 - \frac{kC }{kC + \sigma^2}\right)^2\sum_{i=1}^p (u_i^T R_{U \cap S}(\tilde{v}_{\ell}))^2  \\
    &= \left(1 - \frac{kC }{kC + \sigma^2}\right)^2 \| R_{U \cap S}(\tilde{v}_{\ell}) \|_2^2
\end{align*}
Now we just need to lower bound $\| R_{U \cap S}(\tilde{v}_{\ell}) \|_2^2$. Recall that $i \in S$ but $i \not \in U$. Then we can decompose
\begin{align*}
     R_{U \cap S}(\tilde{v}_{\ell}) &=  R_{U \cap S}(\sum_{j \in S} \W_{\ell j} v_j )\\
    &= \sum_{j \in S \setminus U} \W_{\ell j} R_{U \cap S}(v_j)\\
    &= \W_{\ell i} R_{S - i}(R_{U \cap S}(v_i)) + \W_{\ell i} \Proj_{S -i}(R_{U \cap S}(v_i)) + 
    \sum_{j \in S \setminus U, j \neq i} \W_{\ell j} R_{U \cap S}(v_j)\\
\end{align*}
The first vector is orthogonal to $\Span(S - i)$, whereas the remaining vectors in the expression live in $\Span(S - i)$. Because $\Span(U \cap S ) \subseteq \Span(S - i) $, we also have that $R_{S - i}(R_{U \cap S}(v_i)) = R_{S - i}(v_i)$, and therefore that 
\begin{align*}
    \|R_{U \cap S}(\tilde{v}_{\ell})\|^2_2 &\geq  \| \W_{\ell i} R_{S - i}(R_{U \cap S}(v_i))  \|_2^2 \\
    &\geq \| \W_{\ell i} R_{S - i}(v_i)  \|_2^2\\
    &\geq \W_{\ell i}^2 m
\end{align*}
where we have recalled from above that $\|R_{S - i}(v_i)\|^2_2 \geq m$. If $\ell$ is such that $|W_{\ell i}| \geq \eta$, then putting everything together gives that 
\begin{equation*}
    (\bSigma_{\ell \ell} - \bSigma_{\ell U}\bSigma_U^{-1}\bSigma_{U\ell}) - (\bSigma_{\ell \ell} - \bSigma_{\ell S}\bSigma_S^{-1}\bSigma_{S\ell}) \geq \left(1 - \frac{kC }{kC + \sigma^2}\right)^2 \eta^2m > 0.
\end{equation*}

\subsection{Proof of \texorpdfstring{\Cref{thm:compromise}}{Theorem}}
The reverse direction is obvious. We show the forward direction. For simplicity we'll assume $X$ is mean zero. For the general case apply the same argument to $X - \mu$. Since $Z \in \R^k$ has covariance $\I_k$, if $Z = \B X$ then $\B \in \R^{k \times p}$ must be rank $k$. Plugging $Z = \B X$ into our factor model equation, we get that $X = \W\B X + \epsilon$, where $\epsilon$ is uncorrelated with $\B X$. By \Cref{cor:uncorrelated_coeffs}, $\W = \bSigma_{X, \B X}\bSigma^{-1}_{\B X} = \bSigma \B^\top(\B \bSigma \B^{\top})^{-1}$. Our expression $X =\W \B X + \epsilon$ follows the definition of regression component decomposition given in \citep{Schonemann1976}. Referencing \citep[Equation 2.6]{Schonemann1976} and the discussion following it, the covariance $\D$ of $\epsilon$ has rank exactly $p - k$. Since $\D$ is diagonal, this means exactly $k$ of its entries are $0$. Let $S \subseteq [p]$ be the subset of indices such if $i \in S$ then $\D_{ii} = 0$ and thus $\epsilon_i = 0$. Then $X_{S} = \W_{S \bullet} \B X$. It clear that since $\epsilon$ is uncorrelated with $\B X$, it must also be uncorrelated with $X_{S}$.
Since $X_{S}$ and $\B X$ have full rank covariances, $\W_{S \bullet}$ must be full rank and thus invertible, and we can write $\B X = \W_{S \bullet}^{-1}X_{S}$. Plugging this into our original factor model we have $X_{-S} = \widetilde{\W}X_{S} + \epsilon$ where $\widetilde{\W} = \W_{-S \bullet}\W_{S \bullet}^{-1}$ and $X_{S}$ is uncorrelated with $\epsilon$.

\subsection{Proof of \texorpdfstring{\Cref{lem:residual_cov_update}}{Lemma}}
By \Cref{lem:coefficient_minimiziation} the residual $R(Y, Z)$ is given by $Y - \bSigma_{YZ} \bSigma_{Z}^+Z$. By standard properties of projections in finite dimensional Hilbert spaces, $R(X, X_{S}) = R(R(X, X_{U}), R(X_{i}, X_{U})) $. We can therefore compute, 
\begin{align*}
\bSigma_{R(X, X_{S})} &= \bSigma_{R(R(X, X_{U}), R(X_{i}, X_{U})))}\\
                     &= \bSigma_{R(X, X_{U})} - \bSigma_{R(X, X_U), R(X_i, X_U)}\bSigma_{R(X_i, X_U) }^{+}\bSigma_{ R(X_i, X_U), R(X, X_U)}\\
                     &= \bSigma_{R(X, X_{U})} - (\bSigma_{R(X, X_U)})_{\bullet i}(\bSigma_{R(X, X_U)})_{i\bullet}/(\bSigma_{R(X, X_U)})_{ii} \cdot I_{(\bSigma_{R(X, X_U)})_{ii} > 0}. 
\end{align*}
Where the second equality follows from \Cref{thm:proj_props}\ref{thm:proj_props_iv}. Also using \Cref{thm:proj_props}\ref{thm:proj_props_iv} we can compute that $(\bSigma_{R(X, X_U)})_{\bullet i}  =  \bSigma_{\bullet i} - \bSigma_{\bullet T}\bSigma_{T}^{+}\bSigma_{T, i}$.

\subsection{Proof of \texorpdfstring{\Cref{lem:finite_sample_validity}}{Lemma}}

Suppose $x^{(1)}, \dots, x^{(n)} \in \R^p$ are drawn from a distribution satisfying (\ref{eq:diag_pcss_model}) for a size-$k$ subset $S$, $k < p$, and population mean $\mu$. 

First we show that the quantiles of $T(S)$ from \eqref{eq:test_stat} are bounded by those of \eqref{eq:null_dist}. We will perform our analysis conditional on $x^{(1)}_{S}, \dots, x^{(n)}_{S}$ (hence, the independence between $X_S$ and $\epsilon$ is crucial). Let $\X_{S} \in \R^{n \times k}$ and $\X_{-S} \in \R^{n \times (p-k)}$ be the matrices with rows $x_{S}^{(i)}$ and $x_{-S}^{(i)}$. Defining $e^{(i)} = (x_{-S}^{(i)} - \mu_{-S}) - \W(x_{S}^{(i)} - \mu_{S})$ and $\hat{e}^{(i)} = (x_{-S}^{(i)} - \hat{\mu}_{-S}) - \hat{\bSigma}_{-S, S}\hat{\bSigma}^{+}_{S}(x_{S}^{(i)} - \hat{\mu}_{S})$, let the matrices $\E, \hat{\E} \in \R^{n \times (p - k)}$ have rows $e^{(i)}$ and $\hat{e}^{(i)}$. Note that $e^{(i)}$ are the unique factors $\epsilon$ for the $i$th sample. Having conditioned on the $x^{(i)}_{S}$, we are exactly in the setting of fixed-X Ordinary Least Squares (OLS). In particular, we are running a multi-response regression of $\X_{-S}$ on $\X_{S}$ where the responses are independent. The $\hat{\E}$ are the residuals from OLS, while $\E$ are the residuals one would get from using the population regression coefficients. With this in mind, let $\H \in \R^{n \times n}$ be the orthogonal projection matrix onto the span of the columns of $\X_{S}$ and $\1 \in \R^n$, a $n$-dimensional vector of ones. By construction, $\H$ has rank $r \leq k + 1$. Therefore,
\begin{align*}
\hat{\bSigma}_{-S} - \hat{\bSigma}_{-S, S} \hat{\bSigma}_{S}^{+}\hat{\bSigma}_{S, -S} &= \frac{1}{n} \hat{\E}^\top\hat{\E}\\
&= \frac{1}{n} \X_{-S}^\top(\I_n - \H)\X_{-S}\\
&= \frac{1}{n} ((\X_{S} - \1\mu_{S}^\top)\W^\top + \1\mu_{-S}^{\top} + \E)^\top(\I_n - \H)((\X_{S} - \1\mu_{S}^\top)\W^\top + \1\mu_{-S}^\top + \E)\\
&= \frac{1}{n} \E^\top(\I_n - \H)\E.
\end{align*}

Let $\N \in \R^{n \times (p-k)}$ and $\M \in \R^{(n - r) \times (p-k)}$ be random matrices with i.i.d $N(0, 1)$ entries and $\boldsymbol{R} \in \R^{n \times n}$ be a rotation such that $\boldsymbol{R}^\top(\I_n - \H)\boldsymbol{R}$ is diagonal matrix with $1$ on the first $n-r$ diagonal entries and $0$ otherwise. Then we can compute
\begin{align*}
n \log \left(\frac{|\Diag(\E^\top(\I_n - \H)\E)|}{|\E^\top(\I_n - \H)\E|} \right) &= n \log \left( \frac{|\Diag(\D^{-1/2}\E^\top(\I_n - \H)\E \D^{-1/2})|}{|\D^{-1/2}\E^\top(\I_n - \H)\E \D^{-1/2}|} \right)\\
            &\overset{d}{=} n \log \left(\frac{|\Diag(\N^\top(\I_n - \H)\N|)}{|\N^\top(\I_n - \H)\N|} \right)\\
            &\overset{d}{=} n \log\left( \frac{|\Diag(\N^\top \boldsymbol{R}^\top(\I_n - \H)\boldsymbol{R} \N|)}{|\N^\top \boldsymbol{R}^\top(\I_n - \H)\boldsymbol{R}\N|} \right)\\
            &\overset{d}{=} n \log \left( \frac{|\Diag(\M^\top \M)|}{|\M^\top \M|} \right)\\
            &\overset{d}{=} n\log \left( \frac{|\Diag(W_{p- k}(I_{p-k}, n -r))|}{|W_{p- k}(I_{p-k}, n -r)|} \right) \\
            &\overset{d}{=} n \sum_{j=2}^{p-k} \log \left(1 + \frac{\tilde{\chi}^2_{j - 1}}{\chi^2_{n - r - j + 1}}\right).
\end{align*}

where $W_{p-k}(I_{p-k}, n - r)$ is a Wishart distribution and $\{ \chi^2_{\ell} \}, \{ \tilde{\chi}^{2}_{\ell}\}$ are mutually independent chi-squared random variables with degrees of freedom specified by their subscript. The final equality in distribution follows from the Bartlett decomposition of Wisharts \citep[Corollary 7.2.1]{anderson1958}. For clarity, we specify that the final distribution is a point mass at zero when $k = p-1$.

Recall that these equalities in distribution are conditional. Since $r \leq k + 1$, it is clear that the quantiles of $n \sum_{j=2}^{p-k} \log \left(1 + \frac{\tilde{\chi}^2_{j - 1}}{\chi^2_{n - k - j}}\right)$ are always at least as large as those of $n \sum_{j=2}^{p-k} \log \left(1 + \frac{\tilde{\chi}^2_{j - 1}}{\chi^2_{n - r - j + 1}}\right)$. Since the conditional quantiles of $T(S)$ are always bounded above by the quantiles of $n \sum_{j=2}^{p-k} \log \left(1 + \frac{\tilde{\chi}^2_{j - 1}}{\chi^2_{n - k - j}}\right)$, the marginal quantiles must be as well. 

Now we can easily establish the final claim. Recall that $Q_{n, p, k}(1-\alpha)$ is the $(1-\alpha)$-quantile of $n \sum_{j=2}^{p-k} \log \left(1 + \frac{\tilde{\chi}^2_{j - 1}}{\chi^2_{n - k - j}}\right)$.  Since $T_k \leq T(S)$, 
\begin{equation*}
    P(T_k  > Q_{n, p, k}(1-\alpha)) \leq P(T(S) > Q_{n, p, k}(1-\alpha)) \leq \alpha.
\end{equation*}

\subsection{Proof of \texorpdfstring{\Cref{lem:asymptotic_dist}}{Lemma}}

First, some preliminaries. Define the function $\vec(\cdot)$ that maps a symmetric matrix $M \in \R^{q \times q}$ to a vector $m \in \R^{q(q+1)/2}$ that contains all it's unique entries. If $m = \vec(M)$, then we will index $m$ by $m_{ij}$ for $i \leq j$ so that $m_{ij} = M_{ij}$. The inverse map is denoted $\vec^{-1}(\cdot)$. We will use the stochastic calculus notation of \citep[Chapter 2]{Vaart} and also freely refer to results from this chapter. 

Suppose we observe $x^{(1)}, \dots, x^{(n)}$ drawn from a distribution that satisfies \eqref{eq:subset_factor_model} for some size-$k$ subset $S$ where $k < p - 1$. In the case that $k = p -1$, $T(S)$ is a point mass at zero and the result is obvious. Recall $\E$ and $\H$ from the proof of \Cref{lem:finite_sample_validity} and consider the function $g(M) = \log(|\Diag(M)|) - \log(|M|)$. Computations in the proof of \Cref{lem:finite_sample_validity} imply that  $T(S) = n  g( \E^T(\I_n-\H)\E /n)$, and it thus suffices to show that  $n g( \E^T(\I_n-\H)\E / n) \rightsquigarrow \chi^2_{(p-k)(p-k-1)/2}$. We will show this result in steps. \newline 

\noindent \textbf{Step One - CLT for $\boldsymbol{\vec(\E^T\E/n)}$: } Let $a = \vec(\E^T\E/n)$ and $\tilde{a} \in \R^{p(p-1)/2}$ be only the entries of $a$ that correspond to the non-diagonal entries of $\E^T\E/n$. We index $\tilde{a}$ as $\tilde{a}_{ij}$, $i < j$. The standard multivariate CLT tells us that $\sqrt{n}\tilde{a} \rightsquigarrow N(0, \V)$ where $\V \in \R^{p(p-1)/2 \times p(p-1)/2}$ is such that $\V_{ij, st} = \D_{ii}\D_{jj}$ if $i=s$, $j=t$ and $0$ otherwise. Furthermore, by the weak law of large numbers $a_{ii} - \D_{ii} = o_{p}(1)$. \newline 

\noindent \textbf{Step Two - Controlling $\boldsymbol{\vec(\E^T\H\E/n)}$: } We want to control $\E^T \H \E/n$, the difference between $\E^T\E/n$ and $\E^T (\I_n - \H)\E/n$. Particularly we want to show that $(\E^T \H \E/n)_{ij} = o_p(n^{-1/2})$, $i \leq j$. To do so it suffices to show that $\| \E^T \H \E\|^2_F = O_p(1)$. Noting that $\H$ is the sum of the outer product of at most $k +1 $ unit vectors, it suffices to show that $\| \E^T v v^T \E\|^2_F = \| \E^Tv \|_2^4 = O_p(1)$ for an arbitrary sequence of unit vectors $v \in \R^n$. To show this, it suffices to show that $\sum_{i=1}^n e^{(i)}_{\ell} v_i = (\E^T v)_{\ell} = O_p(1)$ for all $1 \leq \ell \leq p$. But $\Var(\sum_{i=1}^n e^{(i)}_{\ell} v_i) = \D_{\ell \ell} \sum_{i=1}^n v_i^2 = \D_{\ell \ell}$ and $\E[\sum_{i=1}^n e^{(i)}_{\ell} v_i] = 0$, so this must be the case by Chebyshev. \newline 

\noindent \textbf{Step Three - CLT for $\boldsymbol{\vec(\E^T(\I_n - \H)\E/n)}$: } Let $a_H = \vec(\E^T(\I_n - \H)\E/n)$ and $\tilde{a}_H \in \R^{p(p-1)/2}$ be only the entries of $a$ that correspond to the non-diagonal entries of $\E^T(\I_n - \H)\E/n$. Step Two implies that $\sqrt{n}(\tilde{a} - \tilde{a}_H) = o_p(1)$, so $\sqrt{n}\tilde{a}_H  \rightsquigarrow N(0, \V)$ by Slutsky's lemma and Step One, where $\V$ is as in Step One. Furthermore, since $a_{ii} - (a_H)_{ii} = o_p(1)$, we also have $(a_H)_{ii} - D_{ii} = o_p(1)$ by Slutsky's. \newline 

\noindent \textbf{Step Four - Delta Method for $\boldsymbol{g(\E^T (\I_n - \H)\E/n)}$: } Define the function $f: \R^{q(q+1)/2} \rightarrow \R$ by $f(m) = g(\vec^{-1}(m))$. We will consider the domain of $f$ to be the open set of $m$ such that $\vec^{-1}(m)$ is a positive definite matrix. On this domain $f$ is $C^{\infty}$. We will take a second order Taylor expansion of the function $f$ around an $d \in \R^{(p-k)(p-k + 1)/2}$ such that $\vec^{-1}(d) = \D$. Note then $d_{ij} = 0$ for $i < j$. We can compute the first and second order partial derivatives of $f$ using \citep[Section 2]{Petersen}:
 
\begin{equation*}
\frac{\partial f(m)}{\partial m_{ij}} =
\begin{cases}
        -2(\vec^{-1}(m))^{-1}_{ij} & i \neq j \\
        m^{-1}_{ii} - (\vec^{-1}(m))^{-1}_{ii} & i = j,
\end{cases}
\end{equation*}

\begin{equation*}
\frac{\partial^2 f(m)}{\partial m_{ij} \partial m_{st}} =
\begin{cases}
        2( (\vec^{-1}(m))^{-1}_{is}(\vec^{-1}(m))_{jt}^{-1} + (\vec^{-1}(m))_{it}^{-1}(\vec^{-1}(m))^{-1})_{js}  &   i \neq j, s \neq t \\
        2(\vec^{-1}(m) )^{-1}_{is}(\vec^{-1}(m) )^{-1}_{js} &  i \neq j, s=t\\
        ((\vec^{-1}(m) )^{-1}_{is})^2 - m^{-2}_{ii} I_{i = s} & i=j, s=t.
        \\
\end{cases}
\end{equation*}
By repeated application of \citep[9.1.2]{Petersen}, it's clear that $f \geq 0$ on its domain. Further $f(d) = 0$. In line with this, all the first order partial derivatives of $f$ are zero at $d$. Many of the second order partial derivatives are also zero. Taylor's theorem  gives the following expansion of $f$ around $d$:
\begin{align*}
    f(m) &= \sum_{i < j} \D^{-1}_{ii} \D^{-1}_{jj}m_{ij}^2 \\
    &\qquad +  \sum_{i_1 \leq j_1, i_2 \leq j_2, i_3 \leq j_3} c_{i_1 j_1 i_2 j_2 i_3 j_3} \frac{\partial f^3(d + t(m - d))}{\partial m_{i_1 j_1} \partial m_{i_2 j_2} \partial m_{i_3 j_3}}  (m_{i_1 j_1} - d_{i_1 j_1}) (m_{i_2 j_2} - d_{i_2 j_2}) (m_{i_3 j_3} - d_{i_3 j_3}).
\end{align*}
for some $t \in [0, 1]$ and constants $c_{i_1 j_1 i_2 j_2 i_3 j_3}$ . It is then clear that 
\begin{equation*}
    n g(\E^T(\I_n - \H)\E/n) = n f(\vec(\E^T(\I - \H)\E/n)) = \sum_{i < j} \D^{-1}_{ii} \D^{-1}_{jj} (\sqrt{n}(a_H)_{ij})^2  + o_p(1),
\end{equation*}
and by continuous mapping theorem $n g(\E^T(\I_n - \H)\E/n) \rightsquigarrow \chi^2_{(p-k)(p-k-1)/2}$.

\subsection{Proof of \texorpdfstring{\Cref{lem:asymptotic_validity}}{Lemma}}

The case that $k = p -1$ is obvious, so we consider $k < p -1$ without loss of generality. 

First we will argue that the $Q_{n, p, k}(1-\alpha)$, the $(1-\alpha)$-quantile of \eqref{eq:null_dist}, converges to $Q_{p, k}(1 - \alpha)$, the $(1-\alpha)$-quantile of a $\chi^2_{(p-k)(p-k-1)/2}$ distribution as the number of samples $n$ tends to infinty. Consider observing $x^{(1)}, \dots, x^{(n)}$ samples drawn from a distribution that satisfies \eqref{eq:subset_factor_model} for a size-$k$ subset $S$, and suppose the principal variable distribution $F$ admits a density. Following the proof of \Cref{lem:finite_sample_validity}, we see that $T(S)$ will have distribution exactly \eqref{eq:null_dist} since $\hat{\bSigma}_{S}$ will be full-rank with probability one. But \Cref{lem:asymptotic_dist} tells us that $T(S) \rightsquigarrow \chi^2_{(p-k)(p-k-1)/2}$. This implies that $Q_{n, p, k}(1-\alpha)$ tends to $Q_{p, k}(1-\alpha)$ as $n \rightarrow 0$. 

Now, by Slutsky's and \Cref{lem:asymptotic_dist} we know that $T(S) - Q_{n, p, k}(1- \alpha) \rightsquigarrow \chi^2_{(p-k)(p-k-1)/2} - Q_{p, k}(1-\alpha)$. Since $0$ is a continuity point of $\chi^2_{(p-k)(p-k-1)/2} - Q_{p, k}(1-\alpha)$ we have that 
\begin{align*}
    \limsup_{n \rightarrow \infty} P(T_k - Q_{n, p, k}(1-\alpha)> 0) &\leq \limsup_{n \rightarrow \infty} P(T(S) - Q_{n, p, k}(1-\alpha)> 0)\\
    &= P(\chi^2_{(p-k)(p-k-1)/2} - Q_{p, k}(1-\alpha) > 0)\\
    &= \alpha.
\end{align*}

\subsection{Proof of \texorpdfstring{\Cref{thm:error_control}}{Theorem}}
\label{thm:error_control:proof}
First we show the asymptotic result. Consider a sample  $x^{(1)}, \dots, x^{(n)}$ from a distribution $P$ and suppose that $k^*$ is the smallest $k$ for which $P$ belongs to the subset factor model \eqref{eq:subset_factor_model}. Note that every distribution belongs to the $(k-1)$-dimensional factor model so $k^* \leq k + 1$. Recall that $\hat{k}$ is the smallest $k$ for which $T_k \leq Q_{n, p, k}(1-\alpha)$. Then, using \Cref{lem:asymptotic_validity}, 
\begin{equation*}
       \limsup_{n \rightarrow \infty} P(\hat{k} > k^*) \leq  \limsup_{n \rightarrow \infty}  P(T_{k^*} > Q_{n, p, k^*}(1-\alpha)) \leq \alpha
\end{equation*}

Now, for the finite sample result. Let $k^*$ be the smallest $k$ for which $P$ satisfies the model \eqref{eq:diag_pcss_model} for some size-$k$ subset $S$, and take $\hat{k}$ as before. It is now possible that $k^* = p$, but $P(\hat{k} > k) = 0$ if so. Thus, without loss of generality, we can suppose that $k^* < p$. Then, from \cref{lem:finite_sample_validity} 

\begin{equation*}
        P(\hat{k} > k^*) \leq  P(T_{k^*} > Q_{n, p, k^*}(1-\alpha)) \leq \alpha.
\end{equation*}

\subsection{Proof of \texorpdfstring{\Cref{lem:subset_recovery}}{Lemma}}

First we show that a distribution $P$ satisfies \eqref{eq:uncorrelated_subset_factor_model} with a size-$k$ set $S$ if and only if, for the population covariance $\bSigma$, $\bSigma_{-S} - \bSigma_{-S, S}\bSigma_{S}^{+}\bSigma_{S, -S}$ is diagonal and positive definite. First we show the forward direction. We know that $X_{-S} - \mu_{-S} = \W(X_{S} - \mu_S) + \epsilon$ where $\Cov(X_S, \epsilon) = \0$. Then \Cref{cor:uncorrelated_coeffs} tells us that $X_{-S} - \mu_{-S} = \bSigma_{-S, S}\bSigma_{S}^+(X_{S} - \mu_S) + \epsilon$. Via this, we can compute that $\bSigma_{S}= \bSigma_{-S, S}\bSigma_{S}^+\bSigma_{S, -S} + \D$ where $\D \succ 0$ so $\bSigma_{S} - \bSigma_{-S, S}\bSigma_{S}^+\bSigma_{S, -S}$ is diagonal and positive definite. For the reverse direction \Cref{cor:uncorrelated_coeffs} tells us that $X_{-S} - \mu_{-S} = \bSigma_{-S, S}\bSigma_{S}^{+}(X_{S} - \mu_{S}) + \epsilon$ where $\Cov(X_S, \epsilon) = 0$. A simple computation shows that $\epsilon$ must have diagonal, positive definite covariance and thus the distribution satisfies \eqref{eq:uncorrelated_subset_factor_model} with a size-$k$ set $S$ . 

Now, consider $n > p$  samples $x^{(1)}, \dots, x^{(n)} \in \R^p$ drawn from a distribution $P$ that satisfies \eqref{eq:uncorrelated_subset_factor_model} for some size-$k$ set $S$ and suppose that $P$ has population covariance $\bSigma$ such that $\bSigma_{S} \succ 0 $. Then it must be the case that $\bSigma \succ 0$. Thus, for every size-$k$ subset $U$, $|\bSigma_{-U} - \bSigma_{-U, U}\bSigma_{U}^{+}\bSigma_{U, -U}| > 0$. Since $\hat{\bSigma}$ converges almost surely to $\bSigma$ as $n\rightarrow \infty$, we have that, for every size-$k$ subset $U$,
\begin{equation}
\label{eq:sample_test_stat}
    \log\left( \frac{|\Diag(\hat{\bSigma}_{-U} - \hat{\bSigma}_{-U, U}\hat{\bSigma}_{U}^{+}\hat{\bSigma}_{U, -U})| }{|\hat{\bSigma}_{-U} - \hat{\bSigma}_{-U, U}\hat{\bSigma}_{U}^{+}\hat{\bSigma}_{U, -U}|} \right)
\end{equation}
converges almost surely to 
\begin{equation}
\label{eq:pop_test_stat}
    \log\left( \frac{|\Diag(\bSigma_{-U} - \bSigma_{-U, U}\bSigma_{U}^{-1}\bSigma_{U, -U})| }{|\bSigma_{-U} - \bSigma_{-U, U}\bSigma_{U}^{-1}\bSigma_{U, -U}|} \right)
\end{equation}
As a consequence of \citep[9.1.2]{Petersen}, \eqref{eq:pop_test_stat} is zero if and only if $\bSigma_{-U} - \bSigma_{-U, U}\bSigma_{U}^{-1}\bSigma_{U, -U}$ is diagonal and otherwise it is strictly greater than zero. Thus, By the discussion above, it is zero when $P$ satisfies \eqref{eq:uncorrelated_subset_factor_model} with $U$, and otherwise it is strictly greater than zero. Since $\hat{S}$ minimizes $T(U)$ it must also minimize \eqref{eq:sample_test_stat}. And since \eqref{eq:sample_test_stat} converges almost surely to \eqref{eq:pop_test_stat}, eventually $\hat{S}$ will be one of the $U$ for which \eqref{eq:pop_test_stat} is zero, meaning $P$ will eventually satisfy \eqref{eq:uncorrelated_subset_factor_model} with $\hat{S}$. 

\subsection{Proof of \texorpdfstring{\Cref{lem:pcss_finite_sample_validity}}{Lemma} }
We adopt the notation from the proof of \cref{lem:finite_sample_validity}. Similarly we perform our analysis conditional on the $x^{(i)}$. Recall that 

\begin{align*}
\hat{\bSigma}_{-S} - \hat{\bSigma}_{-S, S} \hat{\bSigma}_{S}^{+}\hat{\bSigma}_{S, -S} &= \frac{1}{n} \E^\top(\I_n - \H)\E.
\end{align*}
Since we have assumed that $F$ has a density, we know that $\rank(\H) = k + 1$. We can then similarly compute 

\begin{align*}
\widetilde{T}(S) &= n \log\left( \frac{(\tr(\boldsymbol{E}^T(\I_n - \H)\boldsymbol{E})/(p-k))^{p-k}}{|\boldsymbol{E}^T(\I_n - \H)\boldsymbol{E}|} \right)\\
            &= n \log\left(\frac{(\tr(\sigma^{-1}\boldsymbol{E}^T(\I_n - \H)\boldsymbol{E}\sigma^{-1})/(p-k))^{p-k}}{|\sigma^{-1}\boldsymbol{E}^T(\I_n - \H)\boldsymbol{E}\sigma^{-1}|} \right)\\
            &\overset{d}{=} n \log \left(\frac{(\tr(W_{p- k}(I_{p-k}, n - k - 1))/(p-k))^{p-k}}{|W_{p- k}(I_{p-k}, n - k -1 )|} \right)\\
            & \overset{d}{=} n\log \left(\left(\frac{\tilde{\chi}^2_{(p-k)(p-k-1)/2} + \sum_{j=1}^{p-k} \chi^2_{n - k - j }}{p-k}\right)^{p-k}\bigg/\left(\prod_{j=1}^{p-k} \chi^2_{n - k - j}\right)\right)
\end{align*}
where $W_{p-k}(I_{p-k}, n - r)$ is a Wishart distribution, $\{ \chi^2_{\ell} \}, \{ \tilde{\chi}^{2}_{\ell}\}$ are mutually independent chi-squared random variables with degrees of freedom specified by their subscript, and again the final equality in distribution follows from the Bartlett decomposition of Wisharts \citep[Corollary 7.2.1]{anderson1958}. Although these equalities in distribution are conditional, since the final distribution does not depend on the $x^{(i)}$, it also holds marginally. Note that the distribution is a point mass at zero when $k = p -1$.

Now we can easily establish the final claim. Recall that $\widetilde{Q}_{n, p, k}(1-\alpha)$ is the $(1-\alpha)$-quantile of 
\begin{equation*}
    n\log \left(\left(\frac{\tilde{\chi}^2_{(p-k)(p-k-1)/2} + \sum_{j=1}^{p-k} \chi^2_{n - k - j }}{p-k}\right)^{p-k}\bigg/\left(\prod_{j=1}^{p-k} \chi^2_{n - k - j}\right)\right)
\end{equation*}
Since $\widetilde{T}_k \leq \widetilde{T}(S)$, 
\begin{equation*}
    P(\widetilde{T}_k  > \widetilde{Q}_{n, p, k}(1-\alpha)) \leq P(\widetilde{T}(S) > \widetilde{Q}_{n, p, k}(1-\alpha)) \leq \alpha.
\end{equation*}

\subsection{Proof of \texorpdfstring{\Cref{prop:pcss_error_control}}{Proposition}}

Let $k^*$ be the smallest $k$ for which $P$ satisfies the model \eqref{eq:diag_pcss_model} for some size-$k$ subset $S$, and recall that $\hat{k}$ is the smallest $k$ for which $\widetilde{T}_k \leq \widetilde{Q}_{n, p, k}(1-\alpha)$. It is possible that $k^* = p$, but then $P(\hat{k} > k) = 0$ if so. Thus, without loss of generality, we can suppose that $k^* < p$. Then, from \cref{lem:pcss_finite_sample_validity},
\begin{equation*}
        P(\hat{k} > k^*) \leq  P(\widetilde{T}_{k^*} > \widetilde{Q}_{n, p, k^*}(1-\alpha)) \leq \alpha.
\end{equation*}

\subsection{Proof of \texorpdfstring{\Cref{lem:pcss_subset_recovery}}{Lemma}}
First we claim that a distribution $P$ satisfies \eqref{eq:uncorrelated_pcss_model} with a size-$k$ set $S$ if and only if, for the population covariance $\bSigma$, $\bSigma_{-S} - \bSigma_{-S, S}\bSigma_{S}^{+}\bSigma_{S, -S}$ is isotropic, i.e., equal to $\eta \I_{p-k}$ for some $\eta > 0$. The argument is identical to that in the proof of \Cref{lem:subset_recovery}.

Now, consider $n > p$  samples $x^{(1)}, \dots, x^{(n)} \in \R^p$ drawn from a distribution $P$ that satisfies \eqref{eq:uncorrelated_pcss_model} for some size-$k$ subset $S$. Further suppose that, at the population level no set of $k$ variables perfectly linearly reconstruct the rest. This implies that $\tr(\bSigma_{-U} - \bSigma_{-U, U}\bSigma_{U}^{+}\bSigma_{U, -U})$ is never zero for any size-$k$ subset $U$. Then, since $\hat{\bSigma}$ converges almost surely to $\bSigma$ as $n\rightarrow \infty$, we can guarantee that 
\begin{equation}
    \label{eq:pcss_sample_test_stat}
    \log\left( \frac{(\tr(\hat{\bSigma}_{-U} - \hat{\bSigma}_{-U, U}\hat{\bSigma}_{U}^{+}\hat{\bSigma}_{U, -U})/(p-k))^{(p-k)} }{|\hat{\bSigma}_{-U} - \hat{\bSigma}_{-U, U}\hat{\bSigma}_{U}^{+}\hat{\bSigma}_{U, -U}|} \right)
\end{equation}
converges almost surely to 
\begin{equation}
    \label{eq:pcss_pop_test_stat}
    \log\left( \frac{ (\tr(\bSigma_{-U} - \bSigma_{-U, U}\bSigma_{U}^{+}\bSigma_{U, -U})/(p-k))^{(p-k)} }{|\bSigma_{-U} - \bSigma_{-U, U}\bSigma_{U}^{+}\bSigma_{U, -U}|} \right) 
\end{equation}
We claim that \eqref{eq:pcss_pop_test_stat} is zero if and only if $\bSigma_{-U} - \bSigma_{-U, U}\bSigma_{U}^{+}\bSigma_{U, -U}$ is of the form $\eta \I_{p-k}$. To see this note that \eqref{eq:pcss_pop_test_stat} is (resp. strictly) larger than zero if and only if 
\begin{equation}
    \label{eq:am_gm}
    \frac{ \tr(\bSigma_{-U} - \bSigma_{-U, U}\bSigma_{U}^{+}\bSigma_{U, -U})/(p-k) }{|\bSigma_{-U} - \bSigma_{-U, U}\bSigma_{U}^{+}\bSigma_{U, -U}|^{1/(p-k)}}
\end{equation}
is (resp. strictly) larger than one. But \eqref{eq:am_gm} is always larger than or equal to one because it is the 
ratio of the arithmetic mean to the geometric mean of the eigenvalues of $\bSigma_{-U} - \bSigma_{-U, U}\bSigma_{U}^{+}\bSigma_{U, -U}$. Equality is only attained when all the eigenvalues are equal, i.e., when  $\bSigma_{-U} - \bSigma_{-U, U}\bSigma_{U}^{+}\bSigma_{U, -U}$ is of the form $\eta \I_{p-k}$, which completes the claim. 

By the discussion above, this quantity will be zero when $P$ satisfies \eqref{eq:uncorrelated_pcss_model} with $U$, and otherwise it will be strictly greater than zero. Since  $\widetilde{S}$ minimizes $\widetilde{T}(U)$, eventually it will be one of the $U$ for which this quantity is zero, and thus eventually $P$ will satisfy \eqref{eq:uncorrelated_subset_factor_model} with $\widetilde{S}$. 

\subsection{Correctness of Subset Search Algorithms}
\label{sec:alg_correctness_appdx}
We provide computations which justify the correctness algorithms in \Cref{sec:algorithms} as well as \Cref{sec:other_algs_appdx}. \newline 

\noindent \textbf{Minimizing $T(\cdot)$:} \citep[9.1.2]{Petersen} and \Cref{lem:residual_cov_update} jointly tell us that $\log|\bSigma_{U+i}| = \log |\hat{\bSigma}_U| + \log((\hat{\bSigma}_{R(X, X_U)})_{ii})$.  Using \Cref{lem:residual_cov_update}, we can compute 
\begin{align*}
&\log|\hat{\bSigma}_{U + i}| + \tr(\log(\Diag(\hat{\bSigma}_{R(X_{-(U + i)},X_{U + i})})))  \\
& \qquad =\log|\hat{\bSigma}_{U}| + \log((\hat{\bSigma}_{R(X, X_U)})_{ii}) + \tr(\log(\Diag((\hat{\bSigma}_{R(X,X_{U + i})})_{-(U + i)}))) \\
& \qquad =\log|\hat{\bSigma}_{U}| + \log((\hat{\bSigma}_{R(X, X_U)})_{ii}) \\
&\qquad \qquad + \tr(\log(\Diag((\hat{\bSigma}_{R(X, X_{U})} -(\hat{\bSigma}_{R(X, X_{U})})_{\bullet i}(\hat{\bSigma}_{R(X, X_{U})})_{i \bullet}/(\hat{\bSigma}_{R(X, X_{U})})_{ii} \cdot I_{(\hat{\bSigma}_{R(X, X_{U})})_{ii} > 0} )_{-(U+i)}))) \\
&\qquad = \log|\hat{\bSigma}_{U}| + \log((\hat{\bSigma}_{R(X, X_U)})_{ii}) + \sum_{j \not \in U + i} \log((\hat{\bSigma}_{R(X, X_U)})_{jj} - (\hat{\bSigma}_{R(X, X_U)})^2_{ij}/(\hat{\bSigma}_{R(X, X_U)})_{ii} \cdot I_{(\hat{\bSigma}_{R(X, X_{U})})_{ii} > 0})
\end{align*}
which is sufficient to imply that the $i \not \in U$ we select minimizes the objective over subsets $U + i$. \newline 

\noindent \textbf{Minimizing $\widetilde{T}(\cdot)$:} Following similar reasoning to above, we can compute
\begin{align*}
&\log|\bSigma_{U + i}| + (p-k)\log(\tr(\bSigma_{R(X_{-(U + i)}, X_{U + i})} )/(p-k) )  \\
& \qquad =\log|\bSigma_{U}| + \log((\bSigma_{R(X, X_U)})_{ii}) + (p-k)\log(\tr(\bSigma_{R(X, X_{U + i})} )) - (p-k)\log(p-k) \\
& \qquad =\log|\bSigma_{U}| + \log((\bSigma_{R(X, X_U)})_{ii}) \\
& \qquad \qquad + (p-k)\log(\tr(\bSigma_{R(X, X_U)} -  (\bSigma_{R(X, X_U)})_{\bullet i}(\bSigma_{R(X, X_U)})_{i \bullet}/(\bSigma_{R(X, X_U)})_{ii} \cdot I_{(\bSigma_{R(X, X_U)})_{ii} > 0} )) - (p-k)\log(p-k) \\
&\qquad  = \log|\bSigma_{U}| + \log((\bSigma_{R(X, X_U)})_{ii}) + (p-k)\log(\tr(\bSigma_{R(X, X_U)}) -  \|(\bSigma_{R(X, X_U)})_{\bullet i}\|_2^2/(\bSigma_{R(X, X_U)})_{ii} \cdot I_{(\bSigma_{R(X, X_U)})_{ii} > 0}  ) \\
&\qquad \qquad- (p-k)\log(p-k)
\end{align*}
which is sufficient to imply that the $i \not \in U$ we select minimizes the objective over subsets $U + i$. \newline

\noindent \textbf{McCabe's First Criterion:} If there exists a $k$ by $k$ principal sub-matrix of $\bSigma$ that is full rank then minimizing $|\bSigma_{-S} - \bSigma_{-S, S}\bSigma_{S}^{+} \bSigma_{S, -S}|$ is equivalent to maximizing $|\bSigma_{S}|$. Consider having a currently selected subset $U$. We know from above that $ |\bSigma_{U + i}| = |\bSigma_{U}|\cdot (\bSigma_{R(X, U)})_{ii}$. This is sufficient to imply that the $i \not \in U$ we select minimizes the objective over subsets $U + i$.\newline 

\noindent \textbf{McCabe's Second Criterion:} Consider having currently selected a subset $U$. For $i \not \in U$ let $\beta = (\bSigma_{R(X, X_{U})})_{\bullet i}$. Then from \Cref{lem:residual_cov_update}:
\begin{align*}
     \tr(\bSigma_{R(X, X_{U + i})}) &= \tr(\bSigma_{R(X, X_{U})}  - \frac{\beta\beta^\top}{\beta_i} \cdot I_{\beta_i > 0})\\
     &= \tr(\bSigma_{R(X, X_{U})})  - \frac{||\beta||_2^2}{\beta_i} \cdot I_{\beta_i > 0}
\end{align*}
which is sufficient to imply that the $i \not \in U$ we select minimizes the objective over subsets $U + i$.\newline

\noindent \textbf{McCabe's Third Criterion:} Consider having currently selected a subset $U$. For $i \not \in U$ let $\beta = (\bSigma_{R(X, X_{U})})_{\bullet i}$. Then from \Cref{lem:residual_cov_update}:
\begin{align*}
 \|\bSigma_{R(X, X_{U + i})}\|^2_{F} &= \|\bSigma_{R(X, X_{U})}  - \frac{\beta\beta^\top}{\beta_i} \cdot I_{\beta_i > 0}\|^2_{F}\\
                                    &= \tr((\bSigma_{R(X, X_{U})}  - \frac{\beta\beta^\top}{\beta_i}  \cdot I_{\beta_i > 0})^\top(\bSigma_{R(X, X_{U})}  - \frac{\beta\beta^\top}{\beta_i} \cdot I_{\beta_i > 0}))\\
                                    &=  \|\bSigma_{R(X, X_{U})}\|^2_{F} + \left[ \frac{\|\beta\|_2^4}{\beta^2_{i}} - \frac{2\beta^\top\bSigma_{R(X, X_{U})}\beta}{\beta_i} \right]\cdot I_{\beta_i > 0}
\end{align*}
which is sufficient to imply that the $i \not \in U$ we select minimizes the objective over subsets $U + i$. \newline 

\noindent \textbf{McCabe's Fourth Criterion: } McCabe's fourth criterion suggests finding a subset $S$ that maximizes $\tr(\bSigma_{S}^+\bSigma_{S, -S}\bSigma_{-S}^+\bSigma_{-S, S})$.

We briefly justify that our generalized objective has the same interpretation. Fix a subset $S$. It is well known that $\tr(\bSigma_{S}^+\bSigma_{S, -S}\bSigma_{-S}^+\bSigma_{-S, S})$ is the sum of the squared canonical correlations when $\bSigma \succ \0$ \citep{McCabe}. When $\bSigma$ is singular, suppose that the ranks of $\bSigma_{S}$ and $\bSigma_{-S}$ are  $r_1$ and $r_2$ respectively. Then let $Q_1$ and $Q_2$ be rotations so that the last $k - r_1$ and $p- k - r_2$ entries of $Q_1 X_{S}$ and $Q_2 X_{-S}$ are zero. Let $Y_1 \in \R^{r_1}$ and $Y_2 \in \R^{r_2}$ be the first $r_1$ and $r_2$ entries of $X_S$ and $X_{-S}$. By the definition of canonical correlations, the sum of the squared canonical correlations between $X_S$ and $X_{-S}$ and $Y_1$ and $Y_2$ are the same, and it is easy to verify that $\tr(\bSigma_{Y_1}^{-1}\bSigma_{Y_1, Y_2}\bSigma_{Y_2}^{-1}\bSigma_{Y_2, Y_1})$ is equal to $\tr(\bSigma_{S}^+\bSigma_{S, -S}\bSigma_{-S}^+\bSigma_{-S, S})$. 

Now, we compute that
\begin{equation*}
    \tr(\bSigma_{S}^+\bSigma_{S, -S}\bSigma_{-S}^{+}\bSigma_{-S, S}) = \tr(\bSigma_S^{+}(\bSigma_S - \bSigma_{R(X_{S}, X_{-S})}) ) = \rank(\bSigma_{S}) - \tr(\bSigma_S^+ \bSigma_{R(X_{S}, X_{-S})})
\end{equation*}
Consider having currently selected a subset $U$. Take a fixed $i \not \in U$ and let $V = U + i$. Take $j, h$ to be as described in the presentation of the modified algorithm. $\bSigma_{V}$ will have rank larger than $\bSigma_{U}$ by one if feature $i$ cannot be exactly reconstructed by the features in $U$ and the same rank as $\bSigma_{U}$ otherwise. So, $\rank(\bSigma_{V}) = \rank(\bSigma_{U}) + I_{(\bSigma_{R(X_{-U}, X_{U})})_{hh} > 0 }$. Taking $\tilde{\beta} = \bSigma_{\bullet i} -  \bSigma_{\bullet, -V}\bSigma_{-V}^{+}\bSigma_{-V, i}$, we know from \Cref{lem:residual_cov_update} that
\[\bSigma_{R(X, X_{-V})} =  \bSigma_{R(X, X_{-U)})} + \frac{\tilde{\beta} \tilde{\beta}^\top}{\tilde{\beta}_i} \cdot I_{\tilde{\beta}_i > 0} \]
From this, defining $\beta = \tilde{\beta}_{V}$, and recalling that $\beta_j = \tilde{\beta}_i$, we have 
\[\bSigma_{R(X_{V}, X_{-V})} =  \bSigma_{R(X_{V}, X_{-U)}} + \frac{\beta \beta^\top}{\beta_j} \cdot I_{\beta_j > 0} \]
Since $i \in -U$, the $j$th row and column of $\bSigma_{R(X_{V}, X_{-U)})}$ are zeros. Putting everything together 
\begin{align*}
    &\rank(\bSigma_{V}) - \tr(\bSigma_{V}^+ \bSigma_{R(X_{V}, X_{-V})})\\
    &=\rank(\bSigma_{U}) + I_{(\bSigma_{R(X_{-U}, X_{U} )})_{hh} > 0} - \tr((\bSigma^{+}_{V})_{-j, -j}\bSigma_{R(X_{U}, X_{-U)})}) - \frac{\beta^\top \bSigma^+_{V} \beta}{\beta_j} \cdot I_{\beta_j > 0}
\end{align*}
which is sufficient to imply that the $i \not \in U$ we select minimizes the objective over subsets $U + i$. 

Now we consider the case where $\bSigma \succ \0$. First, $\bSigma_{S}^{+} = \bSigma_{S}^{-1}$ so our objective simplifies to minimizing $\tr(\bSigma^{-1}_S \bSigma_{R(X_{S}, X_{-S})})$. In this setting, \citep{Khan} tells us that $\bSigma^{-1}_{-V} = (\bSigma_{-U}^{-1})_{-h, -h} - (\bSigma_{-U})_{-h, h}(\bSigma_{-U})_{h, -h}/(\bSigma_{-U})_{hh}$.  Consider $\beta$ from before. Since $\bSigma \succ \0$, $\beta_j > 0$, so $\bSigma_{R(X_{V}, X_{-V})} =  \bSigma_{R(X_{V}, X_{-U)})}  + \beta\beta^\top/\beta_j$. Also from \citep{Khan}, letting $\alpha = \bSigma_{U}^{-1}\bSigma_{Ui}$, $\delta = \bSigma_{ii} - \bSigma_{iU}\alpha$,  we know that $(\bSigma^{-1}_{V})_{-j, -j} = \bSigma^{-1}_{U} - \alpha\alpha^\top/\delta$, $(\bSigma_V^{-1})_{-j, j} = -\alpha/\delta$ and $(\bSigma_V^{-1})_{j, j} = 1/\delta$. Again, recalling that the $j$th row and column of $\bSigma_{R(X_{V}, X_{-U)})}$ are zeros because $i \in  -U$ we can compute 
\begin{align*}
\tr(\bSigma^{-1}_{V}\bSigma_{R(X_{V}, X_{-V})}) &= \tr( (\bSigma^{-1}_{U} - \alpha\alpha^\top/\delta)(\bSigma_{R(X_{U}, X_{-U})} + \beta_{-j}\beta_{-j}^\top/\beta_{j} )  -\alpha\beta_{-j}^\top/\delta) - \alpha^\top\beta_{-j}/\delta + \beta_j/\delta ) \\ 
&= \tr(\bSigma^{-1}_{U} \bSigma_{R(X_{U}, X_{-U})}) +  \frac{\beta_{-j}^{T}\bSigma^{-1}_{U}\beta_{-j}}{\beta_{j}} - \frac{(\alpha^{T}\beta_{-j})^2}{\delta \beta_{j}} - \frac{\alpha^\top\bSigma_{R(X_U, X_{-U})}\alpha + 2\alpha^\top\beta_{-j} - \beta_{j}}{\delta } 
\end{align*}
which is sufficient to imply that the $i \not \in U$ we select minimizes the objective over subsets $U + i$. 

% Completely changed for second revision 

\subsection{Proof of \texorpdfstring{\Cref{prop:coefficient_estimation_error}}{Proposition}}

Fixing a size-$k$ subset $U$ and some variable $i \not \in U$ we show finite sample $L^2$ norm concentration for the coefficients $\hat{\bSigma}_{iU}\hat{\bSigma}_{U}^+$ from the regression of $X_i$ on $X_U$. The rest of the result then follows trivially from a union bound. We will use the same notation as in \Cref{sec:high_dim_consistency_appdx} and refer to results from there as well. The one difference is that the $c_i$ represent positive constants that are independent of $n$, $p$, and also $k$.

First we re-configure 

\begin{align*}
    \|\hat{\bSigma}_{U}^+\hat{\bSigma}_{Ui} - \bSigma_{U}^+\bSigma_{Ui} \|_2 &= \|\hat{\bSigma}_{U}^+\hat{\bSigma}_{Ui} - \bSigma_{U}^+\hat{\bSigma}_{Ui} + \bSigma_{U}^+\hat{\bSigma}_{Ui} - \bSigma_{U}^+ \bSigma_{Ui} \|_2\\
    &\leq \|\hat{\bSigma}_{U}^+ - \bSigma_{U}^+\|_2 \| \hat{\bSigma}_{Ui}\|_2 + \|\hat{\bSigma}_{Ui} -  \bSigma_{Ui} \|_2 \|\bSigma_{U}^+\|_{2}\\
    &\leq \|\hat{\bSigma}_{Ui} - \bSigma_{Ui}\|_2 \|\hat{\bSigma}_{U}^+ - \bSigma_{U}^+\|_2 + \|\bSigma_{Ui}\|_2 \|\hat{\bSigma}_{U}^+ - \bSigma_{U}^+\|_2 + \|\hat{\bSigma}_{Ui} -  \bSigma_{Ui} \|_2 \|\bSigma_{U}^+\|_{2}\\
    &\leq  \|\hat{\bSigma}_{Ui} - \bSigma_{Ui}\|_2 \|\hat{\bSigma}_{U}^+ - \bSigma_{U}^+\|_2 +  B \|\hat{\bSigma}_{U}^+ - \bSigma_{U}^+\|_2 +  c_1\|\hat{\bSigma}_{Ui} -  \bSigma_{Ui} \|_2  
\end{align*}
where in the last line we have used \Cref{ass:invertibility_coeff} to bound $\|\bSigma_{U}^+\|_2 =  \|\bSigma_{U}^{-1}\|_2$ and bounded $\|\bSigma_{Ui}\|_2$ by $B = \max_{j \in -U}\|\bSigma_{Uj}\|_2$. Fixing $t \in (0, 1)$, it suffices to get tail bounds for two things:\newline 

\textbf{Tail bound for $\|\hat{\bSigma}_{iU} -  \bSigma_{iU}\|_2$ : } Because of \Cref{ass:sub_gaussian_coeff} we can apply our sub-exponential concentration result \eqref{eq:sub_exp_concentration} and get a tail bound for $\|\hat{\bSigma}_{iU} -  \bSigma_{iU}\|_2$. 
\begin{align*}
    P(\|\hat{\bSigma}_{iU} -  \bSigma_{iU}\|_2 > t ) &= P(\sum_{j \in U} (\hat{\bSigma}_{ij} - \bSigma_{ij})^2 > t^2)  \\
    &\leq \sum_{j \in U} P(\hat{ |\bSigma}_{ij} - \bSigma_{ij}| > t/k^{1/2}) & \text{(union bound)}\\
    &\leq k \exp(-c_2 \min(t/k^{1/2}, t^2/k) n ) & \text{(sub-exponential concentration)}\\
    &\leq k \exp\left(\frac{-c_2 t^2 n}{k} \right) & \text{($ t < 1$)}
\end{align*}

\textbf{Tail bound for $\|\hat{\bSigma}_U^+ -  \bSigma_U^+ \|_2$ : } To get concentration for $\|\hat{\bSigma}_U^+ -  \bSigma_U^+ \|_2$ we note that, on the event that $\hat{\bSigma}_U^+$ is invertible, we can use \Cref{ass:invertibility} to see that 
\begin{align*}
    \|\hat{\bSigma}_U^+ -  \bSigma_U^+ \|_2 &= \|\hat{\bSigma}_U^{-1} -  \bSigma_U^{-1} \|_2 \\
                                            &= \|\bSigma_U^{-1/2}(\bSigma_U^{1/2}\hat{\bSigma}_U^{-1}\bSigma_U^{1/2} -  \I_k)\bSigma_U^{-1/2} \|\\
                                            &\leq \|\bSigma_U^{-1} \|_2 \| \bSigma_{U}^{1/2}\hat{\bSigma}_U^{-1}\bSigma_{U}^{1/2} - \I_k\|_2 \\
                                            &\leq c_3 \| \bSigma_{U}^{1/2}\hat{\bSigma}_U^{-1}\bSigma_{U}^{1/2} - \I_k\|_2 \\
                                            &= c_3 \max_{j \in [k]} | s_j^{-2}( n^{-1/2}\X_{\bullet U} \bSigma_{U}^{-1/2})  - 1 |
\end{align*}
We recall the concentration result \cite[Theorem 4.6.1]{Vershynin}. Let $M$ be the sub-Gaussian norm of $X_{U}$. Then for any $x \geq 0$
\begin{equation*}
    P( \max_{j \in [k]} |s_i(\X_{\bullet U}\bSigma_U^{-1/2}) - n^{1/2}| > cM^2(k^{1/2} + x) ) \leq 2\exp(-x^2),
\end{equation*}
which means more generally for any $x$ that 
\begin{equation*}
    P( \max_{j \in [k]} |s_i(\X_{\bullet U}\bSigma_U^{-1/2}) - n^{1/2}| > x ) \leq 2\exp(- (x/(cM^2) - k^{1/2})_+^2  ),
\end{equation*}

Fix some $c_4 > 0$. Then we can apply this result to see that 
\begin{align*}
 &P(\|\hat{\bSigma}_U^+ -  \bSigma_U^+ \|_2 > t)  \\
 &\leq P\left(\|\hat{\bSigma}_U^+ -  \bSigma_U^+ \|_2 > t, \max_{i \in [k]} | s_j(n^{-1/2}\X_{\bullet U}\bSigma_U^{-1/2}) - 1| < c_4 \right)\\
 &\qquad + P\left(\max_{j \in [k]} | s_j(n^{-1/2}\X_{\bullet U}\bSigma_U^{-1/2}) - 1| \geq c_4\right)  \\
 &\leq P\left(\max_{i \in [k]} | s^{-2}_j(n^{-1/2}\X_{\bullet U}\bSigma_U^{-1/2}) - 1| > c_5 t, \max_{i \in [k]}| s_j(n^{-1/2}\X_{\bullet U}\bSigma_U^{-1/2}) - 1| < c_4 \right)\\
 &\qquad + P\left(\max_{j \in [k]} | s_j(n^{-1/2}\X_{\bullet U}\bSigma_U^{-1/2}) - 1| \geq c_4\right)  \\
 & & \hspace{-380pt} \text{($\hat{\bSigma}_{U}$ is invertible on the event)}\\
 &\leq P\left(\max_{j \in [k]} | s_i(n^{-1/2}\X_{\bullet U}\bSigma_U^{-1/2}) - 1| > c_6 t, \max_{j \in [k]}| s_i(n^{-1/2}\X_{\bullet U}\bSigma_U^{-1/2}) - 1| < c_4 \right)\\
 &\qquad + P\left(\max_{j \in [k]} | s_i(n^{-1/2}\X_{\bullet U}\bSigma_U^{-1/2}) - 1| \geq c_4\right)  \\
 & & \hspace{-300pt} \text{(\Cref{lem:inverse_local_lip} with $a=c_4$)}\\
 &\leq P\left(\max_{j \in [k]} | s_i(\X_{\bullet U}\bSigma_U^{-1/2}) - n^{1/2}| >  c_6 n^{1/2}t\right) + P\left(\max_{j \in [k]} | s_i(\X_{\bullet U}\bSigma_U^{-1/2}) - n^{1/2}| \geq c_4 n^{1/2}  \right) \\
 &\leq  2 \exp( - (c_6 n^{1/2} t/(cM^2) - k^{1/2} )_+^2 ) + 2 \exp( -(c_4 n^{1/2}/(cM^2) - k^{1/2} )_+^2 ) \\
 & & \hspace{-150pt}\text{(above singular value concentration) }\\
 &\leq 4 \exp( -(c_7 n^{1/2} t/M^2 - k^{1/2})_+^2 ) 
  & & \hspace{-150pt}\text{($t < 1$) }\\
\end{align*}

Putting everything together, we can use a union bound to see that

\begin{align*}
&P(\|\hat{\bSigma}_{iU}\hat{\bSigma}_{U}^+ - \bSigma_{iU} \bSigma_{U}^+\|_2  > t) \\
     &\leq P( \|\hat{\bSigma}_{iU} - \bSigma_{iU}\|_2 > \frac{t^{1/2}}{\sqrt{3}}) +  P( \|\hat{\bSigma}_{U}^+ - \bSigma_{U}^+\|_2 > \frac{t^{1/2}}{\sqrt{3}}) \\
     &\qquad + P( \|\hat{\bSigma}_{U}^+ - \bSigma_{U}^+\|_2 > \frac{t}{3 B})   +  P(\|\hat{\bSigma}_{iU} -  \bSigma_{iU} \|_2  > \frac{t}{3 c_1})\\
     &\leq  2P( \|\hat{\bSigma}_{U}^+ - \bSigma_{U}^+\|_2 > c_9  t/\max(B, 1) )   +  2P(\|\hat{\bSigma}_{iU} -  \bSigma_{iU} \|_2  > c_{10}t  )\\
     &=  8 \exp(-(c_{11}n^{1/2}t/(\max(BM^2, M^2) ) - k^{1/2})_+^2   ) + 2k \exp\left(-(c_{12} n^{1/2}t/k^{1/2})^2 \right) \\
     &=  8 \exp(-k(c_{11}n^{1/2}t/(k^{1/2} \max(BM^2, M^2) ) - 1)_+^2   ) + 2k \exp\left(-(c_{12} n^{1/2}t/k^{1/2} - 1)_+^2 \right) \\
     &\leq c_{13}k \exp(-(c_{14}n^{1/2}t/(k^{1/2} \max(BM^2, M^2, 1) ) - 1)_+^2   )\\
     &= c_{13}k \exp\left( -\left( \frac{c_{14}t n^{1/2}}{k^{1/2} \max(BM^2, M^2, 1)} - 1 \right)_+^2  \right )
\end{align*}

Since the final bound has no dependence on $i$, we have by a union bound over $i \not \in U$ that 

\begin{equation*}
    P(\|\hat{\bSigma}_{\bullet U}\hat{\bSigma}_{U}^+ - \bSigma_{\bullet U} \bSigma_{U}^+\|_{2 \rightarrow \infty}  > t) \leq c_{13}pk \exp\left( -\left( \frac{c_{14}t n^{1/2}}{k^{1/2} \max(BM^2, M^2, 1)} - 1 \right)_+^2  \right )
\end{equation*}

Since one can easily show using  \Cref{ass:sub_gaussian_coeff} that $B \leq c_{14} \sqrt{k}$ and $M \leq c_{15}\sqrt{k}$, we also have the bound

\begin{equation*}
    P(\|\hat{\bSigma}_{\bullet U}\hat{\bSigma}_{U}^+ - \bSigma_{\bullet U} \bSigma_{U}^+\|_{2 \rightarrow \infty}  > t) \leq c_{13}pk \exp\left( -\left( \frac{c_{16}t n^{1/2}}{k^2} - 1 \right)_+^2  \right ),
\end{equation*}
where the right hand side has no dependence on the subset $U$. By union bounding over all subsets $U$, we get

\begin{equation*}
    P( \max_{U \subseteq [p] : |U| = k} \|\hat{\bSigma}_{\bullet U}\hat{\bSigma}_{U}^+ - \bSigma_{\bullet U} \bSigma_{U}^+\|_{2 \rightarrow \infty}  > t) \leq c_{13}p^{k+1}k \exp\left( -\left( \frac{c_{16}t n^{1/2}}{k^2} - 1 \right)_+^2  \right ).
\end{equation*}

\subsection{Proofs of Finding Best Subset}
\label{sec:subset_correctness_appdx}
\textcolor{revision}{We give proofs that our greedy (\Cref{alg:greedy}) and swapping (\Cref{alg:swap}) algorithms find the best subset in certain settings. Our language suggests that there is one best subset, but all the arguments hold if multiple subsets are tied for having the lowest CSS objective value.}

\subsubsection{Diagonal Covariance}
\textcolor{revision}{If $\bSigma$ is a diagonal matrix, then the CSS objective for a subset $S$ is simply $\sum_{j \not \in S} \bSigma_{jj}$. It is clear this is minimized when $S$ is the set of variables with the largest variances.}\newline  

\noindent \textbf{Greedy:} \textcolor{revision}{\Cref{lem:residual_cov_update} implies that our greedy algorithm will select the remaining variable with highest variance on each iteration.} \newline 

\noindent\textbf{Swapping:} \textcolor{revision}{\Cref{lem:residual_cov_update} implies that our swapping algorithm can always improve the objective by swapping out a variable in the selected set that does not have one of the top $k$ variances for one that does.}  

\subsubsection{Perfect Reconstruction}

\textcolor{revision}{If $\bSigma$ is such that there exists a set of $k$ variables that perfectly linearly reconstruct the remaining, then all the variables are in the span of these $k$ variables. This means all the variables live in a $q$-dimensional linear subspace where $q \leq k$.} \newline 

\noindent \textcolor{revision}{\textbf{Greedy}: On the $i$-th iteration of our greedy algorithm, if the dimension of the subspace spanned by the previous selections is $ < q$, then there must be at least one remaining variable that is not yet perfectly linearly reconstructed by our selections so far (otherwise the variables would live in a linear subspace of dimension $< q$). By \Cref{lem:residual_cov_update} our greedy algorithm will select one such variable on the $i$th iteration, and therefore the dimension of the subspace spanned by our selections must increase by one (because we've added something linearly independent to our currently selected subset). Thus, by the end of the $k$th iteration we must have selected a subset that spans a $q$-dimensional subspace. Our selected variables must necessarily act as a basis of the $q$-dimensional subspace that all the variables live in, and the greedy algorithm will thus achieve an objective value of zero.} \newline 

\noindent \textcolor{revision}{\textbf{Swapping}: For our swapping algorithm, if there is a size-$k$ subset $S$ that spans a subspace of dimension $<q \leq k$, then there must be (1) a variable $j_1$ in that size-$k$ subset that is perfectly linearly reconstructed by the remaining $k-1$ variables and (2) some variable $j_2$ not in the subset that we cannot perfectly linearly reconstruct with variables in the subset (otherwise all our variables would live in a $< q$ dimensional subspace). \Cref{lem:residual_cov_update} tells us that swapping variable $j_1$ out of $S$ and $j_2$ into $S$ improves the CSS objective. It also increases the dimension of the subspace spanned by the variables in $S$ by one. Thus, our swapping algorithm will swap variables out until the dimension of the subspace spanned by variables in the selected subset is $q$, and by the same reasoning as for our greedy algorithm, the swapping algorithm will eventually achieve a CSS objective of zero.}

\subsubsection{Block Diagonal}

\textcolor{revision}{If $\bSigma$ is a block diagonal correlation matrix with blocks $M_i \in \mathbb{R}^{q \times q}$, and each block has a row/column with square $L^2$ norm $ > q/2$, then the best subset is one that consists of the variable from each block that has the largest row/column norm in the block. To see why, consider a subset $S$ that does not have a representative from each block. In particular, it must be missing a representative from some block $M_{i_1}$ and have at least two representatives from some block $M_{i_2}$. If we remove all the representatives from block $M_{i_2}$ then \Cref{lem:residual_cov_update} tells us that the CSS objective can increase by at most $q$ (because only $q$ unit variance variables are in the block, and those variables are independent of all the variables in the other blocks). If we then add back in the two variables with largest row/column norm from block $M_{i_1}$ and $M_{i_2}$ then \Cref{lem:residual_cov_update} tells us that the CSS objective will decrease by strictly more than $q$. This new subset has less variables than our original subset and a lower CSS objective. Thus, the best subset must have exactly one representative from each block. \Cref{lem:residual_cov_update} implies it should be the variable with the largest row/column norm from each block.} \newline 

\noindent \textcolor{revision}{\textbf{Greedy: } We can prove the greedy algorithm finds the optimal subset by induction. On the first iteration, it is trivially true that the greedy algorithm will select a variable from a block it hasn't selected from yet. \Cref{lem:residual_cov_update} tells us that, for whatever block it selects from, it will select the variable with largest row/column norm. Suppose now that up to the $(i-1)$st iteration, our greedy algorithm has never selected two variables from the same block and has always has selected the variable in each block with the largest row/column norm. Then on the $i$th iteration, if the greedy algorithm selects a variable from a new block with largest row/column norm, \Cref{lem:residual_cov_update} tells us that it will reduce the CSS objective by strictly more than $q/2$. If it selects a variable from a block it has previously selected from however, it can reduce the CSS objective by only strictly less than $ q/2$, because a variable can only further reduce the reconstruction error for variables in its own block, and \Cref{lem:residual_cov_update} tells us that the cumulative reconstruction error for a block that already has a selected representative is strictly less than $q/2$. Thus on the $i$th iteration our greedy algorithm will select a variable from a new block, and \Cref{lem:residual_cov_update} guarantees that it will select a variable that has largest row/column norm in said block. This is sufficient to imply that the greedy algorithm will find the best subset.}

\end{appendix}

\end{document}